\documentclass[lettersize, journal]{IEEEtran}
\usepackage{amsmath,amsfonts}
\usepackage{algorithmic}
\usepackage[ruled,vlined,algo2e]{algorithm2e}
\usepackage{algorithm}
\usepackage{array}
\usepackage[caption=false,font=normalsize,labelfont=sf,textfont=sf]{subfig}
\usepackage{textcomp}
\usepackage{stfloats}
\usepackage{url}
\usepackage{verbatim}
\usepackage{graphicx}
\usepackage{cite}
\usepackage{multirow}
\usepackage{xcolor}
\usepackage{placeins}
\usepackage{balance}
\usepackage[hidelinks,hypertexnames=false]{hyperref}
\newcommand{\figureBelowMargin}{\vspace{0ex}}

\DeclareMathOperator*{\argmin}{arg\,min} 
\newtheorem{theorem}{\bf Theorem}[section]

\newtheorem{definition}{\bf Definition}

\begin{document}

\markboth{IEEE Transactions on Knowledge and Data Engineering}%
{Zhong \MakeLowercase{\textit{et al.}}: EnhanceGraph}

\title{EnhanceGraph: A Continuously Enhanced Graph-based Index for High-dimensional Approximate Nearest Neighbor Search}

\author{Xiaoyao~Zhong, Jiabao~Jin, 
	Peng~Cheng,~\IEEEmembership{Member,~IEEE,} Mingyu~Yang, Haoyang~Li, Zhitao~Shen, Jingkuan~Song, and~Heng~Tao~Shen~\IEEEmembership{Fellow,~IEEE,}
	\IEEEcompsocitemizethanks{
		\IEEEcompsocthanksitem X. Zhong is with the School of Computer Science and Technology, Tongji University, and Ant Group, Shanghai, China, Email:  zhongxiaoyao.zxy@antgroup.com.\protect\\
		\IEEEcompsocthanksitem J. Jin and Z. Shen are with Ant Group, Shanghai, China, Emails: \{jinjiabao.jjb, zhitao.szt\}@antgroup.com.\protect\\
		\IEEEcompsocthanksitem P. Cheng, J. Song and H.T. Shen are with
		the School of Computer Science and Technology, Tongji University, Shanghai, China, Emails:
		cspcheng@tongji.edu.cn; jingkuan.song@gmail.com; shenhengtao@hotmail.com. \protect\\
		\IEEEcompsocthanksitem M. Yang is with Ant Group, Shanghai, China, and HKUST (GZ), Guangzhou, China, Email: yangming.ymy@antgroup.com.\protect\\
		\IEEEcompsocthanksitem H. Li is with PolyU, Hong Kong SAR, China, Email: haoyang-comp.li@polyu.edu.hk.\protect\\
		Corresponding authors: P. Cheng and H. T. Shen.
	} 
}

\maketitle

\begin{abstract}
	Recently, Approximate Nearest Neighbor Search in high-dimensional vector spaces has garnered considerable attention due to the rapid advancement of deep learning techniques. We observed that a substantial amount of search and construction logs are generated throughout the lifespan of a graph-based index. However, these two types of valuable logs are not fully exploited due to the static nature of existing indexes. We present the EnhanceGraph framework, which integrates two types of logs into a novel structure called a conjugate graph. The conjugate graph is then used to improve search quality. Through theoretical analyses and observations of the limitations of graph-based indexes, we propose several optimization methods. For the search logs, the conjugate graph stores the edges from local optima to global optima to enhance routing to the nearest neighbor. For the construction logs, the conjugate graph stores the pruned edges from the proximity graph to enhance retrieving of $k$ nearest neighbors. Across public and real-world industrial datasets, EnhanceGraph improves search accuracy; the largest observed Recall@1 increase is from 41.74\% to 93.42\%, with limited supplementary-stage overhead.
\end{abstract}


\section{Introduction}

\IEEEPARstart{W}{ith} the rapid development of deep learning in various fields, it has become a common trend to use vectors for representing entities. Many applications now require finding the most similar points of a given query point in a large database, known as nearest neighbor search problem, which is widely used in scenarios such as facial recognition~\cite{FaceRecognition}, recommendation systems~\cite{recommendations}, and information retrieval~\cite{Apple2023filter}. Due to the ``Curse of Dimensionality''~\cite{verleysen2005curse, indyk1998approximate}, finding accurate nearest neighbors in high-dimensional data sets is often computationally intensive. Then, Approximate Nearest Neighbor Search (ANNS) has gained significant attention as it can retrieve the nearest neighbors with acceptable deviations under limited space and time constraints.

There have been many ANNS methods recently, such as graph-based methods~\cite{Harwood2016FANNG,fu2019NSG, Gollapudi2023filterDISKANN, Malkov2020hnsw},  tree-based methods~\cite{arya1998optimal, ram2019kdtree, Chen2021SPANN, Apple2023filter} and hashing-based methods~\cite{lv2007multi, Li2018LearningHash, zhao2023LSHAPG}. Numerous studies~\cite{Li2019DPG, wang2021survey} demonstrate that graph-based algorithms show superior search performance. Graph-based methods first construct a \textit{proximity graph}, where each node represents a base data point in the database, and the edges connect similar points. To find the nearest neighbors of a query point, they usually pick a start point on the graph index, then greedily traverse to the target nearest point to the query point. In addition, $k$-Nearest Neighbors ($k$-NNs) Graph~\cite{Hajebi2011GNNS, Jin2014IEH} is used to maintain the connections of similar points, where a node $u$ is connected to a node $v$ with a directed edge if $v$ is one of the $k$-NNs of $u$.

\begin{figure}[t!]\centering
	\scalebox{0.6}[0.6]{\includegraphics{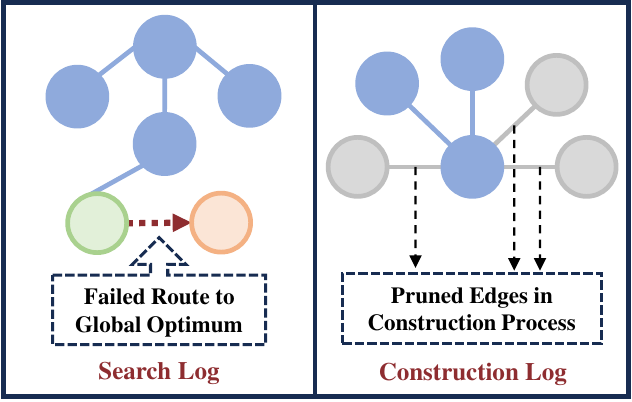}}
	\caption{\small Example of Search and Construction Log. We use the search log to add edges that can route from local optimum (searched approximate nearest neighbor) to the global optimum (exact nearest neighbor), enhancing recall of the nearest neighbor. The construction log maintains pruned nearest neighbors to further improve recall of nearest neighbors.}
	\label{fig:intuition}
\end{figure}

To reduce space usage and improve search efficiency, recent studies apply Relative Neighborhood Graph (RNG) ~\cite{Godfried1980RNG, Li2019DPG, Harwood2016FANNG, Malkov2020hnsw} and Monotonic Relative Neighborhood Graph (MRNG) ~\cite{fu2019NSG,fu2019NSSG, Jayaram2019diskann, Gollapudi2023filterDISKANN} edge pruning strategies to remove edges from the $k$-NNs Graph when building the proximity graph index.
However, due to the pruning of edges, the queries on the proximity graph may return results with some deviations or even fail to retrieve the nearest neighbors. In addition, the graph-based indexes will stay static after construction. From our observations on facial recognition services, many users experience repeated failures to be recognized using HNSW~\cite{Malkov2020hnsw}. In other words, the graph-based indexes built with the existing methods suffer from ``\textit{flaws}'' that prevent them from accurately retrieving the nearest neighbors of the queried points. \textit{How to fix the flaws such that the accuracy of the graph-based indexes can be improved with acceptable space costs is a long-standing problem for the ANNS topic.} According to our observations, the search log and construction log have not been well used to refine the graph-based indexes in the existing studies.

Specifically, the \textit{search log} is not utilized due to the static graph structure. The graph will remain \textit{static} once all dynamic operations (e.g., insert or delete) are completed. If a historical search query cannot find the nearest neighbor on the existing static index, then other similar future queries  will also fail to find it. However, a vast amount of historical query data exists in many industrial applications. Through analysis of a large amount of data, we have observed that similar query points tend to converge to the same local optimum (i.e. the nearest neighbor in search result). It allows us to leverage the search log from existing historical queries or generated queries, providing an opportunity for refining of the graph-based index.

In addition, a significant amount of \textit{construction log} is not effectively utilized for index optimization. During the construction process, each base point in the database is searched for its $k$-NNs. With edge pruning strategies, not all $k$-NNs are included in proximity graph. However, they have the potential to be found as $k$-NNs. The existing graph-based methods can hardly localize the accurate $k$-NNs, which is mainly due to the inability to distinguish between edges that contribute to the search process. Simply removing edges directly increases the uncertainty of locating the $k$-NNs.

\noindent \underline{\textit{Our Idea:}}
In the existing studies, the graph-based indexes typically remain static once built. We find that the search log can be used to detect flaws in the graph structure, and the construction log can be used to supplement the missing $k$-NNs for the proximity graph. In this paper, we propose a framework, namely \textit{EnhanceGraph}, to utilize search and construction logs to enhance graph-based indexes with an auxiliary conjugate graph, enabling a significant improvement on query accuracy with an acceptable increase in space costs.

Specifically, as shown in Figure \ref{fig:intuition}, in EnhanceGraph, we use the conjugate graph to store the pruned edges of proximity graph that are valuable for the search process.
When constructing the graph index, we first add the pruned edges from proximity graph to the conjugate graph.  Then, we heuristically generate $k$ training queries for each base data point on the proximity graph, which can be used to detect flaws in the graph structure (i.e., local optima). Using the generated queries along with historical search log, we add the missing edges that can route local optima to global optimum (i.e., the nearest neighbor) in the conjugate graph such that the flaws can be fixed.
For new queries, we take a small step in the conjugate graph to route from  local optima to  global optima. This repair reduces repeated convergence to the same local optimum for logged failures and future queries with the same local-optimum pattern. Once we find the global optimum with high probability, the conjugate graph can further return other nearest neighbors accurately and efficiently.
To summarize, we make the following contributions:
\begin{itemize}
	\item We introduce the EnhanceGraph framework, which is the first to leverage the search and construction logs to continuously enhance the search process in Section \ref{sec:framework}.
	\item We discuss how to address flaws on graph-based indexes and the optimization opportunities based on observations and theoretical analysis in Section \ref{sec:solution1}.
	\item We propose two optimizations based on the conjugate graph to improve accuracy in search process in Section \ref{sec:solution2}.
	\item We conduct extensive experiments on both public datasets and an anonymous industrial dataset in Section \ref{sec:experimental}.
\end{itemize}

\section{Preliminary}
\label{sec:definition}

\begin{definition} ($k$-NNs Search)
	Given a dataset $D$ and a query point $x_q$, along with a positive integer $k$, the $k$-NNs search aims to find a set $NN_k(x_q)$ containing $k$ points, such that for all $x_r \in NN_k(x_q)$ and all $x_b \in D \setminus NN_k(x_q)$, the condition $dis(x_r, x_q) \leq dis(x_b, x_q)$ holds. Here, $dis(x_1, x_2)$ denotes the distance between two points in a specific metric space, such as Euclidean or Cosine distance.
	
\end{definition}
However, due to the high dimensionality of the data, it is difficult to find accurate $k$-NNs efficiently. Recent studies \cite{gao2023adsample, mrq, sindi} focus on Approximate Nearest Neighbors Search (ANNS), which only requires searching approximate $k$-NNs. For ANNS, the evaluation of search results is based on the recall rate. Let $NN_k(x_q)$ represent the accurate set of $k$-NNs, and $ANN_k(x_q)$ represent the approximate set of $k$-NNs. The recall rate is then defined as $Recall@k = \frac{|ANN_k(x_q) \cap NN_k(x_q)|}{k}$. The nearest neighbor of the query point $x_q$ within the dataset $D$ is denoted as $x_g$, where $x_g = \arg\min_{x_b \in D} dis(x_b, x_q)$.

\begin{definition} (Proximity Graph)
	Given a dataset $D$, the proximity graph can be denoted as $G = (V, E)$, where each node $v \in V$ represents a base point $x_b \in D$.
\end{definition}
In general, each node in a proximity graph is connected to points that are close to it.  

\section{EnhanceGraph Framework}
\label{sec:framework}

\subsection{Search on Proximity Graph}
\begin{algorithm}[t!]
	\DontPrintSemicolon
	\KwIn{proximity graph $G$, initial nodes $I$, query point $x_q$, search parameters $L$, $k$}
	\KwOut{Approximate $k$-NNs of $x_q$ in G, visited nodes $V$}
	initialize a candidate set $C$ as a maximum-heap with size of $L$\;
	
	add the nodes in $I$ into $C$ \;
	
	\While{$C$ has un-expanded nodes} {
		$c \leftarrow $ closest un-expanded nodes in $C$ \;
		
		$V \leftarrow V \cup \{ c \}$ \;
		
		calculate the distance between query $x_q$ and each neighbors of $c$  \;
		
		insert the neighbors into maximum-heap $C$ and $V$ \;
	}
	\Return{$k$ nearest points in $C$, visited points $V$} \;
	\caption{Greedy Search}
	\label{algo:greedy}
\end{algorithm}

Algorithm \ref{algo:greedy} illustrates the greedy search~\cite{Jayaram2019diskann, fu2019NSG, Malkov2020hnsw} process of finding the approximate $k$-NNs on a proximity graph. To begin the search, we initialize a maximum heap of size $L$ as the candidate set (Line 1).  In most cases, $k$ is fixed, whereas users can adjust the heap size $L$ to balance accuracy and efficiency. The heap is ordered based on distance to facilitate the quick retrieval of the $k \le L$ nearest points. Next, we initiate the greedy search from the given initial nodes (Line 2) and iteratively explore each point $c$ in the candidate set $C$ that has not yet been expanded, i.e., whose neighbors have not been visited (Lines 3-5). If the distance of some neighbors is less than that of the node with the largest distance in the current candidate set, the node is popped out and the neighbors are inserted into the candidate set (Lines 6-7). Finally, when all nodes in the candidate set have been expanded, we return the $k$ nearest nodes in the candidate set along with the set of visited nodes $V$.

\begin{definition} (Local Optimum of Greedy Search)
	Given a dataset $D$, a proximity graph $G$ constructed on $D$, and a query point $x_q$. Let $ANN_k(x_q)$ represent the search result of conducting GreedySearch on $G$. Then the local optimum is $x_l = \argmin_{x \in ANN_k(x_q)} dis(x, x_q)$.
	\label{def:local_optimum}
\end{definition}

\begin{figure}[t!]\centering
	\scalebox{0.6}[0.6]{\includegraphics{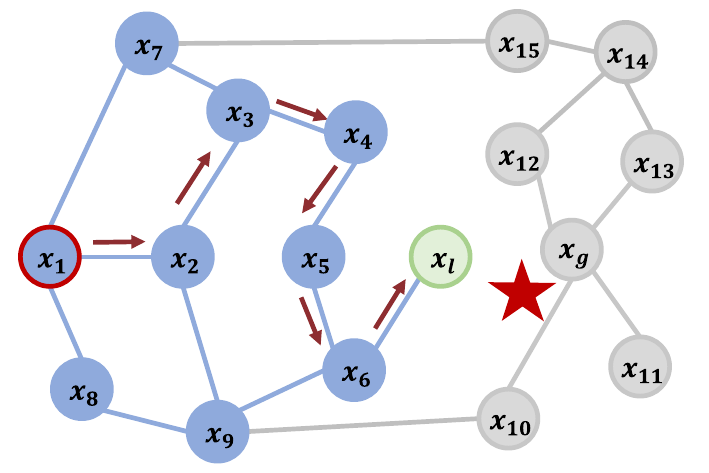}}\vspace{-1ex}
	\caption{\small Example of Searching on Proximity Graph.}
	\label{fig:example_2_information}
\end{figure}

\textit{Example 1. Figure \ref{fig:example_2_information} shows an example of a proximity graph $G$ with $17$ base points $x_l, x_g, x_1, \ldots, x_{15}$. The red pentagram represents the query. The blue circles represent the visited points, while the gray circles represent points that cannot be searched. The red arrow shows the routing process. We demonstrate an example search process on the graph: $GreedySearch(G, {x_1}, x_q, 4, 4)$. We first expand the starting point $x_1$ by inserting its three neighbors $x_2, x_7, x_8$ into candidate list $C$. Currently, the $C=\langle x_2, x_7, x_8, x_1 \rangle$. Then, we expand $x_2$ and insert $x_3, x_9$. Since the capacity limit of the candidate list is $4$, $x_1$ and $x_8$ are popped out. Currently, the $C=\langle x_3, x_9, x_2, x_7 \rangle$. Then, when expanding $x_3$, $x_4$ is inserted and $x_7$ is popped out, even if it has a closer neighbor $x_{15}$. Finally, the search process early converges to a local optimum $x_l$ with $C = \langle x_l, x_6, x_4, x_5 \rangle$. Then, the nearest neighbor $x_g$ and other $k$-NNs $x_{10}, x_{11}, x_{12}, x_{13}$ cannot be found.}

\textit{Why not avoid the dilemma by choosing a better starting point (e.g., $x_{12}$)?} Many studies~\cite{zhao2023LSHAPG, Malkov2020hnsw} propose different starting point selection strategies. However, they still can't guarantee that the search process will always converge to the global optimum, either theoretically or experimentally. We replicate it experimentally in Section \ref{sec:experimental}. The state-of-the-art algorithm HNSW~\cite{Malkov2020hnsw} still hardly achieve high $Recall@1$, especially when the candidate size is limited.

\textit{Why not avoid the dilemma by adding more edges to the graph (e.g., $(x_6,x_g)$)?} Naively adding edges in proximity graph can severely decrease the search speed. It is because each iteration of the greedy search requires the computation of distances to a large number of irrelevant points. Actually, many existing studies \cite{fu2019NSG, Malkov2020hnsw, Jayaram2019diskann} use edge selection strategies to avoid adding redundant edges in the graph. Thus, routing edges and non-routing edges serve different purposes. Routing edges guide greedy search, whereas non-routing edges can be checked after the proximity-graph search reaches a local neighborhood.

\begin{figure*}[t!]\centering\vspace{5ex}
	\scalebox{0.45}[0.45]{\includegraphics{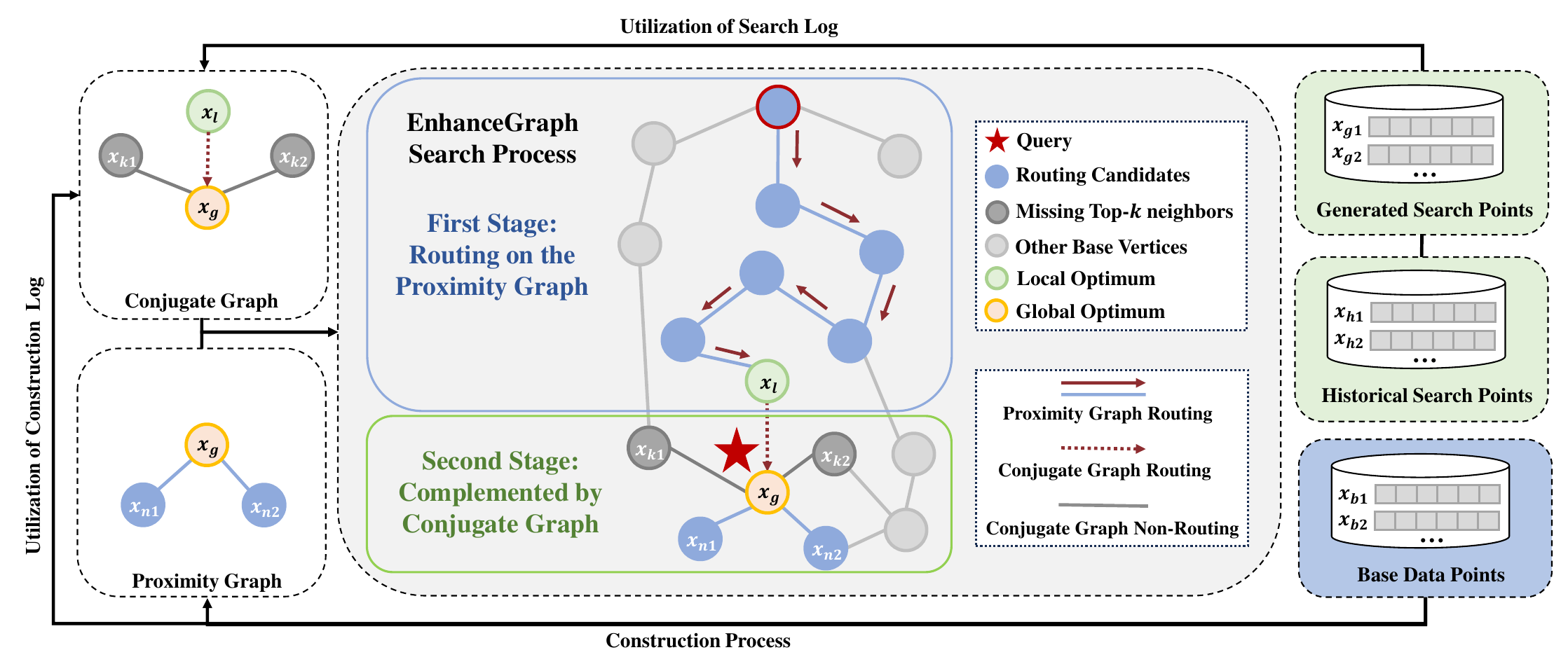}}\vspace{-1ex}
	\caption{\small A Pipeline of the EnhanceGraph Framework. We utilize both search and construction log to  construct a conjugate graph. A query is initially routed through the proximity graph. Subsequently, both routing and non-routing edges in the conjugate graph are employed to navigate from the local optimum in the proximity graph to the global optimum and to complement the missing $k$-NNs.}
	\label{fig:framework}
\end{figure*}

\subsection{Routing Edge and Non-routing Edge}

We distinguish the edges of the proximity graph into two categories: \textit{routing edges} and \textit{non-routing edges}. Routing edges connect to the next nearest neighbor in candidate set and are used to quickly locate the region near the query during the search process for finding the final nearest neighbor. Non-routing  edges do not change the next nearest neighbor in candidate set and are used to find the $k$-NNs. However, during the incremental edge pruning process in the construction of proximity graph, it is inevitable to encounter situations where these edges are not recognized.

\noindent \underline{\textit{Absence of Routing Edges. }} Proximity-graph indexes prune candidate edges to make top-1 routing more efficient~\cite{fu2019NSG, Malkov2020hnsw, Jayaram2019diskann}. Such pruning can remove close edges that are less useful for routing, weakening the graph as a full $k$-NNs approximation. As a result, the pruned graph does not guarantee finding the nearest neighbor for queries of non-base points. Some edges may be redundant for base point searches but important for non-base points.

\noindent \underline{\textit{Absence of Non-Routing Edges. }} Even if we can accurately find the global nearest neighbor of the queried point, $Recall@k$ still cannot be guaranteed. Because the current index construction algorithms first greedily search for the approximate $k$-NNs of the current base point and then prune edges according to specific criteria. For example, the pruning strategies may remove the longest edges of all possible triangles (i.e., RNG~\cite{Godfried1980RNG} strategy), or ensure that the angle between any two edges is greater than 60 degrees (i.e., MRNG~\cite{fu2019NSG} strategy). These non-routing edges only affect search performance during the routing process. In fact, we can utilize them in another search phase. Once we have routed to the vicinity of the query, we can use non-routing edges to quickly locate accurate $k$-NNs.

\subsection{Our Framework}

\noindent \underline{\textit{The Utilization of Search Log. }}
There are many methods~\cite{peng2023taumg, Malkov2020hnsw, fu2019NSSG} attempting to preserve more routing edges. However, after the dynamic update (e.g., insertion or deletion) of the index is completed, the index remains static. This leads to certain failed queries consistently failing without any opportunity for correction, a problem that is difficult to identify in generalized methods. On the contrary, in our framework, we adopt a data-driven approach using the \textit{search log}. If we have historical search logs, we can identify the local optimum that most queries converge to. A local optimum arises due to the absence of routing edges from the local optimum to the global optimum. Recurrent local optima can be associated with verified targets, providing explicit repair candidates for future similar queries. Therefore, the search log can be utilized to continuously optimize the graph-based indexes. A useful search-log record identifies the converged local optimum $x_l$ and a verified target $x_g$; $G'$ can keep multiple repair candidates and re-rank them by distance to the current query.

\noindent \underline{\textit{The Utilization of Construction Log. }} Determining which edges are non-routing is relatively easier. To improve Recall@$k$ is to increase the overlap between the graph-based indexes and the full $k$-NNs graph (i.e., $k$-NNs graph contains all non-routing edges). However, the existing edge pruning strategies remove many close neighbors to improve the search speed and decrease the space cost. As a consequence, non-routing edges are inevitably absent. Therefore, the key issue is how to construct an approximate $k$-NNs graph that retains more important non-routing edges and utilizes them in the specific search phase. We can leverage the \textit{construction log} to help us achieve this. During the graph construction process, each base point is repeatedly searched on the graph for its $k$-NNs. However, the search results are only used for pruning. Instead, we can store the pruned edges of each base point into a conjugate graph to complement the proximity graph. These pruned candidates supplement final top-$k$ results without increasing the routing fanout of the base graph.

\begin{algorithm}[t]
	\DontPrintSemicolon
	\KwIn{proximity graph $G$, conjugate graph $G'$, initial nodes $I$, query point $x_q$, search parameters $L$, $k$}
	\KwOut{Approximate $k$-NNs of $x_q$ in G, visited nodes $V$}
	$R, V \leftarrow$ GreedySearch($G, I, x_q, L, k$)\;
	
	$x_l \leftarrow $ the nearest point of $x_q$ in $R$ \;
	
	$x_g \leftarrow $ the nearest point of $x_q$ in $G'[x_l]\cup \{x_l\}$ \;
	
	$V' \leftarrow V \cup G'[x_g] \cup \{x_g\}$ \;
	
	$R' \leftarrow $ top-$k$ nearest points of $x_q$ in $ G'[x_g] \cup \{x_g\} \cup R$ \;

	\Return{Approximate $k$-NNs $R'$, visited nodes $V'$} \;
	\caption{Greedy Search with Conjugate Graph}
	\label{algo:robust_greedy}
\end{algorithm}

\noindent \underline{\textit{Framework. }} In this paper, we propose a framework, namely EnhanceGraph, to utilize both construction and search log to enhance a deployed graph-based index through an auxiliary sidecar graph. As shown in Fig. \ref{fig:framework}, we first use the construction log to build the proximity graph and conjugate graph. Specifically, each base point is searched for its $k$-NNs during the construction process. Then, we insert pruned edges into the conjugate graph and insert the remaining edges into proximity graph. We use generated and historical search logs to further improve the conjugate graph. Specifically, we connect the local optima with their corresponding global optima in the search log. Then, the proximity graph is used for searching a preliminary result, and the conjugate graph is used as enhancement for the result.

As shown in Algorithm \ref{algo:robust_greedy}, we utilize the conjugate graph to enhance the greedy search process. First, we search the proximity graph for preliminary nearest neighbors and the local optima (Lines 1-2). Next, we utilize the connections of those local-global optima pairs that are stored in the conjugate graph based on search log. It is possible to retrieve the global optimum from the neighbors of $G'[x_l]$ (Line 3). When several verified targets are attached to the same local optimum, Line 3 selects the target nearest to the current query. Once we reach the global optimum of the query point, we use its neighbors $G'[x_g]$ in the conjugate graph as the complement of the $k$-NNs (Lines 4-5). Insertions are handled by the underlying graph index and do not invalidate existing $G'$ records. For deletions, the underlying index can use a mark-delete interface and query-time validity check~\cite{hnswlib}; streaming graph indexes similarly filter deleted identifiers and clean stale edges during consolidation or in-place updates~\cite{freshdiskann,ipdiskann}. EnhanceGraph follows this boundary by skipping marked-deleted identifiers in $G'$ and reclaiming stale sidecar edges during periodic compaction.

\section{Opportunities of Enhancement}
\label{sec:solution1}

In this section, we examine the flaws and potential for enhancement in graph-based indexes. First, we then perform a theoretical analysis of the local optimum's location. Based on this theorem and heuristic, we identify several opportunities to enhance the graph structure. Then, we use the Voronoi Diagram to represent the relationship between query points and base points. Based on it, we suggest a log-generation method for identifying cases that are likely to lead to search misses.

\begin{figure}[t!]\centering\vspace{2ex}
	\subfloat{
		\scalebox{0.2}[0.2]{\includegraphics{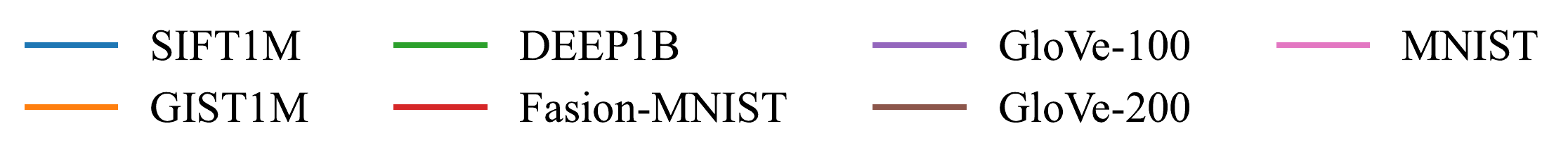}}}\hfill\\\vspace{0ex}
	\addtocounter{subfigure}{-1}
	\subfloat[][{\scriptsize Rank of Local Optimum}]{
		\scalebox{0.18}[0.18]{\includegraphics{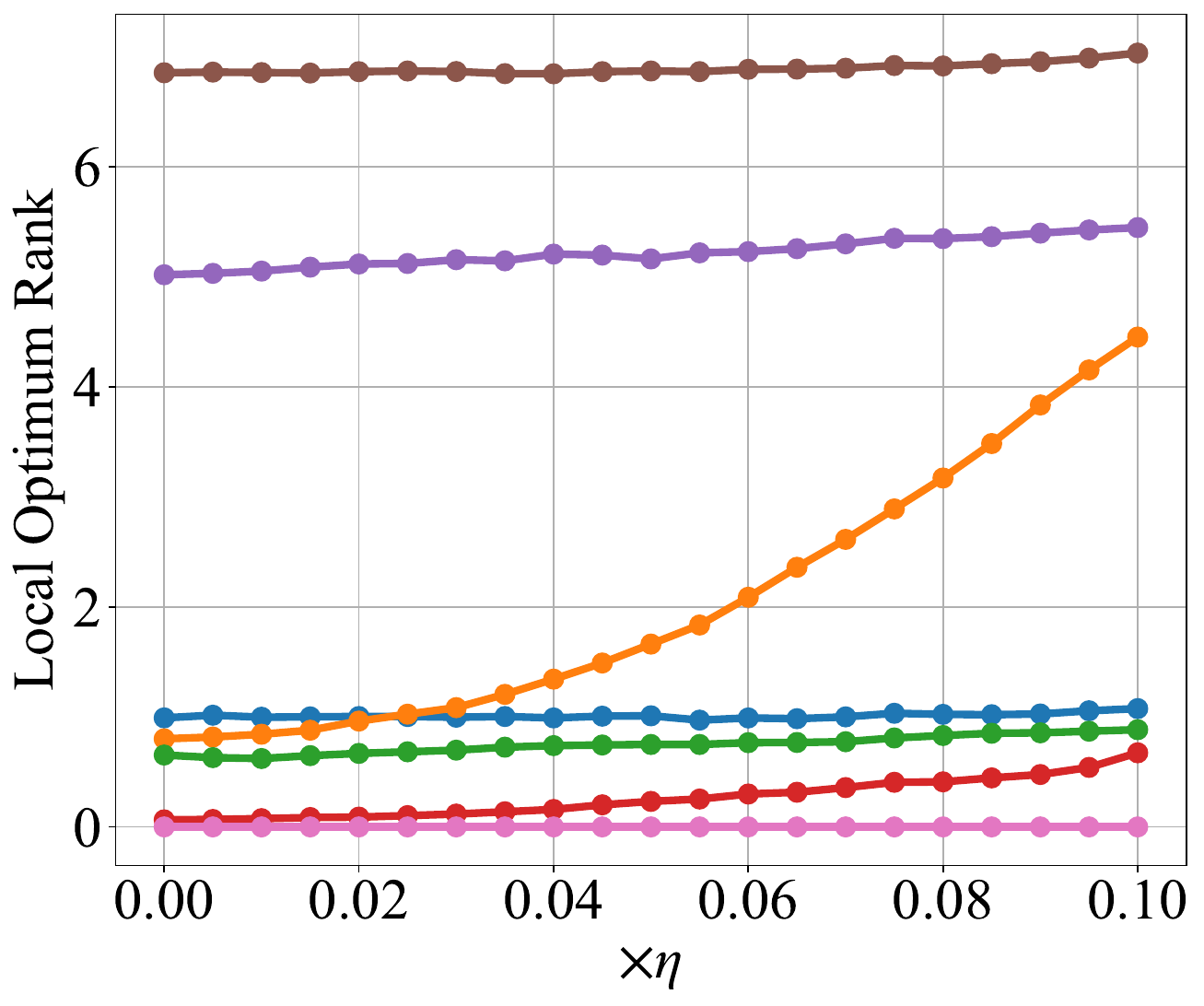}}
		\label{subfig:rank}
	}
	\hspace{-1em}
	\subfloat[][{\scriptsize 20-NNs Overlap Rate}]
	{
		\scalebox{0.18}[0.18]{\includegraphics{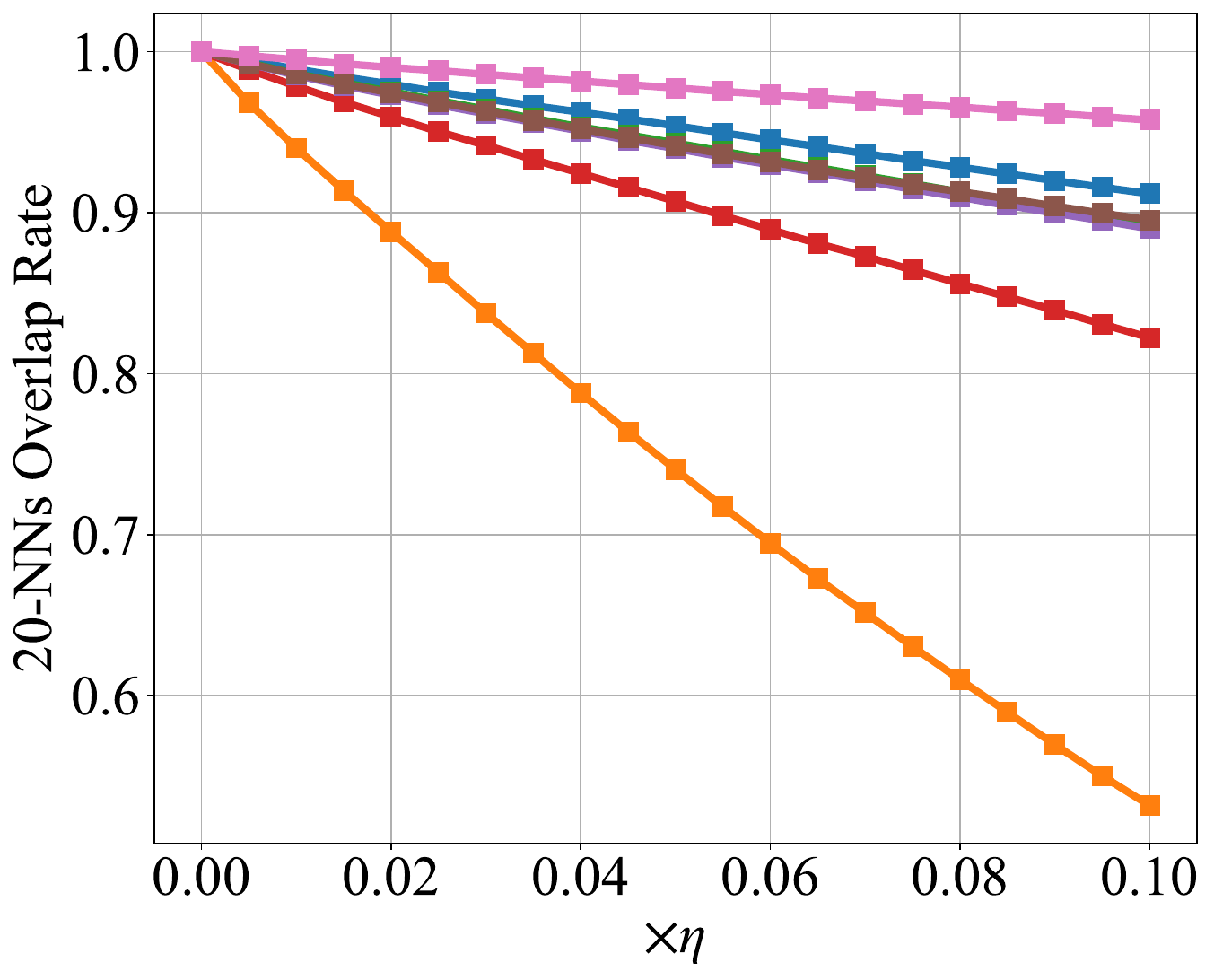}}
		\label{subfig:overlap}
	}
	\caption{\small Rank of Local Optimum and 20-NNs Overlap Rate (Observation 1 and Observation 2). ``Rank'' means the rank of local optimum in the nearest neighbors of global optimum, i.e., the rank of nearest neighbor of global optimum is 1.}
	\label{fig:rank_overlap}
\end{figure}

\subsection{Observations and Opportunities}
In real-world applications (e.g., facial recognition), there are users who fail many times to search their accurate $k$-NNs (e.g., their own facial images). Existing indexes are trapped in these situations due to their \textit{static nature}, which leads to subsequent queries repeating the same mistakes. \textit{Note that, here we say the existing graph indexes have static nature referring to that their graph edges are forbidden to be changed after construction.} However, the abundant search log has not been effectively utilized to detect graph flaws and refine its structure. We heuristically guide how to utilize the search log through several observations and opportunities.

\begin{theorem}
	Given the query point $x_q$, it's accurate $k$-NNs set $NN_{k}(x_q)$ and approximate $k$-NNs set $ANN_{k}(x_q)$, we assume $x_l$ is the local optimum of the approximate $k$-NNs of query point $x_q$. If $Recall@k > 0$, we have $x_l \in NN_{k}(x_q)$.
	
	\label{theo:local_opt_locate}
\end{theorem}
\noindent \textbf{Proof. }
We prove it by contradiction. Assume that there's a point $x_l \notin NN_{k}(x_q)$ that is the local optimum. Given that $Recall@k > 0$, there must be another point $x_l' \in ANN_{k}(x_q)$ in the search result that satisfies $x_l' \in NN_{k}(x_q)$. Since $x_l' \in NN_{k}(x_q)$, we have $dis(x_l', x_q) < dis(x, x_q), \forall x \notin NN_{k}(x_q)$, where $x$ is another base point. Since $x_l \notin NN_{k}(x_q)$, we have  $dis(x_l', x_q) < dis(x_l, x_q)$. According to Definition \ref{def:local_optimum}, we have $x_l$ is not the local optimum, contradicting our initial assumption.

\textbf{Opportunity 1: Local optimum is likely the one of $k$-NNs of the global optimum.}  Any query point $x_q$ is in the neighborhood of its nearest neighbor $x_g$. Since they are both surrounded by the neighbors of $x_g$, we have another heuristic: \textit{the neighbors of query point $x_q$ and the neighbors of its nearest neighbor $x_g$ are highly overlapping.} According to the theorem \ref{theo:local_opt_locate}, as long as the $Recall@k > 0$, the local optimum $x_l$ of the search process must be in the accurate $k$-NNs of the query point $x_q$. Combining the heuristic and theorem, we can infer that $x_l \in NN_k(x_g)$ with high probability.

\noindent \underline{\textit{Observation 1.}} Figure \ref{subfig:rank} shows the trend of ranks, where a smaller rank indicates local optimum is closer to the global optimum. For this observation, we construct probing points by calculating the mean $\eta$ of each dimension of base points and adding positive noise (e.g., $0.06 \times \eta$) to them. Next, we conduct searches on the GIST1M and GLoVe-100 datasets using the VAMANA~\cite{Jayaram2019diskann} index to collect generated search logs. We find that a significant proportion of the searched local optima are neighbors to the global optimum, and they often constitute the closer neighbors (i.e., rank less than $7$).

\noindent \underline{\textit{Observation 2.}} Figure \ref{subfig:overlap} shows the trend of the overlap rate, which represents the proportion of identical neighbors in the 20-NNs of both the query and the base points. In most cases, there is a considerable portion (i.e., greater than $60\%$) of overlap. Besides, the overlap rate decreases as the similarity of the constructed points decreases (i.e., moves to the right of the x-axis).

\textbf{Opportunity 2: Similar points may converge to the same local optimum.} Most graph-based indexes have a fixed starting point for searching to ensure a Monotonic Decrease Path~\cite{fu2019NSG} from that point to any base point~\cite{fu2019NSG,fu2019NSSG,Gollapudi2023filterDISKANN}. It means that similar points may have similar search paths and thus converge to the same local optimum.

\begin{figure}[t!]\centering\vspace{2ex}
	\subfloat{
		\scalebox{0.2}[0.2]{\includegraphics{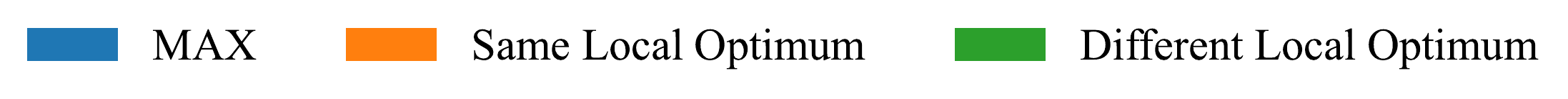}}}\hfill\\\vspace{0ex}
	\addtocounter{subfigure}{-1}
	\subfloat[][{\scriptsize VAMANA(GLoVe-200)}]{
		\scalebox{0.18}[0.18]{
			\includegraphics{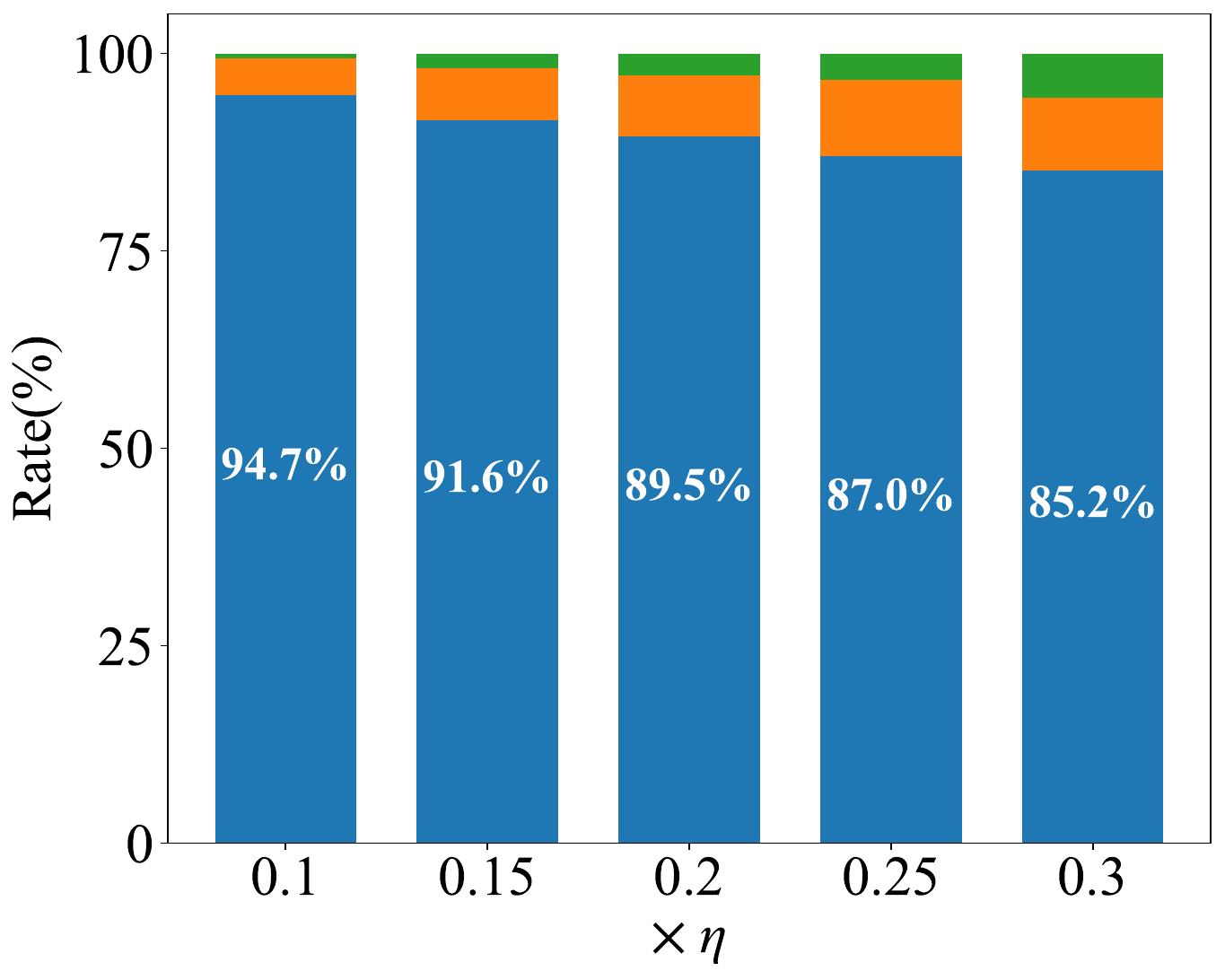}
		}
	}
	\hspace{-1em}
	\subfloat[][{\scriptsize VAMANA(GIST1M)}]{
		\scalebox{0.18}[0.18]{
			\includegraphics{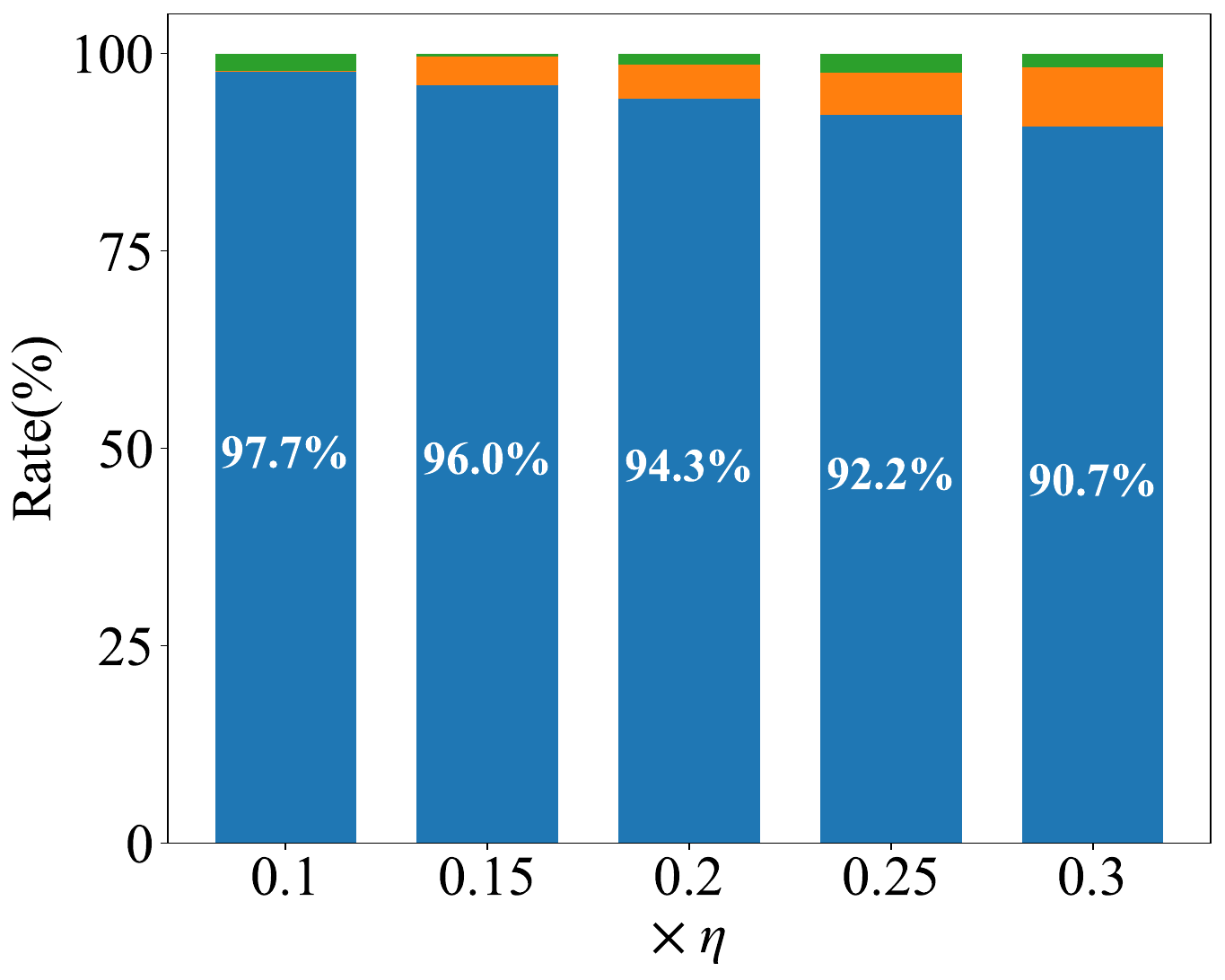}
		}
	}
	\caption{\small Same Local Optimum Rate of Similar Point (Observation 3).
	}
	\label{fig:same_local_optimum_rate}
\end{figure}

\noindent \underline{\textit{Observation 3.}} As shown in Figure \ref{fig:same_local_optimum_rate}, we construct generated points by adding noise sampled from a uniform distribution (e.g. $U(-0.2 \times \eta, 0.2\times\eta)$) and collect generated search logs by performing greedy searches on these constructed points. The blue and orange bars indicate constructed points converging to some same local optima. The blue bar indicates the percentage of the most constructed points converging to one same local optimum.
For example, assume we have 10 constructed points converging to base points with IDs (1, 1, 1, 1, 1, 1, 1, 2, 2, 3). There are two local optima (i.e., IDs 1 and 2) that have at least two constructed points converging to them. The ``MAX'' local optimum ID is 1, with a corresponding percentage of $7/9$, shown in the blue bar. The percentage of $2/9$ for the other local optimum (i.e., ID 2) is displayed in the orange bar and the percentage of $1/9$ for the single local optimum (i.e., ID 3) is displayed in the green bar. We find that a large number of similar points (up to 97.7\%) converge to the same local optimum on two datasets.

\noindent \underline{\textit{Discussion.}} Together with theoretical analyses and data analyses, we conclude that the local optima of similar points are tending to be the $k$-NNs of corresponding base points. Since query points with the same global optimum are similar, we can utilize historical search logs to enhance the performance of online queries to avoid future repeating failures of them. Regrettably, the recall rate requirement for real-world applications (e.g., facial recognition) is usually very high. Consequently, historical search logs may be inadequate, as they only reveal a limited number of local optima. Therefore, we propose an efficient log-generation method to identify flaws in the graph index (i.e., potential local optima). The following subsection describes this log-generation procedure.

\subsection{Generated Search Logs}
\label{appendix:self-generated}

In addition to utilizing historical search logs, we can also construct points and collect generated search logs using heuristic methods. Generated search logs can also continuously optimize the graph by identifying potential flaws. Thus, we first investigate which types of points are more likely to encounter graph flaws by  Voronoi Diagram~\cite{fortune2017voronoi, lee1980delaunay, Aurenhammer1991Voronoi}.

\begin{definition} (Voronoi Diagram~\cite{fortune2017voronoi, lee1980delaunay, Aurenhammer1991Voronoi})
	Given a dataset $D$ consisting of points in $\mathbb{R}^d$ with $d$-dimensional real numbers, the Voronoi cell of any point $x_g \in D$ is represented as $V_D(x_g)=\{x\in \mathbb{R}^d, dis(x_g, x) \le dis( x_b, x), \forall x_b\in D \}$, where each point $x_g$ is noted as a Voronoi site. The set of all Voronoi cells of $D$ is its Voronoi diagram $Vor(D)$.
\end{definition}

\begin{figure}[t!]\centering\vspace{2ex}
	\scalebox{0.5}[0.5]{\includegraphics{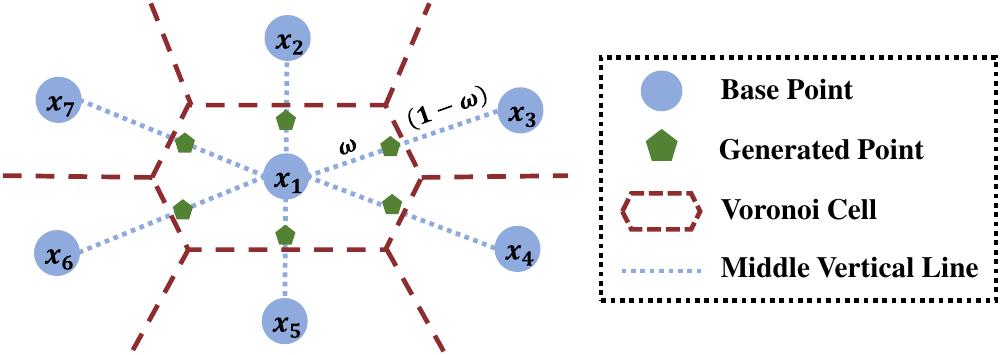}}\vspace{-1ex}
	\caption{\small Example of Constructed Points in Voronoi Diagram.}\vspace{-2ex}
	\label{fig:voronoi_diagram}
\end{figure}

According to this definition, each Voronoi cell $V_D(x_g)$ is a subspace in $\mathbb{R}^d$. Any point that lies within the subspace have $x_g$ as its closest Voronoi site. It means that for any query within this cell, its  nearest neighbor should be the point $x_g$.

We consider each point $x_b$ in the dataset $D$ as a Voronoi site. In a high-dimensional space, each base point and its $k$-NNs are likely to be \textit{locally homogeneous}~\cite{goldman2010locally}. This means that the cells where the $k$-NNs are uniformly distributed in various directions of the cell $V_D(x_b)$. Additionally, for most $k$-NNs of $x_b$, the cells are adjacent to the cell $V_D(x_b)$.

\noindent \underline{\textit{Heuristic.}} We use the following heuristic for log generation: \textit{points near the boundary between two adjacent Voronoi cells are more likely to encounter flaws.} It's because these points have similar distances to the two Voronoi sites, making it easier to deviate from the optimal path during the search process and enter another cell. The base point is guaranteed to be adjacent to the two cells belonging to its nearest neighbor. The $k$-NNs with larger rankings (i.e., larger $k$) may fail to meet the homogeneity requirement.

\noindent \underline{\textit{Approximation.}} To address this problem, our simple approximation is to only consider the $k$-NNs with smaller ranks (i.e., smaller $k$) for point generation. Simultaneously, this approach avoids comparing with distant neighbors. Under the widely used edge pruning strategies~\cite{Godfried1980RNG,fu2019NSG}, there's a high probability to prune the longest edge in any triangle. Thus, in the original proximity graph, the distant neighbors will not have an edge connected to the base point.

Given the base point $x_b$ and its nearest neighbor point $x_k \in NN_k(x_b)$, the constructed point is:
\begin{equation}
	x_e = \omega \cdot x_b + (1 - \omega) \cdot x_k
	\label{eq:generated_vector}
\end{equation}

\noindent where $\omega$ is a parameter used to control the distance between the constructed point and the base point.

As shown in Figure \ref{fig:voronoi_diagram}, the constructed point lies near the perpendicular bisector of the line connecting two base points. To ensure that $x_e \notin V_D(x_k)$, we control $\omega > 0.5$. In most cases (i.e., when the two cells are adjacent), we can ensure that $x_e \in V_D(x_b)$. In other words, given the base point $x_b$, this construction identifies error-prone cases based on its $k$-NNs. Further, by searching on the graph, we can obtain generated search logs to enhance the recall for all query points whose global optimum is the given base point.

\section{Log-based Search Enhancement}
\label{sec:solution2}
In this section, we introduce a log-based enhancement method. We first use generated and historical search logs to update the conjugate graph. Then, with the updated conjugate graph, we implement one-step optimization on the nearest neighbor search. After achieving a high $Recall@1$, the construction log enhanced conjugate graph can quickly complement the result of $k$-NNs search.

\subsection{Optimization on the Nearest
	Neighbor Search}

According to Theorem \ref{theo:local_opt_locate} and Observations 1-3, we can infer that local optima are likely to be among the $k$-NNs of a base point. We can then group queries based on the local optima they converge to. Since the number of queries greatly exceeds $k$, many queries will converge to the same local optimum. This suggests that the routing edge from the local optimum to the global optimum is missing. To address this, we can construct a conjugate graph with edges connecting local optima to their corresponding global optima. After the first-stage search concludes, the conjugate graph directs all queries that converged to local optima towards the global optimum.

\begin{algorithm}[t]
	\DontPrintSemicolon
	\KwIn{proximity graph $G$, conjugate graph $G'$, initial nodes $I$, search parameters $L_2$, $k$, generate parameter $\omega$, $k_g$}
	\KwOut{Updated Conjugate Graph $G'$}
	initialize $E$ as an empty set \;
	
	$E \leftarrow$ historical search result $\{x_l,x_g \}$ with parameter $L_2$\;
	
	\ForEach{$x_b \in G$} {
		$ANN_k(x_b)\leftarrow$ approximate  $k_g$-NNs\;
		
		\ForEach{$x_k \in ANN_k(x_b)$}{
			$x_e \leftarrow \omega \cdot x_b + (1-\omega) \cdot x_k$ \;
			
			$x_l \leftarrow$ GreedySearch($G,I,x_e,L_2,k$) \;
			
			$x_g \leftarrow \argmin_{x \in ANN_k(x_b)\cup \{x_b\}} dis(x, x_e)$ \;
			
			\If{$x_l \ne x_g$} {
				$E \leftarrow E \cup \{x_l, x_g\}$ \;
			}
		}
	}
	\ForEach{$\{ x_l,x_g \} \in E$}{
		insert edge $(x_l, x_g)$ in $G'$ \;
		
	}
	
	\Return{$G'$} \;
	\caption{Conjugate Graph Update}
	\label{algo:conjugate_graph_evolve}
\end{algorithm}

We use two categories of search logs: generated search logs and historical search logs. Following Theorem \ref{theo:local_opt_locate} and Observations 1-3, we collect generated search logs for each base point to identify potential flaws. Historical search logs complement this source by covering observed failures that are not captured when constructing generated search logs. Note that, the global optima of the search log can be obtained through both user complaints and asynchronous computations~\cite{wu2020complaint, koh2017understanding}. For example, in facial recognition payment systems ~\cite{tang2022revisiting, lines2020no}, a failure query may require the user to provide additional information, such as a phone number, to complete the payment process. When historical search logs contain verified failures, we use those logged cases to update $G'$ for the covered failures. When such logs are unavailable or sparse, generated search logs supplement them by exposing additional potential local-optimum failures.

As shown in Algorithm \ref{algo:conjugate_graph_evolve}, the first step is to add the historical search log to the experience set $E$ (Lines 1-2). The experience set is used to record specific local optima found by greedy search with parameter $L_2$ and their corresponding global optima. Note that we emphasized $L_2$, which is the search parameters of the log that converge to the local optimum. When the query is performed, the best enhancement with conjugate graph can be achieved only when search parameter $L = L_2$. When $L \ne L_2$, the search process may converge to a different local optimum. Thus, the routing edges in conjugate graph cannot be utilized. Therefore, we treat $L_2$ as the search budget under which the repair log is collected.

Then, we use the log-generation method described in Section~\ref{appendix:self-generated} to obtain more local optima. Firstly, we collect generated search logs using $k_g$-NNs of each base point, where $k_g$ is the parameter to adjust the number of generated search logs.
Since we can hardly search the accurate $k_g$-NNs for each base point, we use the approximate $k_g$-NNs as replacement (Lines 4-5). It can be quickly obtained by using the construction log stored in the conjugate graph, which we will discuss in Algorithm \ref{algo:construct_graph}. For each neighbor in $k_g$-NNs, we construct $x_e$ with Equation \ref{eq:generated_vector} (Line 6).

After constructing $x_e$, we perform a greedy search to obtain potential local optimum (Line 7). Since $x_e \in V_D(x_b)$ and $V_D(x_b)$ is surrounded by the Voronoi cell of its $k$-NNs, we can quickly obtain an approximate global optimum from these cells (Line 8). Then, we add generated search logs to the experience set $E$ (Lines 7-10). Finally, each element in the experience set $E$ is treated as a directed edge and added to the conjugate graph $G'$ (Lines 11-12). These directed edges provide supplementary repair candidates for nearest-neighbor search under the assumptions and observations discussed above.

Note that, Algorithm \ref{algo:conjugate_graph_evolve} is an approximation based on Opportunity 1. In fact, a small value of $k_g$ is sufficient to achieve higher $Recall@1$. This is because Opportunity 1 states that the local optimum found by the query is likely to be in the $k$-NNs of the global optimum in a high probability. In other words, it is highly probable that $x_l \in NN_k(x_g)$ for the added edges $(x_l,x_g)$. Furthermore, since $k_g$ is small, the number of added edges is also small. Let $M$ be the maximum rank of the local optimum in the $NN_k(x_g)$. Then, the size of added edges is bounded by $O(Mn)$, where $n$ is the number of base points.

\noindent \underline{\textit{Discussion.}} When recall is high, improving it with traditional graph-based index optimization becomes difficult. EnhanceGraph targets local-optimum failures in the current index and applies a supplementary stage after proximity-graph search. This is because we only access the conjugate graph at the end of the search on proximity graph. In the simple one-to-one repair case, the supplementary stage mainly checks candidates associated with the local optimum and the selected global optimum; with multiple repair candidates, the actual extra checks are measured empirically. This overhead is measured separately from the end-to-end Recall-QPS tradeoff in Appendix C.

\subsection{Optimization on $k$-NNs Search} \label{subsec:topk_optimize}

\begin{algorithm}[t]
	\DontPrintSemicolon
	\KwIn{proximity graph $G$, conjugate graph $G'$
		initial nodes $I$, search parameters $L_1$, $k$, maximum degree $R$}
	\KwOut{Proximity Graph $G$, Conjugate Graph $G'$}
	
	\tcp{Phase 1: construct proximity graph}
	\ForEach{$x_b \in G$}{
		$A_b, V_b \leftarrow$ GreedySearch($G,I,x_b,L_1,k$)  \;
		
		$N_b \leftarrow$ Prune($V_b, R$) \;
		
		\ForEach{$x_n \in N_b$}{
			add edges between $x_b$ and $x_n$ in $G$ \;
			
			\If{$|G[x_n]| > R$} {
				$G[x_n] \leftarrow$ Prune($G[x_n], R$)\;
			}
		}
		
		$G'[x_b] \leftarrow $  $A_b$ \;
	}
	
	\tcc{Phase 2: construct conjugate graph}
	
	\ForEach{$x_b \in G$} {
		$G'[x_b] \leftarrow G'[x_b] - G[x_b]$ \;
	}
	
	\Return{$G$, $G'$}
	\caption{Proximity and Conjugate Graph Construction}
	\label{algo:construct_graph}
\end{algorithm}

The $k$-NNs graph can guarantee $Recall@1$ as it's a Monotonic Search Network (MSNET)~\cite{fu2019NSG} (i.e., it can always find a monotonic decreasing path between any two base points). It can also guarantee $Recall@k$ by fully connecting each base point with its nearest neighbors. However, the short edges making forward steps during the greedy search in Algorithm \ref{algo:greedy} are also short on a $k$-NNs graph. In other words, the minimal distance between the query and points in the candidate set decreases slowly, resulting in a large number of iterations in the greedy search. Additionally, due to the high degree of each node, more points need to be explored in each iteration. They both increase the search overhead due to additional distance computations.

Some existing graph-based indexes~\cite{Jayaram2019diskann, fu2019NSG, Malkov2020hnsw} approximate RNG or MRNG by adopting the same edge pruning strategies to remove redundant edges. They strive to carefully select edges in the graph indexes to reduce the number of iterations and shrink the fan out of each iteration. They use incremental insertion for construction, similar to the Algorithm \ref{algo:construct_graph}. Initially, a greedy search is performed for each newly inserted point (Lines 1-2), as shown in Algorithm \ref{algo:greedy}. Then, for the visited list $V$, redundant edges are eliminated based on a specific edge pruning strategy (Line 3). For example, the pruning strategies prune the longest edges of all possible triangles (RNG~\cite{Godfried1980RNG} strategy), or ensure that the angle between any two edges is greater than 60 degrees (MRNG~\cite{fu2019NSG} strategy). Finally, the remaining edges are inserted into the graph (Lines 4-7).

During the construction process, approximate $k$-NNs are found for each base point (Line 2). However, this construction log is not further utilized. Only the pruned set of edges $N_b$ is kept for insertion into the proximity graph (Phase 1). The pruned edges may reduce the navigation efficiency, but they are valuable as non-routing edges. We store them in the conjugate graph $G'$ (Line 8). Finally, we remove redundant edges that already exist in the proximity graph to improve storage efficiency (Lines 9-10).

We use $ANN_k(x_b) $ $= G[x_b] \cup G'[x_b]$ as an approximate $k$-NNs, which has two usages in our framework. First, during generating error-prone points, we use $ANN_k(x_b)$ as the approximate $k$-NNs. Second, improve $Recall@k$ when $Recall@1$ is close to 100\%. As mentioned before, we can assume that there is a large overlap of the $k$-NNs between the query and its corresponding global optimum. Therefore, when the search reaches the global optimum with high probability, the neighbors stored in the conjugate graph can be used to complement the final top-$k$ result.

\noindent \underline{\textit{Discussion.}} Note that, unlike traditional indexes for table data, in-memory indexes~\cite{fu2019NSG,Malkov2020hnsw,peng2023taumg} for vectors need to load all vectors into memory for DCO (Distance Comparison Operation). Disk-based indexes~\cite{Jayaram2019diskann,Chen2021SPANN} also require significant IO operations to load vectors from the disk. The memory footprint of the graph structure is typically much smaller than that of vectors. For example, the memory footprint of a graph index with a maximum degree of 32 is approximately equivalent to adding 32 extra dimensions to each vector. However, for many applications, such as RAG (Retrieval Augmented Generation)~\cite{rag}, the dimension of vectors is large (e.g., commonly used OpenAI embedding models use 1536-dimensional or higher vector representations~\cite{openai_embedding}). Thus, the extra memory footprint of conjugate graph is also acceptable even when memory is limited.

\section{Experimental Study}
\label{sec:experimental}

\subsection{Experiment Setups}

\begin{table}[t]
	\begin{center}
		{
			\caption{\small Datasets.}\vspace{-2ex} \label{tab:datasets}
			\begin{tabular}{l|l|l|l|l}
				{\bf Dataset\quad} & {\bf Dimension } & {\bf  \#Base\quad} & {\bf  \#Query} & {\bf  Metric\quad} \\ \hline \hline
				GIST1M & 960 & 1000K & 1K & Euclidean \\
				
				MNIST & 784 & 60K & 10K & Euclidean \\
				
			    Fashion-MNIST & 784 & 60K & 10K & Euclidean \\
				
				FACE & 256 & 800K & 800K & Inner Product \\
				
				GloVe-200 & 200 & 1183K & 10K & Angular \\
				
				SIFT1M & 128 & 1000K & 10K & Euclidean \\
				
				GloVe-100 & 100 & 1183K & 10K & Angular \\
				
				DEEP1B & 96 & 9990K & 10K & Angular \\
				
				\hline
			\end{tabular}
		}
	\end{center}
\end{table}

\begin{figure*}[t!]\centering
	\subfloat{
		\scalebox{0.35}[0.35]{\includegraphics{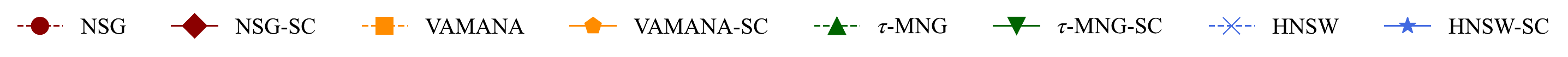}}}\hfill\\
	\addtocounter{subfigure}{-1}
	\figureBelowMargin
	
	\subfloat[][{\scriptsize QPS v.s. Recall@1}]{
		\scalebox{0.18}[0.18]{\includegraphics{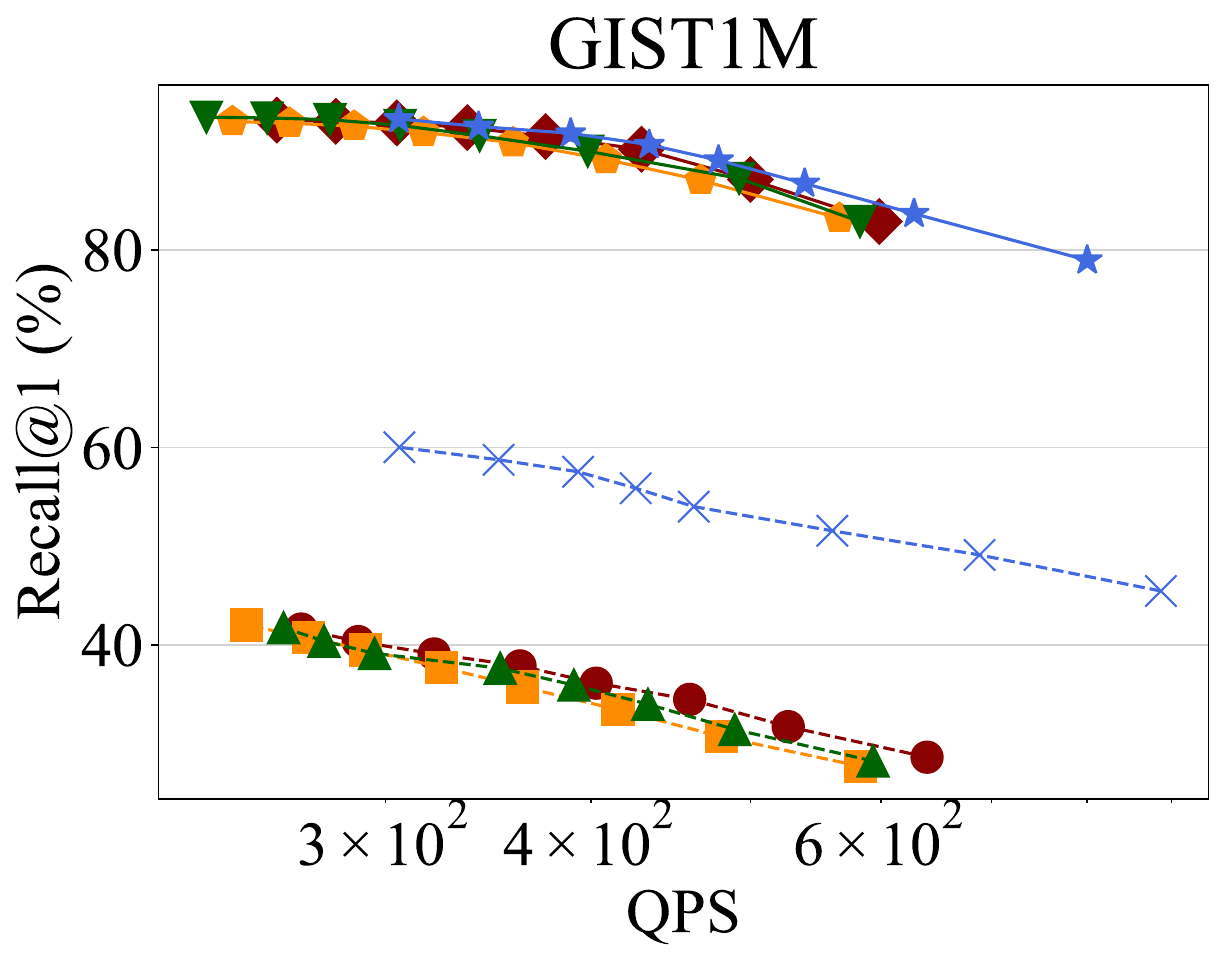}}
		\label{subfig:gist-960-euclidean_K1_var1}}
	\subfloat[][{\scriptsize QPS v.s. Recall@1}]{
		\scalebox{0.18}[0.18]{\includegraphics{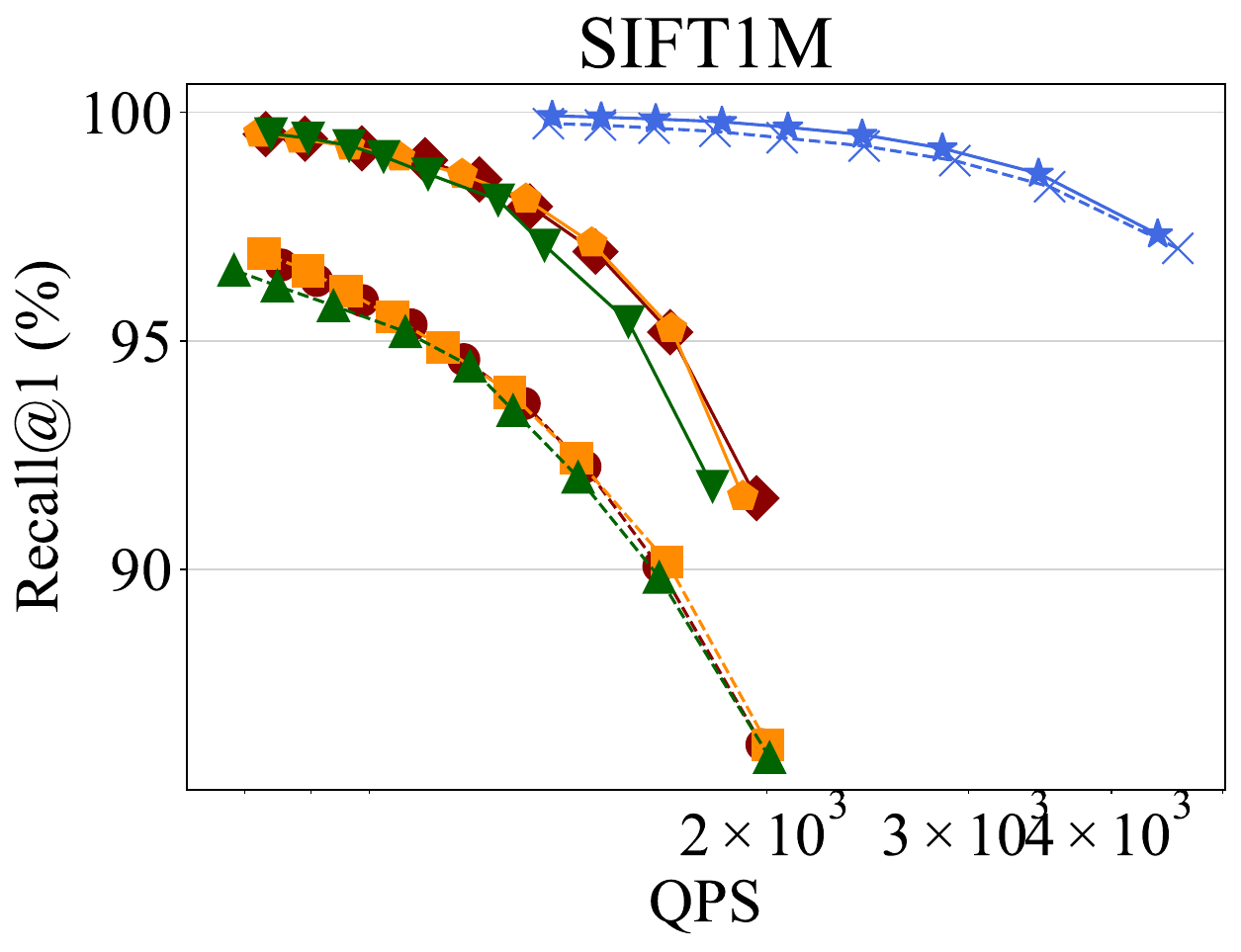}}
		\label{subfig:sift-128-euclidean_K1_var1}}
	\subfloat[][{\scriptsize QPS v.s. Recall@1}]{
		\scalebox{0.18}[0.18]{\includegraphics{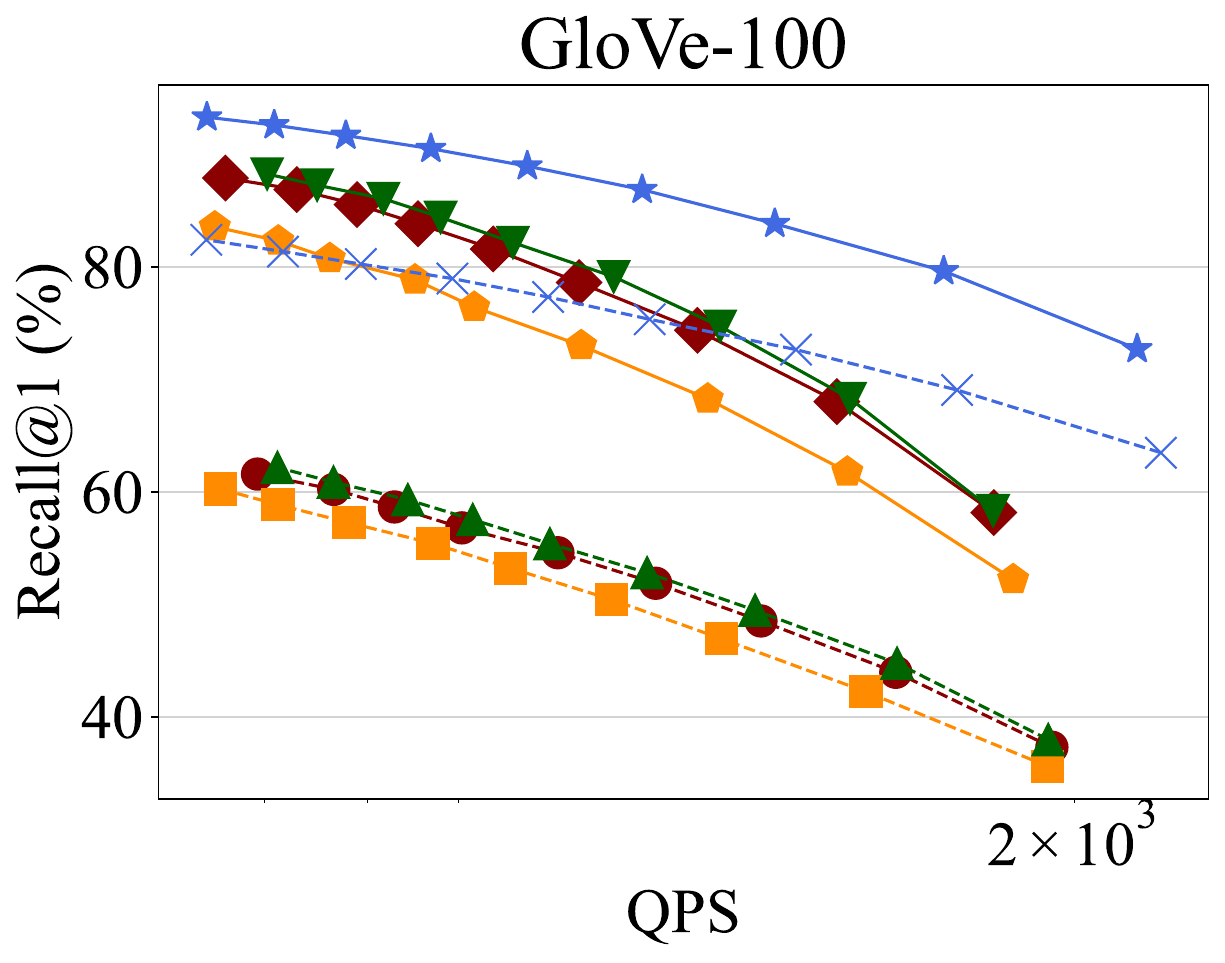}}
		\label{subfig:glove-100-angular_K1_var1}}
	\subfloat[][{\scriptsize QPS v.s. Recall@1}]{
		\scalebox{0.18}[0.18]{\includegraphics{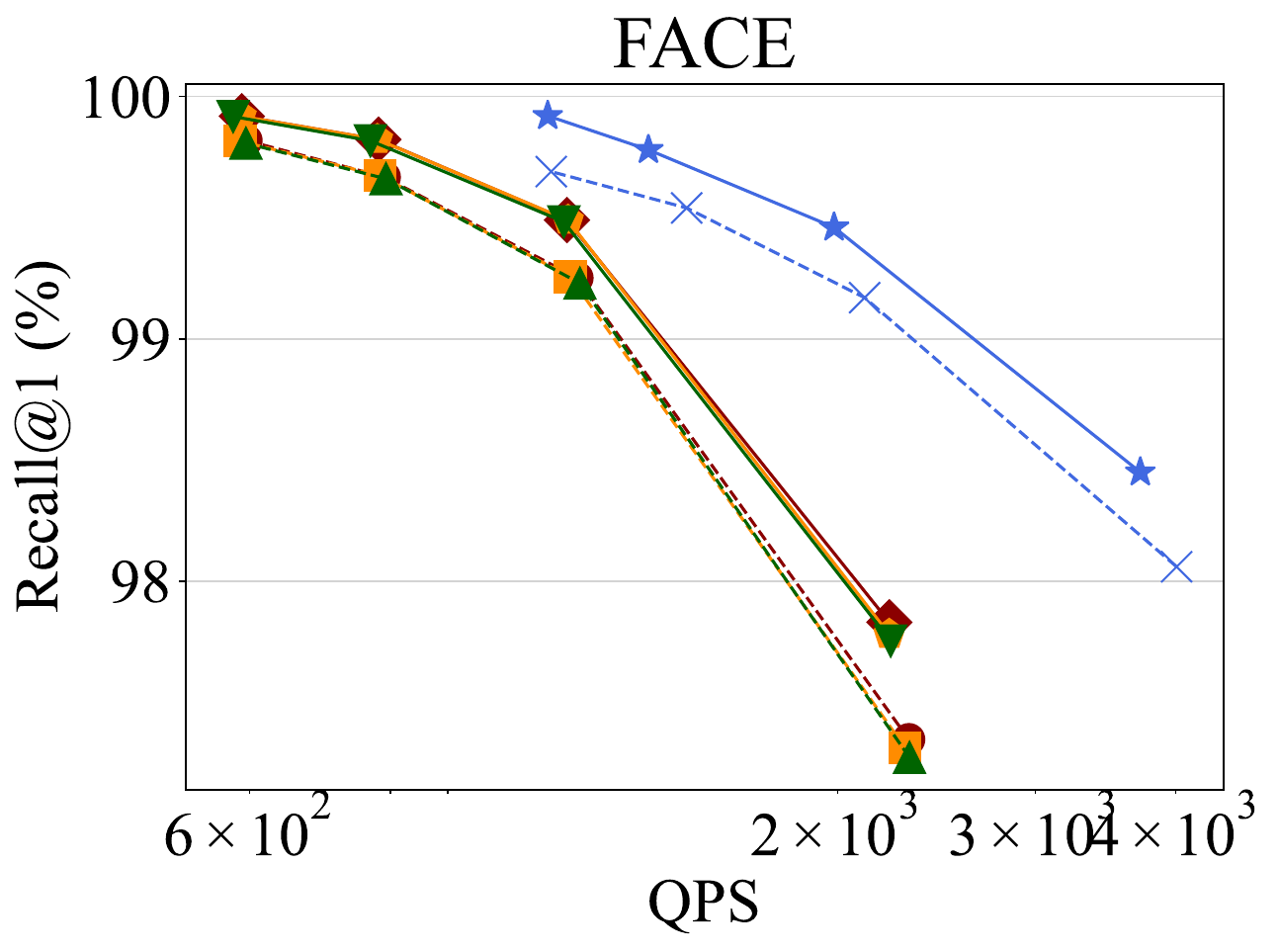}}
		\label{subfig:face_b1t4q1_5000000_float_K1_var1}}
	
	\figureBelowMargin

	\subfloat[][{\scriptsize QPS v.s. Recall@10}]{
		\scalebox{0.18}[0.18]{\includegraphics{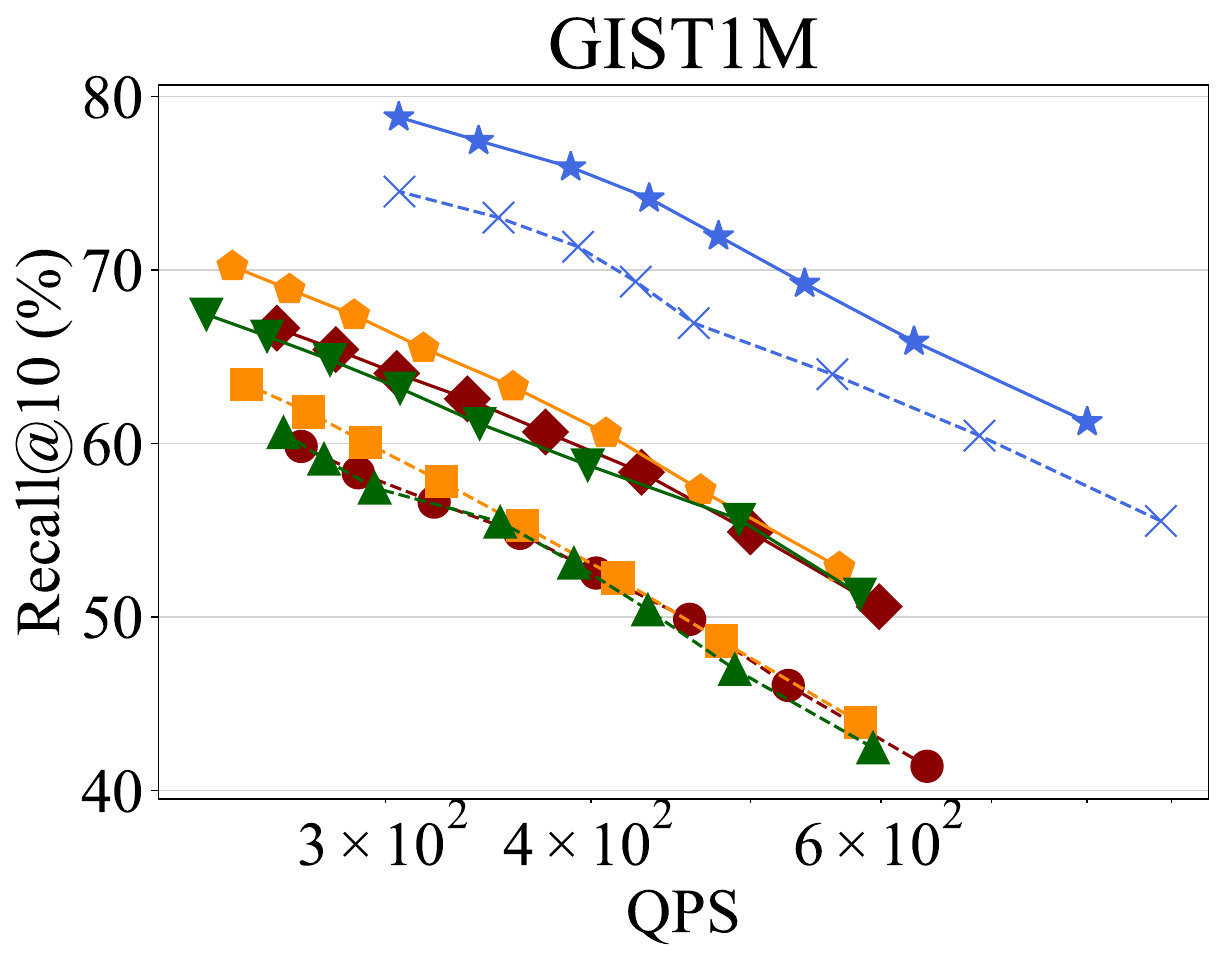}}
		\label{subfig:gist-960-euclidean_K10_var1}}
	\subfloat[][{\scriptsize QPS v.s. Recall@10}]{
		\scalebox{0.18}[0.18]{\includegraphics{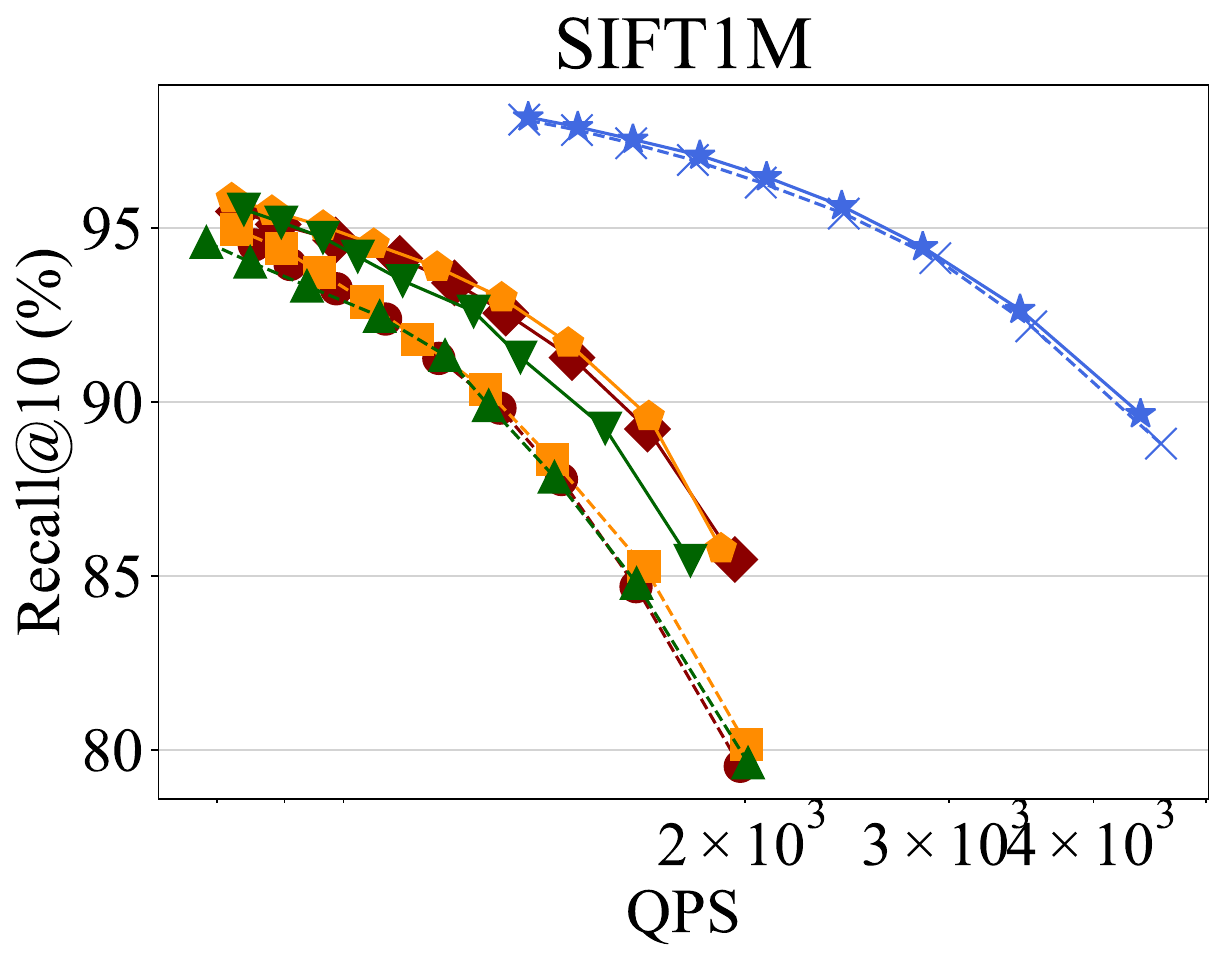}}
		\label{subfig:sift-128-euclidean_K10_var1}}
	\subfloat[][{\scriptsize QPS v.s. Recall@10}]{
		\scalebox{0.18}[0.18]{\includegraphics{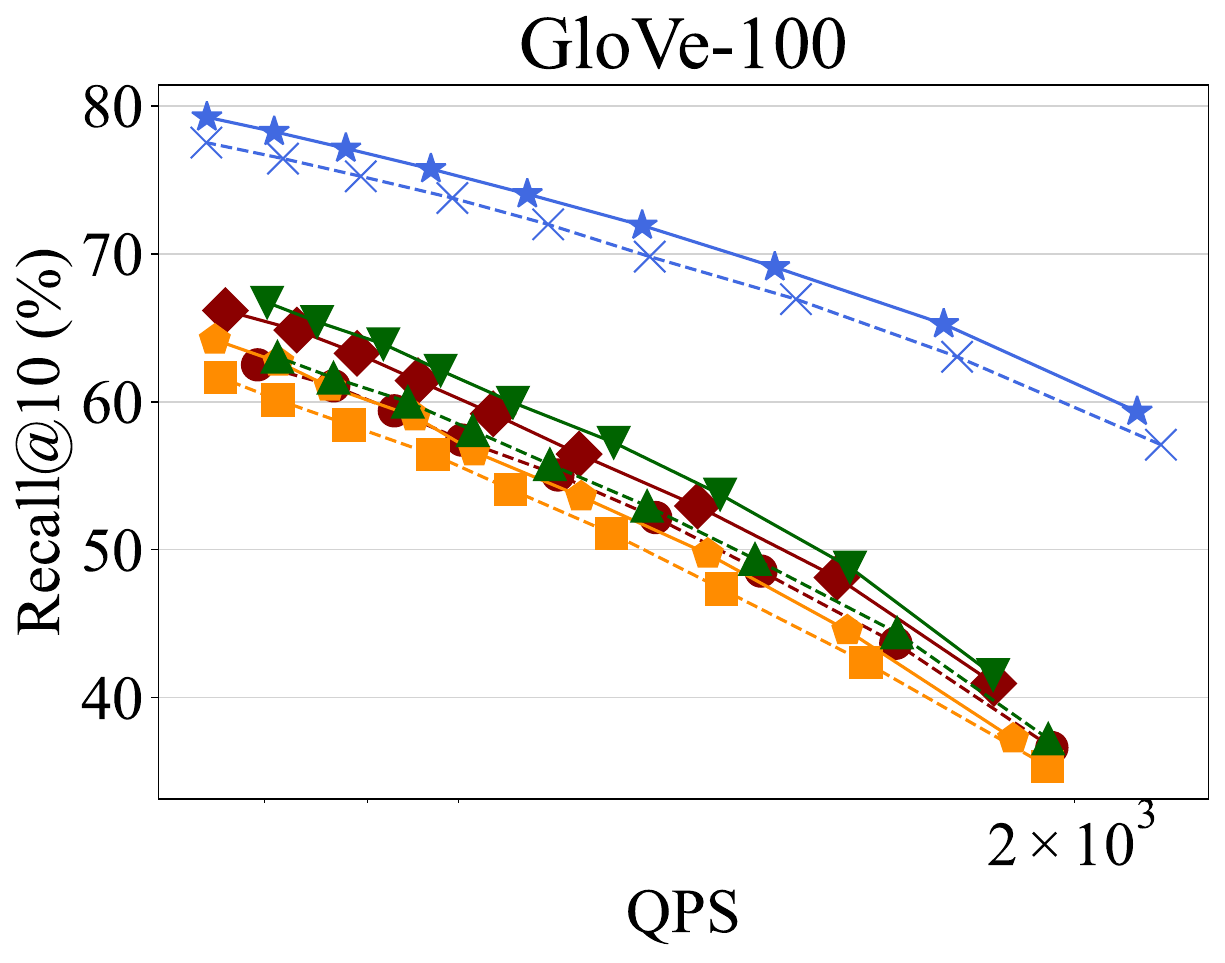}}
		\label{subfig:glove-100-angular_K10_var1}}
	\subfloat[][{\scriptsize QPS v.s. Recall@10}]{
		\scalebox{0.18}[0.18]{\includegraphics{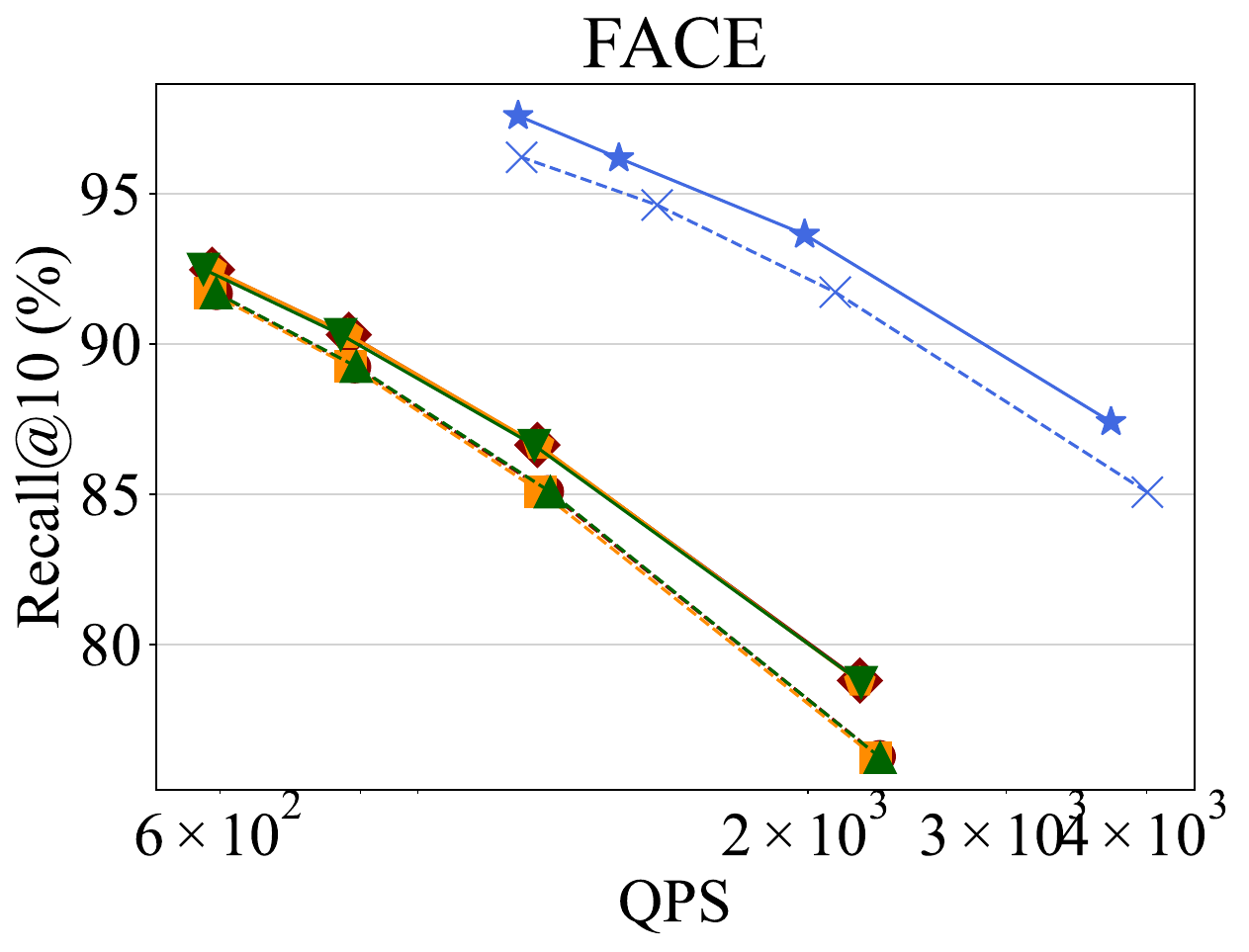}}
		\label{subfig:face_b1t4q1_5000000_float_K10_var1}}
	
	\figureBelowMargin
	
	\subfloat[][{\scriptsize QPS v.s. Recall@1}]{
		\scalebox{0.18}[0.18]{\includegraphics{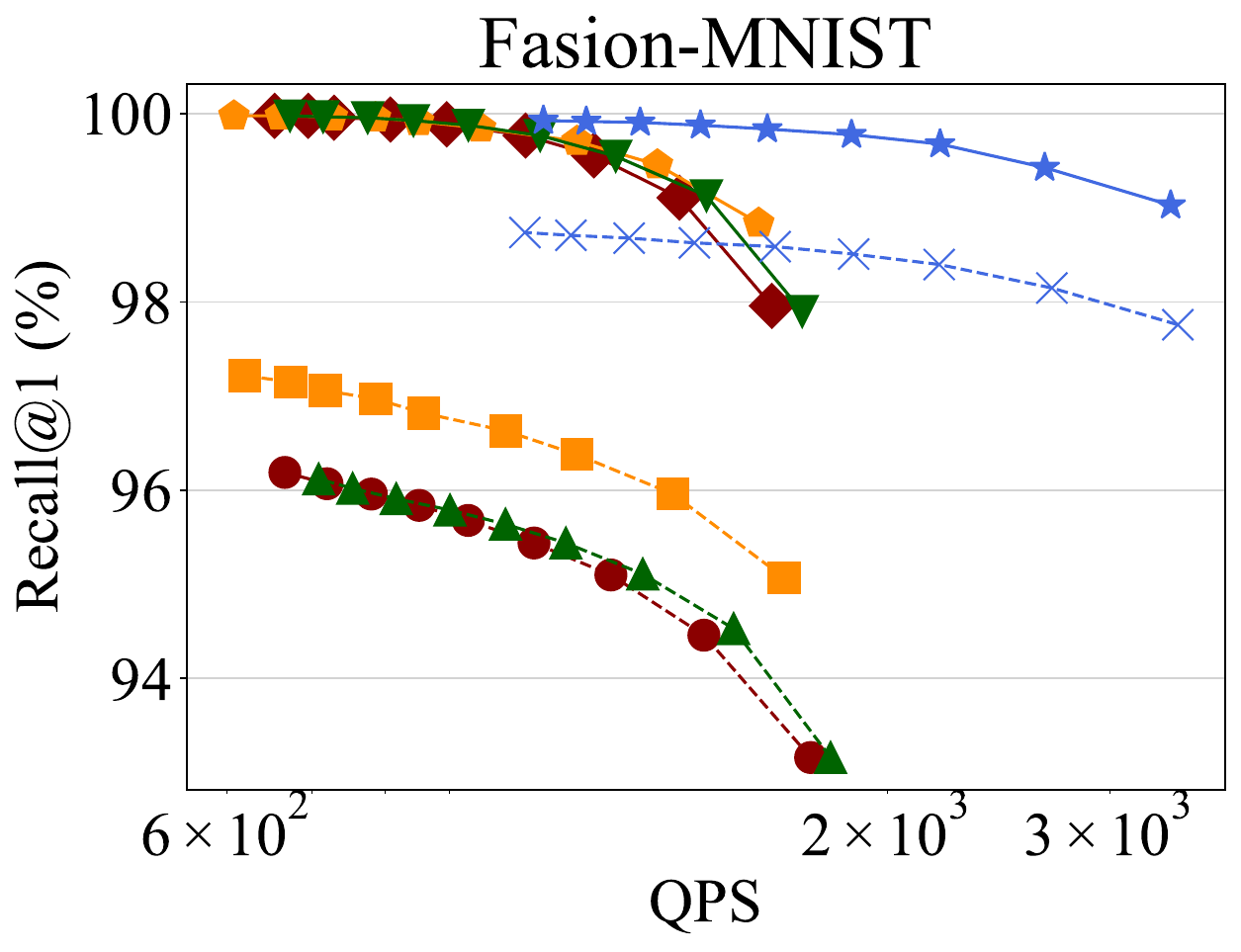}}
		\label{subfig:fashion-mnist-784-euclidean_K1_var1}}
	\subfloat[][{\scriptsize QPS v.s. Recall@1}]{
		\scalebox{0.18}[0.18]{\includegraphics{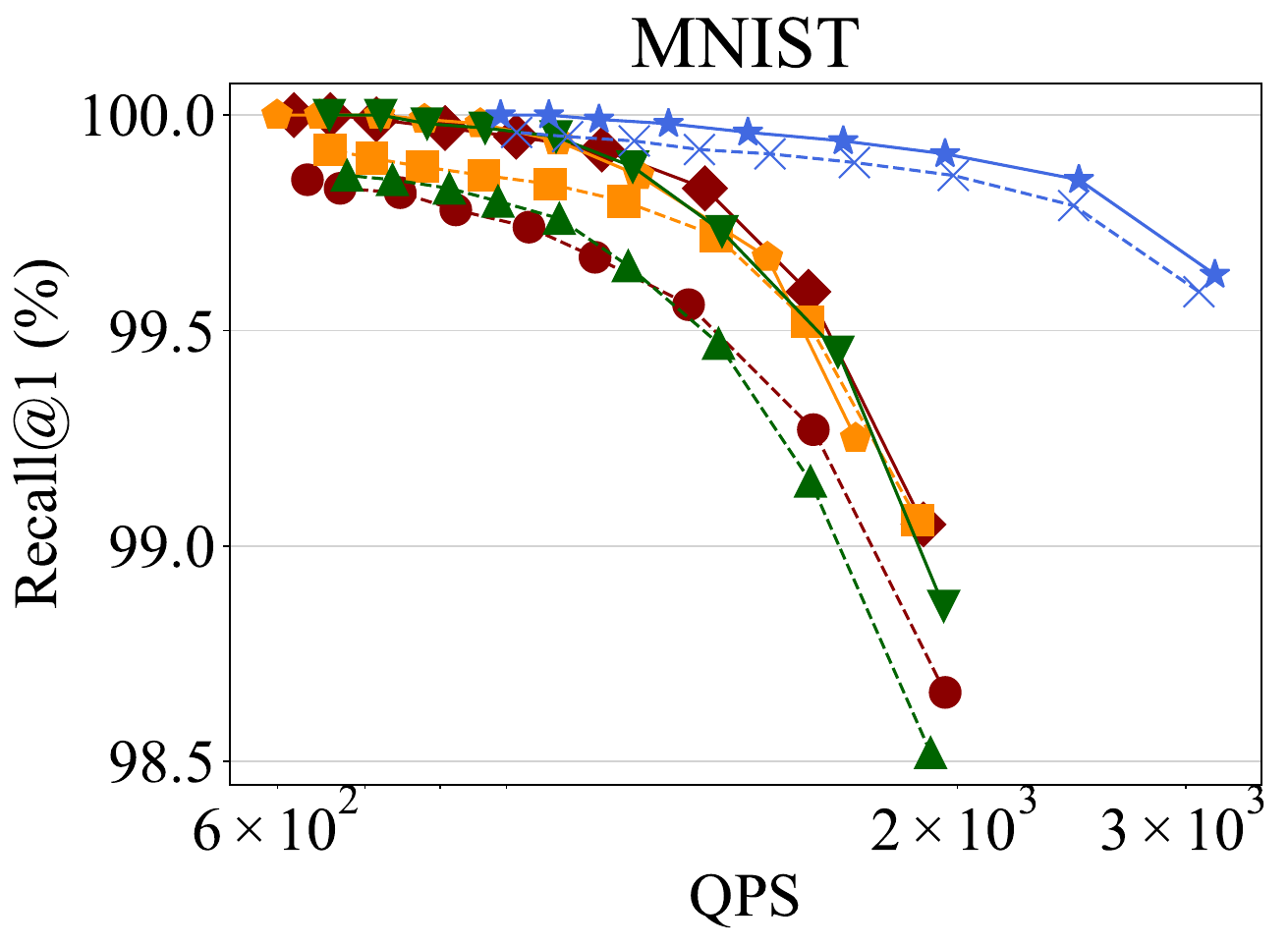}}
		\label{subfig:mnist-784-euclidean_K1_var1}}
	\subfloat[][{\scriptsize QPS v.s. Recall@1}]{
		\scalebox{0.18}[0.18]{\includegraphics{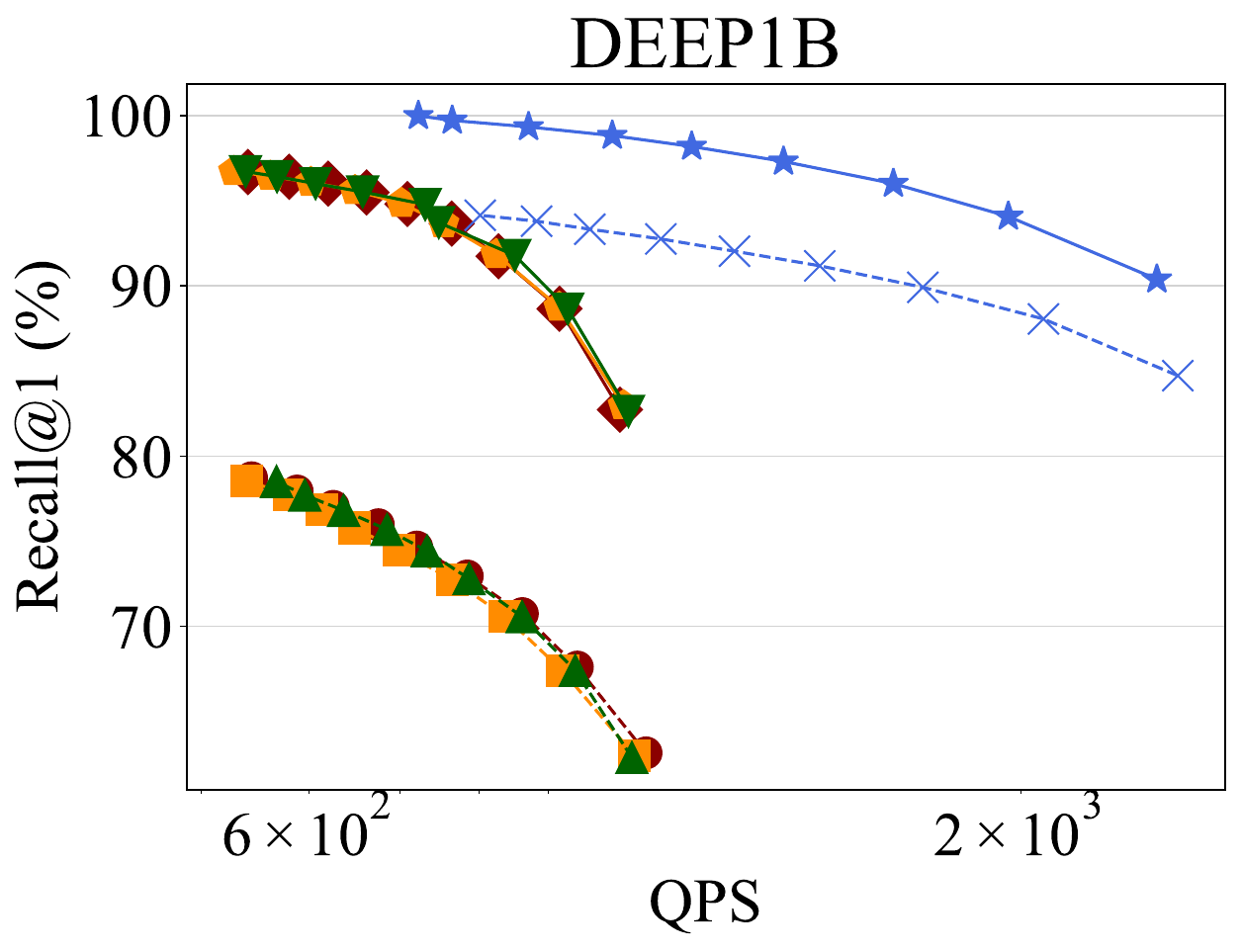}}
		\label{subfig:deep-image-96-angular_K1_var1}}
	\subfloat[][{\scriptsize QPS v.s. Recall@1}]{
		\scalebox{0.18}[0.18]{\includegraphics{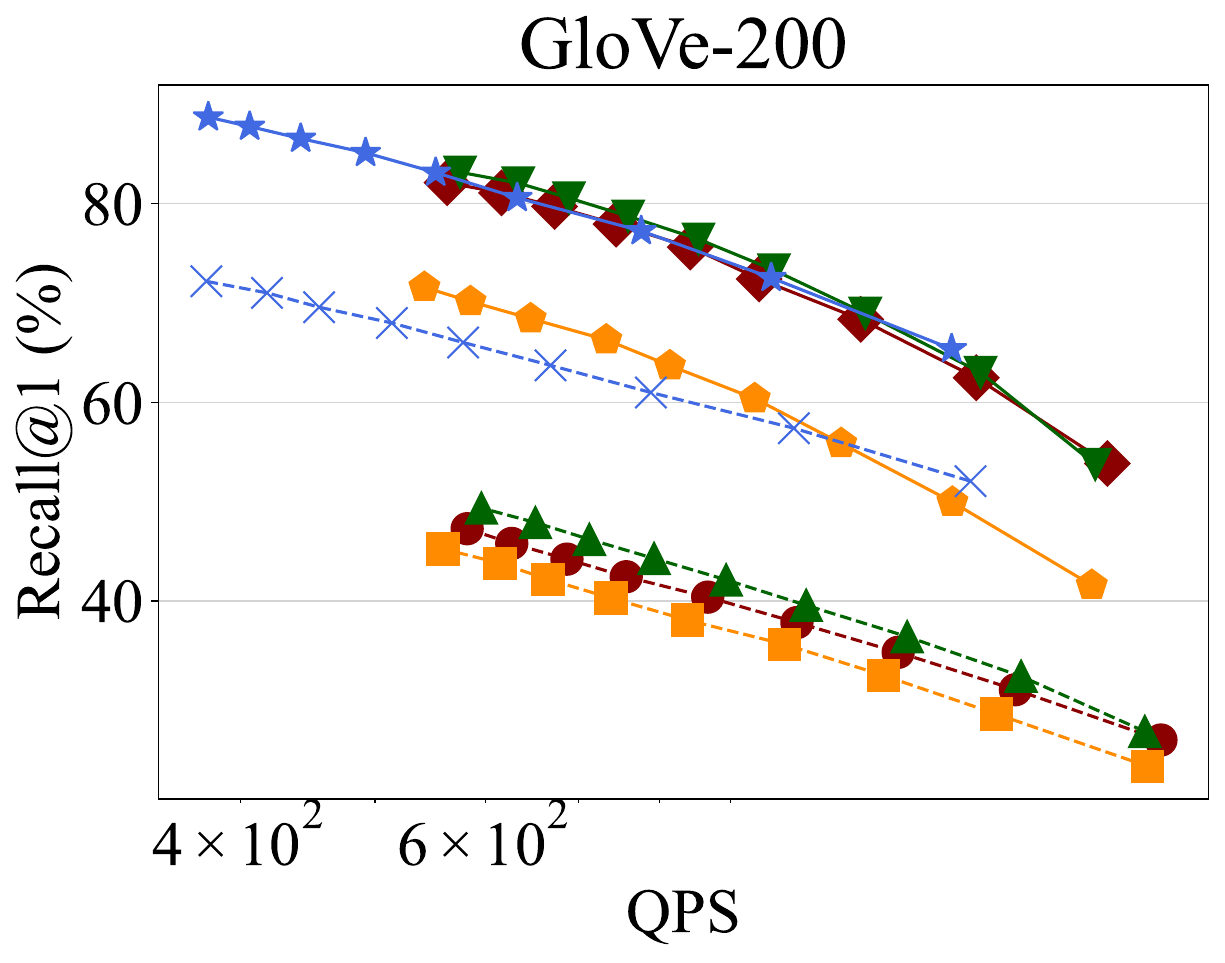}}
		\label{subfig:glove-200-angular_K1_var1}}
	\figureBelowMargin
	
	\subfloat[][{\scriptsize QPS v.s. Recall@10}]{
		\scalebox{0.18}[0.18]{\includegraphics{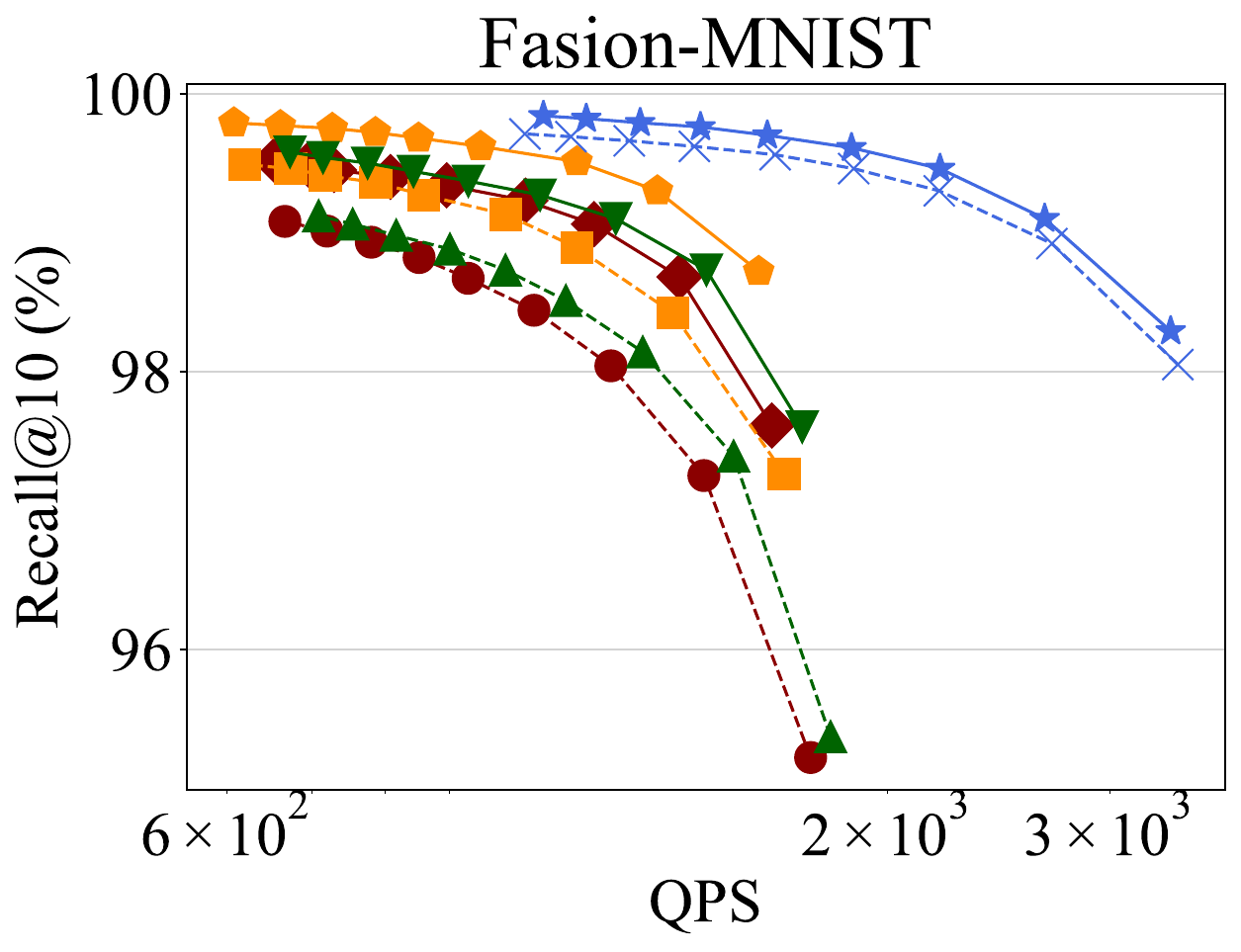}}
		\label{subfig:fashion-mnist-784-euclidean_K10_var1}}
	\subfloat[][{\scriptsize QPS v.s. Recall@10}]{
		\scalebox{0.18}[0.18]{\includegraphics{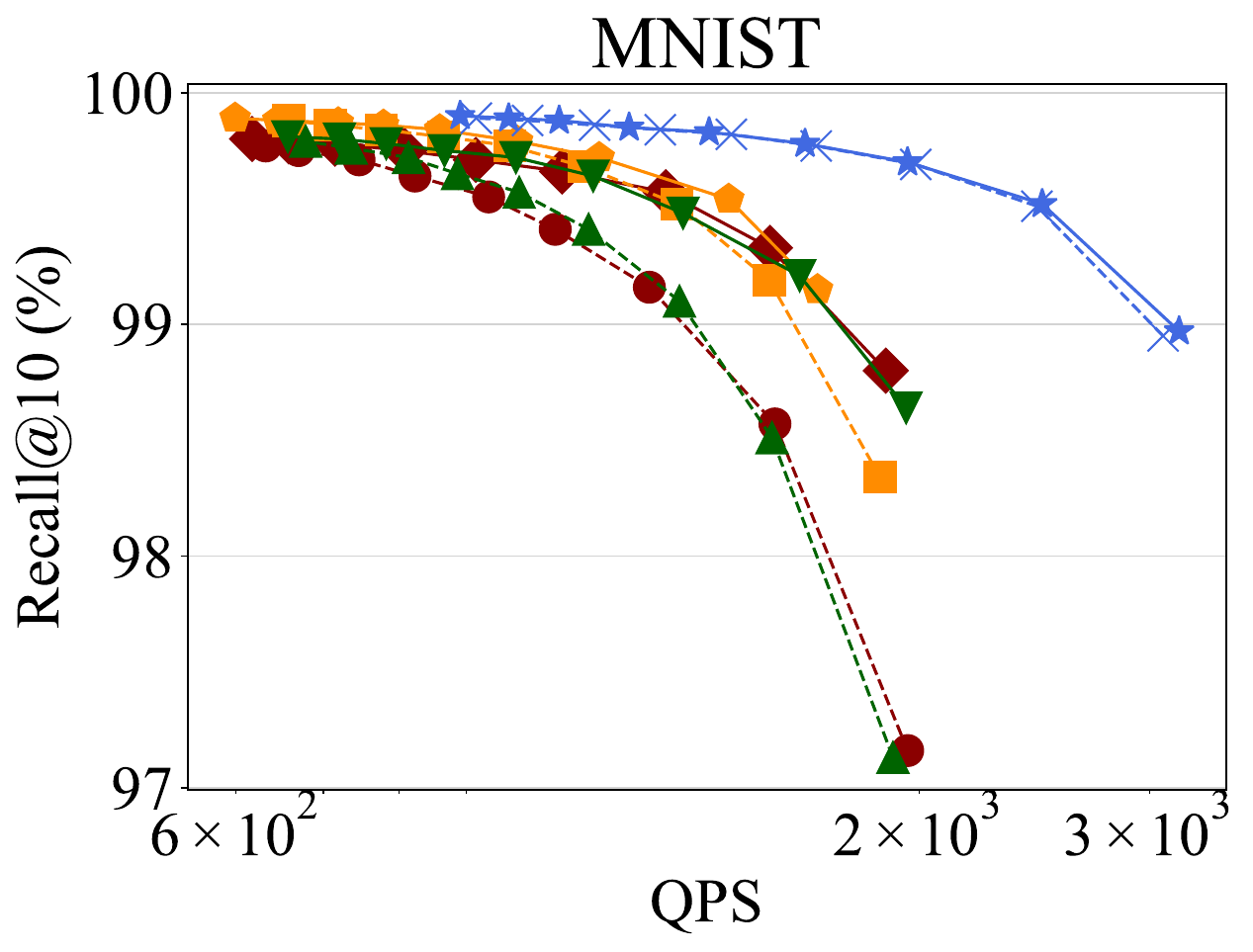}}
		\label{subfig:mnist-784-euclidean_K10_var1}}
	\subfloat[][{\scriptsize QPS v.s. Recall@10}]{
		\scalebox{0.18}[0.18]{\includegraphics{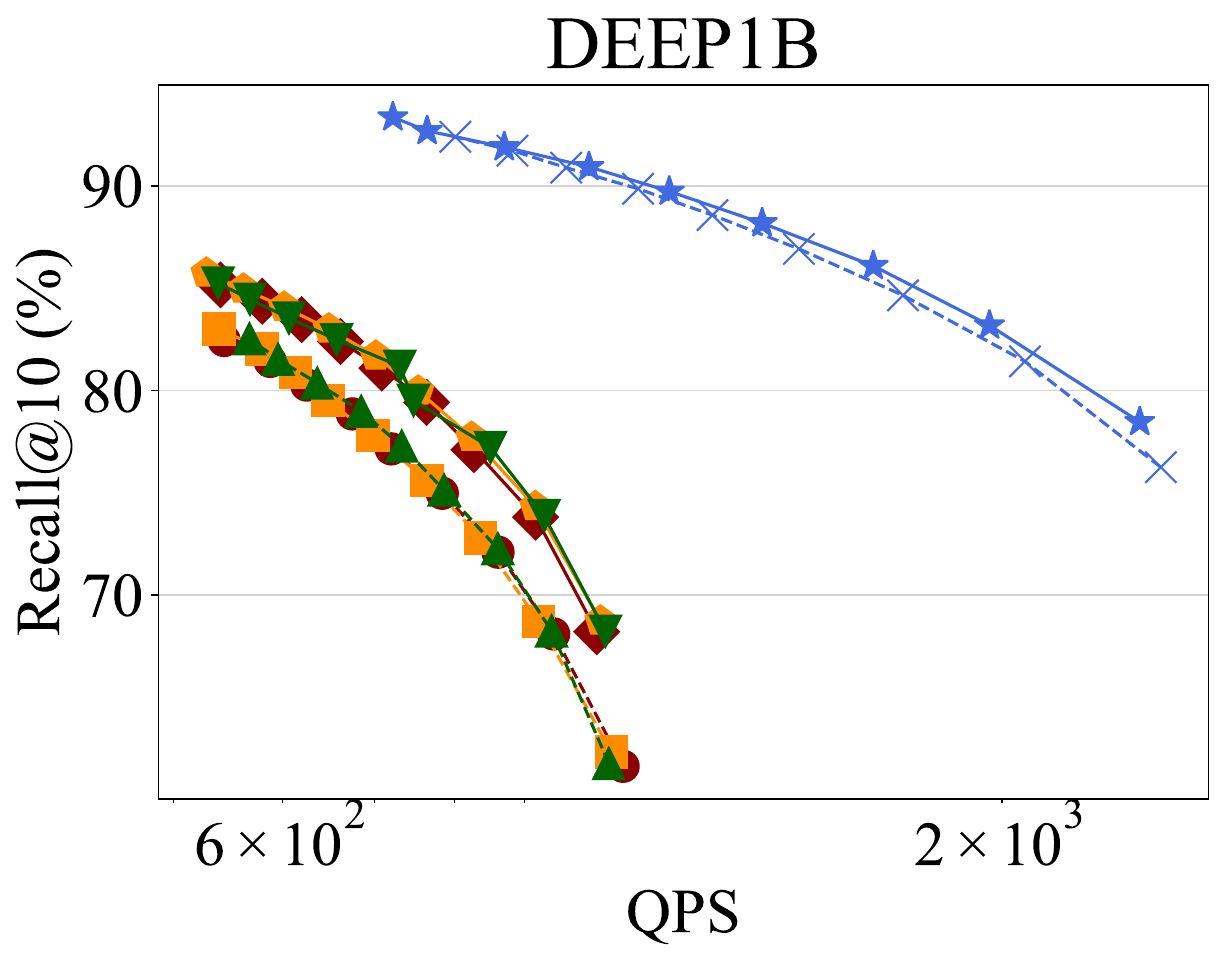}}
		\label{subfig:deep-image-96-angular_K10_var1}}
	\subfloat[][{\scriptsize QPS v.s. Recall@10}]{
		\scalebox{0.18}[0.18]{\includegraphics{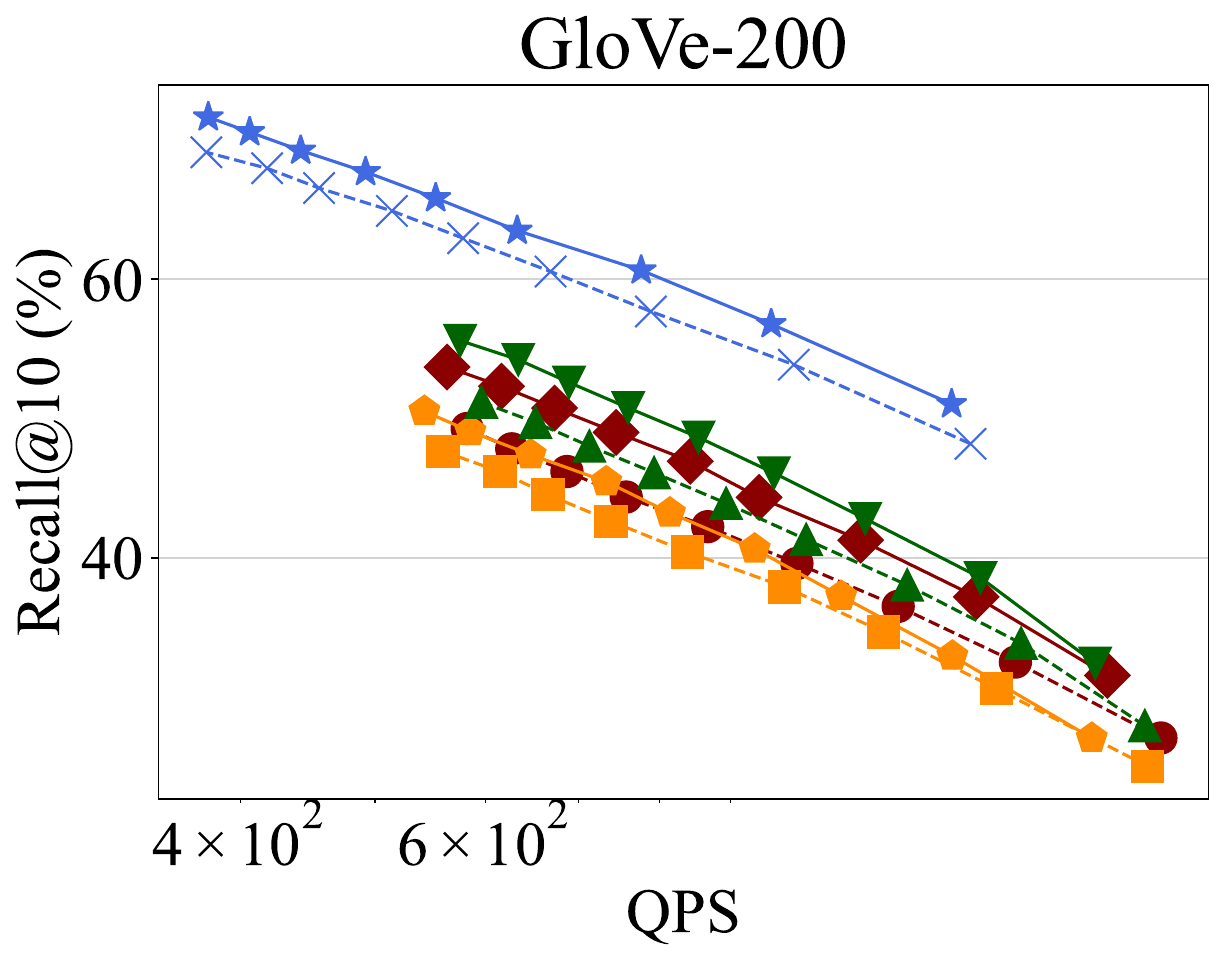}}
		\label{subfig:glove-200-angular_K10_var1}}

	\caption{\small Time-Accuracy Tradeoff (Varying Algorithms).}
	\label{fig:var1}
\end{figure*}

\noindent \underline{\textit{Datasets.}} We use seven real-world public datasets that are widely adopted by ANN Benchmarks~\cite{annbenchmark} and existing studies~\cite{Malkov2020hnsw, ddc, peng2023taumg}. To demonstrate the effectiveness of our approach, we also use an industrial dataset called FACE from a real-world application in Ant Group~\cite{antgroup}. The main characteristics of datasets are summarized in Table \ref{tab:datasets}. \textit{Dimension} is the vector dimension. \textit{\#Base} refers to the number of base points, which are used to construct the graph index. \textit{\#Query} is the number of queries used to evaluate the performance of algorithms. \textit{Metric} is the distance computation function. Note that the data size is also widely adopted in many in-memory index studies~\cite{wang2021survey, Malkov2020hnsw, fu2019NSG, peng2023taumg}. In practice, disk-based indexes~\cite{Jayaram2019diskann, Chen2021SPANN} are commonly used to handle larger datasets.

\noindent \underline{\textit{Compared Algorithms.}}
We integrate EnhanceGraph in four state-of-the-art graph-based indexes:
\begin{itemize}
	\item \textit{\textit{NSG}}~\cite{fu2019NSG}: It is built using the MRNG~\cite{fu2019NSG} strategy for edge pruning.
	\item \textit{\textit{VAMANA.}}~\cite{Jayaram2019diskann}: It is built based on MRNG strategy. It uses more slack pruning criteria compared to MRNG and a random graph for initialization.
	\item \textit{\textit{$\tau$-MNG}}~\cite{peng2023taumg}: It uses a proximity graph for initialization. Its edge pruning strategy includes all edges with a distance less than $3\tau$. It keeps long edges that are sufficiently long for navigation.
	\item \textit{\textit{HNSW.}}~\cite{Malkov2020hnsw}: It is constructed hierarchically using the RNG~\cite{Godfried1980RNG} edge pruning strategy. The bottom layer serves as the complete proximity graph. The higher layer samples a subset of points to construct the graph, which acts as a highway in search process.
\end{itemize}

The conjugate graph proposed in this paper can be applied to any graph indexing. We use the algorithm name appending ``-SC'' to indicate that both search and construction logs are used to construct the conjugate graph (e.g., VAMANA-SC).

\begin{table}[t!]
	\begin{center}
		{
			\caption{\small Experimental Settings.} \label{tab:settings}\vspace{-1ex}
			\begin{tabular}{l|l}
				{\bf \qquad \qquad \quad Parameters} & {\bf \qquad Values} \\ \hline \hline
				Proximity graph construction parameter $L_1$ & 20, 40, 60, 80, \textbf{100}\\
				
				Conjugate graph update parameter $L_2$ & 20, 40, 60, 80, \textbf{100}\\
				
				Log-generation position parameter $\omega$& \textbf{0.51}, 0.6, 0.7, 0.8, 0.9\\
				
				Generated search-log count $k_g$ per base point &  \textbf{5} \\
				\hline
			\end{tabular}
		}
	\end{center}
\end{table}

\noindent \underline{\textit{Environment.}}
All experiments are implemented in C++ and compiled with  -Ofast optimization. The experiments on public datasets are conducted on a single server equipped with a Intel(R) Xeon(R) Gold 6258R CPU@2.70GHz with 28 CPU cores and 1024 GB RAM. The experiments on FACE dataset are conducted on a single server equipped with a Intel(R) Xeon(R) Platinum 8163 CPU@2.50GHz with 96 CPU cores and 512GB RAM.

\noindent \underline{\textit{Measurements.}}
We use the recall rate~\cite{fu2019NSG,Jayaram2019diskann, Malkov2020hnsw} (i.e., the ratio of searched nearest neighbors to the accurate nearest neighbors) to measure the accuracy of the algorithm. Here, \textit{Recall@$k$} is the recall of the $k$-NNs. We  use query-per-second (\textit{QPS}), the number of requests that can be processed in each second, to measure efficiency. We measure the in-memory index, so the vectors data also needs to be loaded into memory. Thus, \textit{Memory Usage} is the total size of the vectors data and the graph index. The \textit{Build Time} is the time taken to construct the graph index and update the conjugate graph.

\noindent \underline{\textit{Implementation.}}
Since public datasets provide limited query points for constructing historical search logs, we create one synthetic historical-search-log entry for each base point. Specifically, we calculate the mean of the absolute value of each dimension of the base points, denoted by $\eta$. Then, for each base point, we add random noise sampled from a uniform distribution $U(-0.5\eta, 0.5\eta)$ to construct a synthetic query point for the corresponding historical search log. The random noise is extensively utilized in computer vision and other learning fields~\cite{rice1944mathematical, NIPS2016_7ce3284b}. We then use the historical and generated search logs to update the conjugate graph. For the FACE dataset, we only use four real historical search logs for each base point to update the conjugate graph. The original index is constructed through incremental inserts on a single core, due to data dependencies between insertions. The update of the conjugate graph runs on 28 cores using the OpenMP library. Besides, we used AVX2 SIMD libraries for optimizing the distance computation.

\begin{table*}[t!]
	\centering
	\small\scriptsize
	\caption{Comparison of Enhancement, Memory Usage and Build Time. (Fixed Search Parameter)\vspace{-1ex}} \label{tab:ablation}
	\begin{tabular}{l|l|l|l|l|l|l|l|l|l|l}
		\hline
		\multirow{2}{*}{\textbf{Algorithm}} & \multicolumn{5}{c|}{\textbf{GIST1M} \quad (Vector Data Size: 3662.11MB)} & \multicolumn{5}{c}{\textbf{GloVe-200} \quad (Vector Data Size: 902.95MB)}
		\\ \cline{2-11}
		& \textbf{QPS} & \textbf{Recall@1} & \textbf{Recall@10} & \textbf{Memory Usage} & \textbf{Build Time}    & \textbf{QPS} & \textbf{Recall@1} & \textbf{Recall@10} & \textbf{Memory Usage} & \textbf{Build Time}
		\\ \hline \hline
		NSG
		&   267.3 &  41.61\% & 59.83\% & \textbf{3688.46MB} & \textbf{2902s}
		&   \textbf{555.7} &  47.29\% & 49.29\% & \textbf{949.92MB} & \textbf{2332s}
		\\
		
		NSG-SC
		&   260.0 &  93.12\% & 66.65\% & 3737.43MB & 3604s
		&   549.8 &  82.09\% & 53.70\% & 1002.23MB & 2866s
		\\ \hline
		
		VAMANA
		&   246.0 &  42.01\% & 63.41\% & 3705.16MB & 3757s
		&   546.4 &  45.26\% & 47.65\% & 958.54MB & 2688s
		\\
		
		VAMANA-SC
		&   244.4 &  92.98\% & 70.21\% & 3750.02MB & 4322s
		&   522.7 &  82.10\% & 52.17\% & 1010.3MB & 3191s
		\\ \hline
		
		$\tau$-MNG
		&   267.4 &  41.74\% & 60.64\% & 3693.93MB & 3029s
		&   535.6 &  49.37\% & 51.18\% & 964.19MB & 2402s
		\\
		
		$\tau$-MNG-SC
		&   260.4 &  \textbf{93.42\%} & 67.44\% & 3738.25MB & 3448s
		&   519.6 &  83.14\% & 55.60\% & 1009.29MB & 2838s
		\\ \hline\hline
		
		KNNG
		&   306.1 &  49.41\% & 52.29\% & 3802.11MB  & 16483s
		&   360.8 &  67.25\% & 64.70\% & 1070.37MB  & 13512s
		\\
		
		HNSW
		&   \textbf{327.4} &  60.01\% & 74.52\% & 3773.46MB & 2944s
		&   388.1 &  72.17\% & 69.12\% & 1034.73MB & 2913s
		\\
		
		HNSW-SC
		&   322.7 &  93.24\% & \textbf{78.81\%} & 3817.11MB & 3420s
		&   380.7 &  \textbf{88.68\%} & \textbf{71.64\%} & 1081.92MB & 3190s
		\\ \hline\hline
	\end{tabular}
\end{table*}

\noindent \underline{\textit{Parameters.}}
We set the maximum degree of the graph to $R = 12$ in the main comparison to stress-test whether EnhanceGraph can repair sparse deployed graphs under a tight graph-degree budget. For the parameter-sensitivity experiments, we set the maximum degree to $R = 32$ to reduce the confounding effect of an overly sparse base graph and to better isolate the trends caused by EnhanceGraph-specific parameters. The parameters $L_2$ and $\omega$ are used only when updating or generating the auxiliary conjugate graph $G'$; they are not native parameters of the VAMANA baseline. As demonstrated in Table \ref{tab:settings}, we adjust three major parameters at different values to examine their influence on the conjugate graph enhancement. We fix $k_g = 5$ to collect $k_g$ generated search logs for each base point.

\subsection{Overall Results}

\noindent \underline{\textit{Overall Results (Time-Accuracy Trade-Off).}}
As shown in Figure \ref{fig:var1}, we examine the Time-Accuracy Tradeoff of four algorithms on each dataset.  The dashed lines in various colors represent different proximity graphs, while the solid lines represent the version that utilized a conjugate graph to enhance (i.e., our EnhanceGraph framework). We found that the conjugate graph boosts the recall rate for each algorithm at the same QPS. In GIST1M, it enhances $Recall@1$ the most from 41.74\% to 93.42\%.

The improvement is mainly due to the substantial number of edges added from the search log in the conjugate graph. These edges help the search process efficiently move from local optima to global optima. Especially on the FACE dataset, our EnhanceGraph framework improves the recall, even when it's already high (i.e., it improves $Recall@1$ from 99.8\% to 99.9\%). However, for some low dimension datasets (e.g. SIFT1M), conjugate graph may hardly improve the recall since it's already high (e.g., more than 99\%). Besides, the conjugate graph improves $Recall@10$. The improvement is attributed to the use of construction log, which preserves numerous pruned $k$-NNs in the proximity graph.

\noindent \underline{\textit{Overall Results (Memory Usage and Build Time).}}

As shown in Table \ref{tab:ablation}, we evaluated the enhancement on GIST1M and GloVe-200 datasets. The conjugate graph is constructed using both search and construction logs  across four graph structures. As a comparison of extra space and enhancement, we constructed a KNNG ($k$-NNs Graph) without using the pruning strategy and set the maximum degree to $R=36$. The search list length of algorithms, except for KNNG, was set at $L=100$. We found that NSG-SC, $\tau-$MNG-SC, and VAMANA-SC improve the $Recall@1$ on GIST1M under the evaluated settings, with the largest gain from 41.74\% to 93.42\%. On the GloVe-200 dataset, the conjugate graph increases $Recall@1$ to over 80\%. The post-search overhead is measured in Appendix C. Figure \ref{fig:var1} also shows that the enhanced variants generally improve recall at comparable QPS in the evaluated operating range.

The extra space cost of constructing the conjugate graph stabilizes across datasets. This is primarily because the size of the conjugate graph mainly depends on the maximum degree $R$, which is a constant in our experimental setup. Although the size of the conjugate graph is comparable to the proximity graph, both are relatively small compared to the size of the data. For instance, a 50MB conjugate graph is still relatively small compared to a 3662.11MB dataset. Among them, the memory usage of KNNG is larger than that of all algorithms except HNSW-SC.

\subsection{Evaluation on Varying Parameters}
\label{sec:appendix_experiment}

\begin{figure}[t!]
	\centering
	\includegraphics[width=0.9\columnwidth]{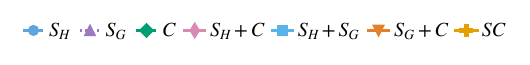}\\[-1ex]
	\subfloat[][{\scriptsize \shortstack{GIST1M $\Delta$Recall@1\\(baseline Recall@1 = 0.724)}}]{
		\includegraphics[width=0.441\columnwidth]{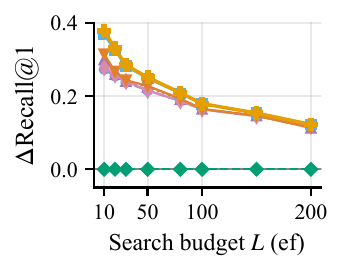}}
	\subfloat[][{\scriptsize \shortstack{GloVe-200 $\Delta$Recall@1\\(baseline Recall@1 = 0.765)}}]{
		\includegraphics[width=0.441\columnwidth]{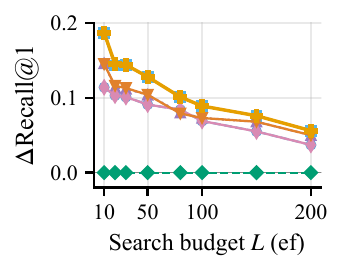}}\\[0.5ex]
	\subfloat[][{\scriptsize \shortstack{GIST1M $\Delta$Recall@10\\(baseline Recall@10 = 0.791)}}]{
		\includegraphics[width=0.441\columnwidth]{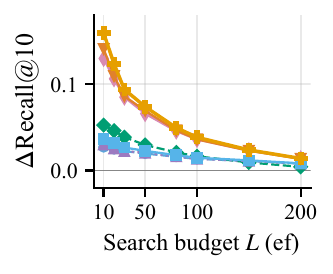}}
	\subfloat[][{\scriptsize \shortstack{GloVe-200 $\Delta$Recall@10\\(baseline Recall@10 = 0.783)}}]{
		\includegraphics[width=0.441\columnwidth]{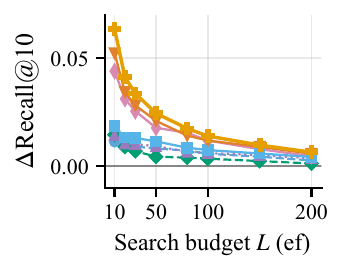}}
	\caption{\small Delta-recall log-source ablation (each panel annotates the omitted baseline recall at $L=50$).}
	\label{fig:r2_log_source_ablation_main}
	\vspace{-2ex}
\end{figure}

\begin{figure}[t!]\centering
	\includegraphics[width=0.98\columnwidth]{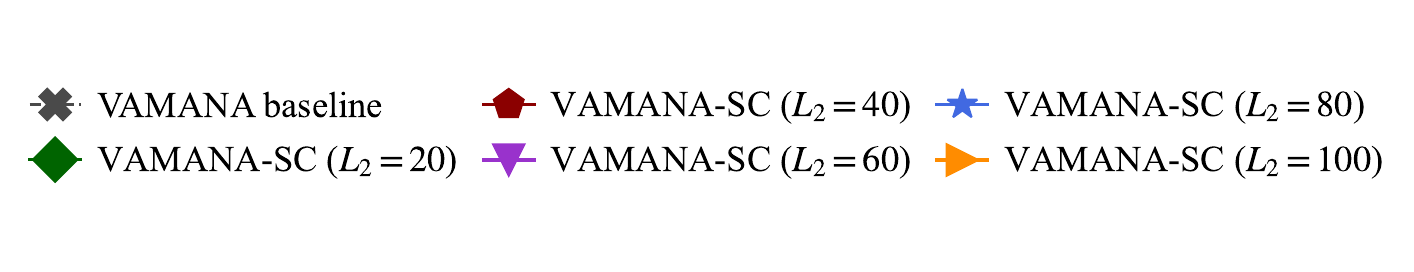}\\[-1ex]
	\subfloat[][{\scriptsize GIST1M Recall@1}]{
		\scalebox{0.18}[0.18]{\includegraphics{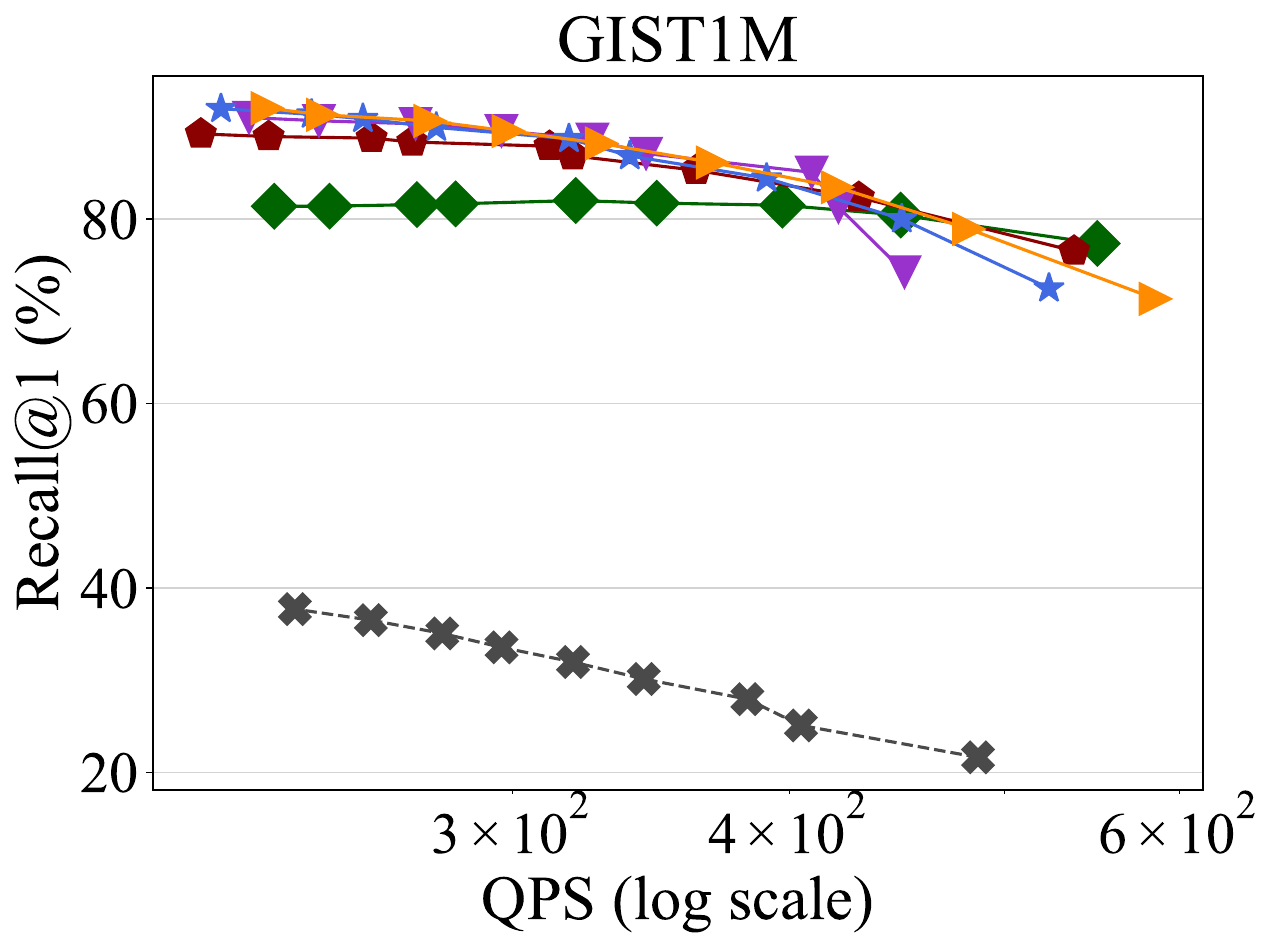}}
		\label{subfig:gist-960-euclidean_K1_var3}}
	\subfloat[][{\scriptsize GloVe-200 Recall@1}]{
		\scalebox{0.18}[0.18]{\includegraphics{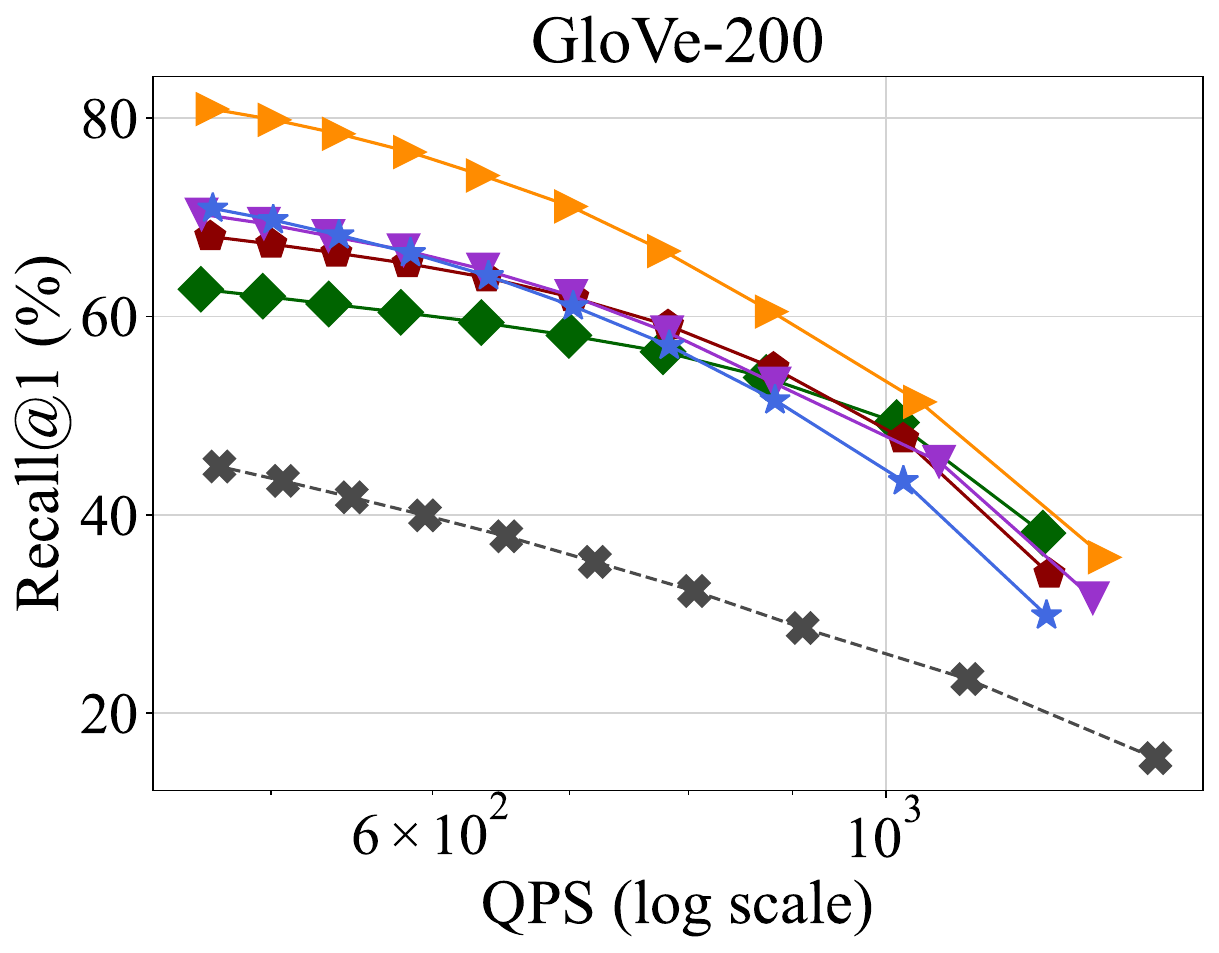}}
		\label{subfig:glove-200-angular_K1_var3}}\\[0.5ex]
	\subfloat[][{\scriptsize GIST1M Recall@10}]{
		\scalebox{0.18}[0.18]{\includegraphics{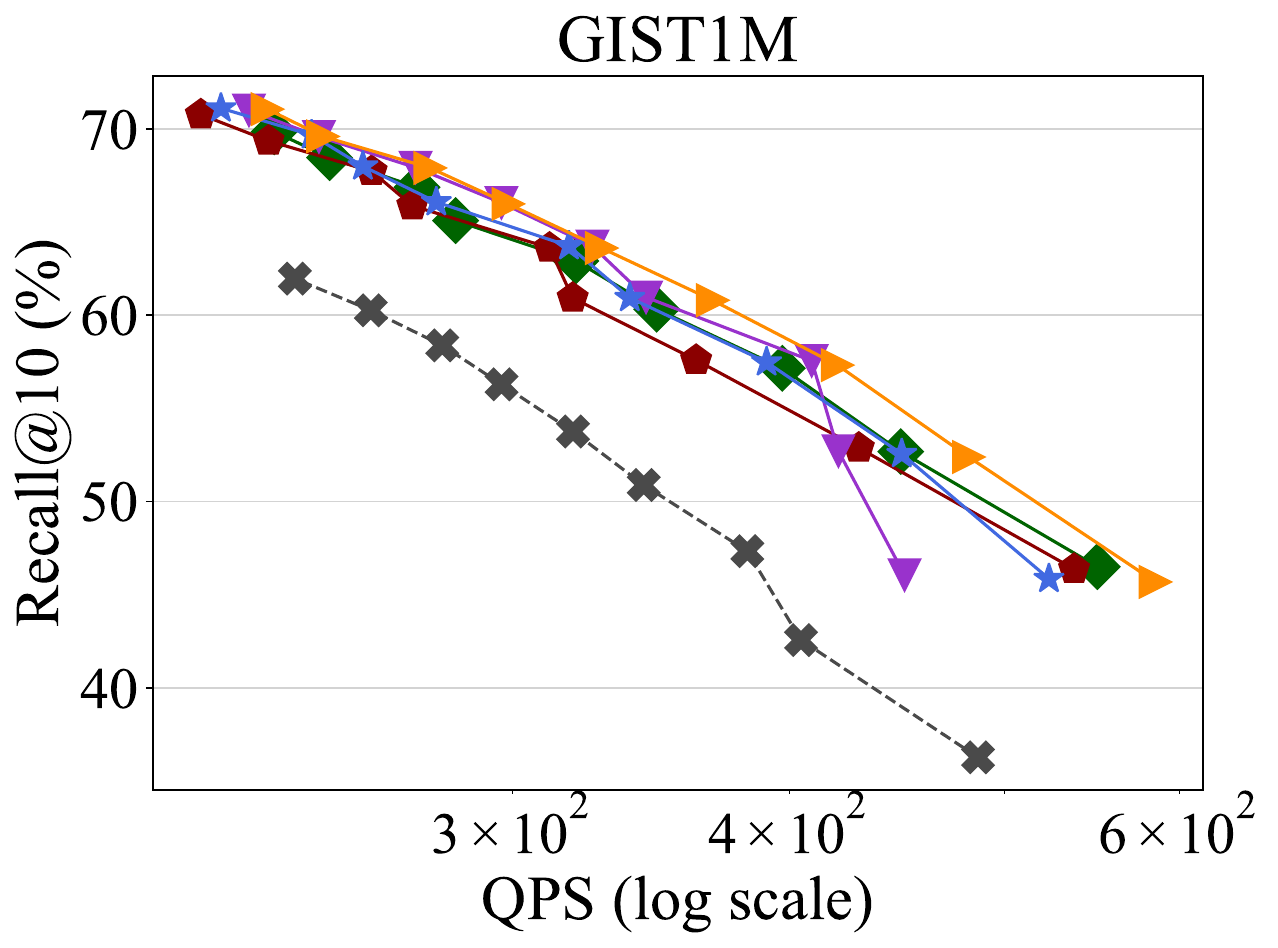}}
		\label{subfig:gist-960-euclidean_K10_var3}}
	\subfloat[][{\scriptsize GloVe-200 Recall@10}]{
		\scalebox{0.18}[0.18]{\includegraphics{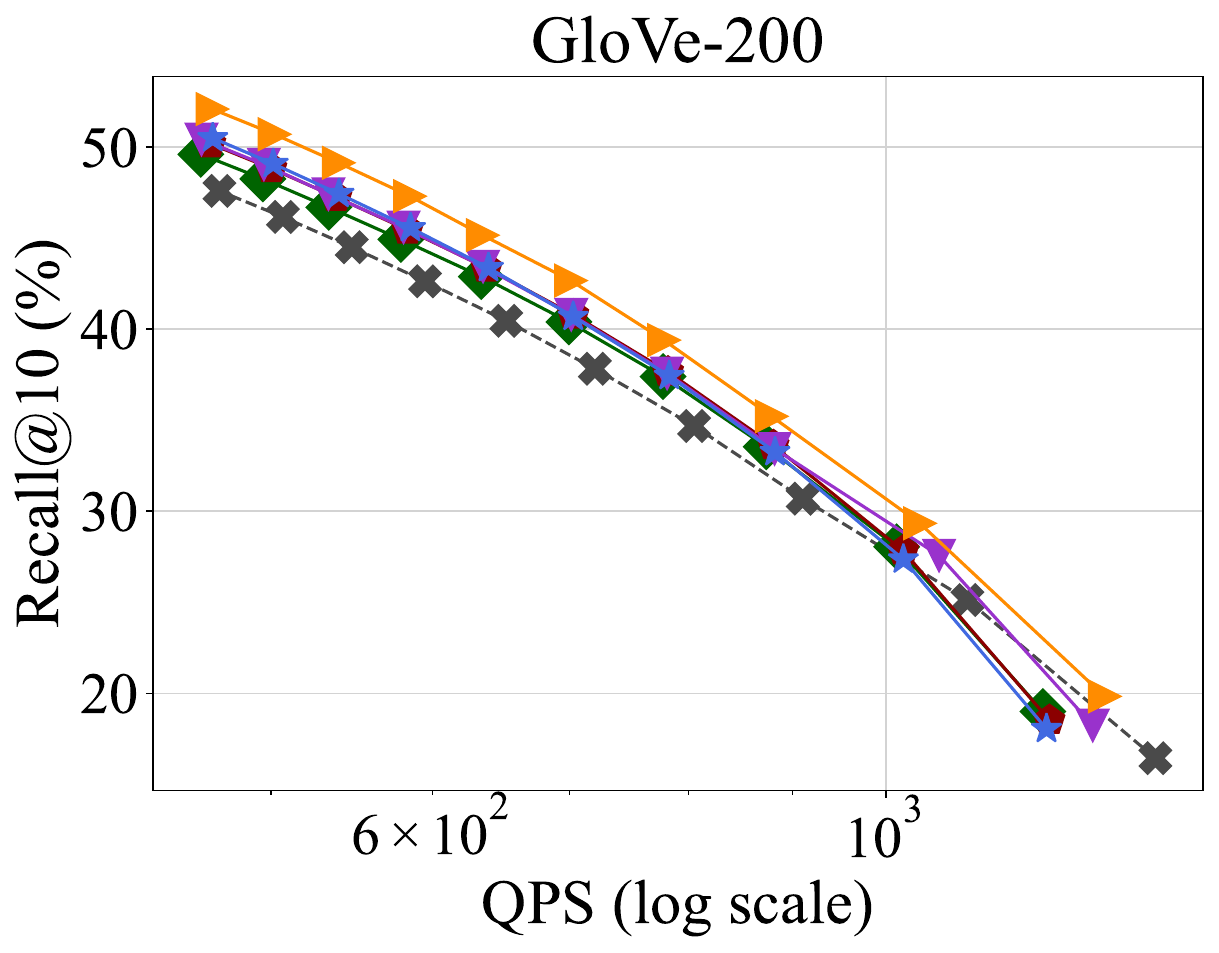}}
		\label{subfig:glove-200-angular_K10_var3}}
	\caption{\small Time-Accuracy Trade-off (Varying EnhanceGraph-specific update parameter $L_2$).}
	\label{fig:var3}
	\vspace{-2ex}
\end{figure}

\begin{figure}[t!]\centering
	\vspace{-3ex}
	\includegraphics[width=0.98\columnwidth]{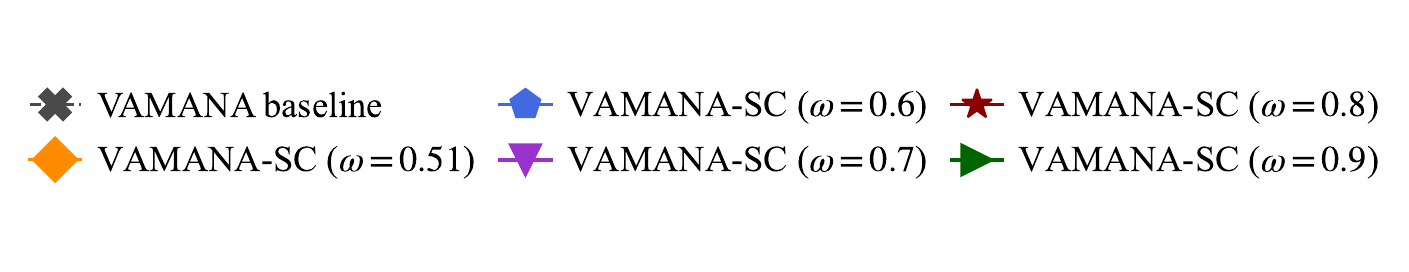}\\[-3ex]
	\subfloat[][{\scriptsize GIST1M Recall@1}]{
		\scalebox{0.18}[0.18]{\includegraphics{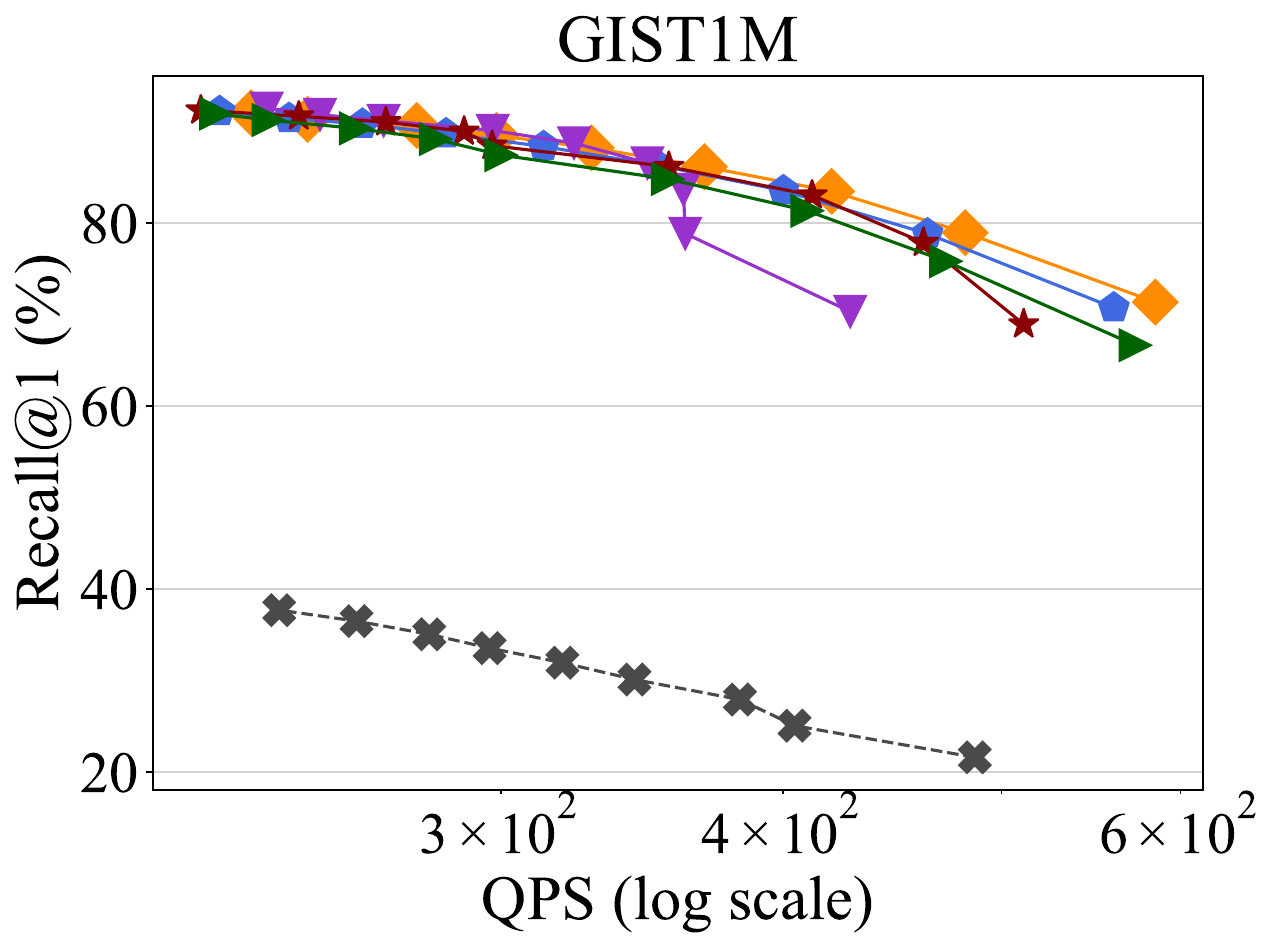}}
		\label{subfig:gist-960-euclidean_K1_var4}}
	\subfloat[][{\scriptsize GloVe-200 Recall@1}]{
		\scalebox{0.18}[0.18]{\includegraphics{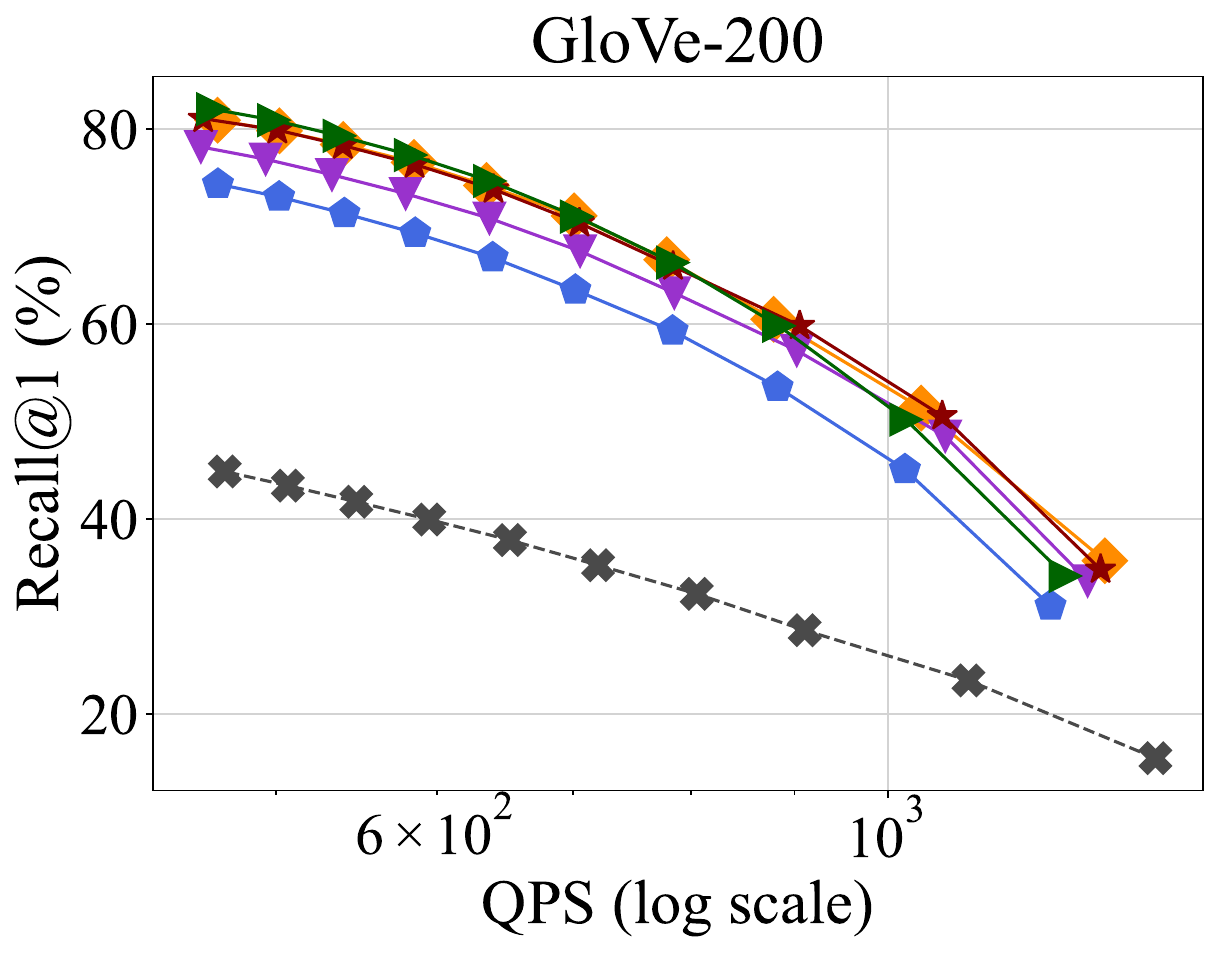}}
		\label{subfig:glove-200-angular_K1_var4}}\\[0.5ex]
	\subfloat[][{\scriptsize GIST1M Recall@10}]{
		\scalebox{0.18}[0.18]{\includegraphics{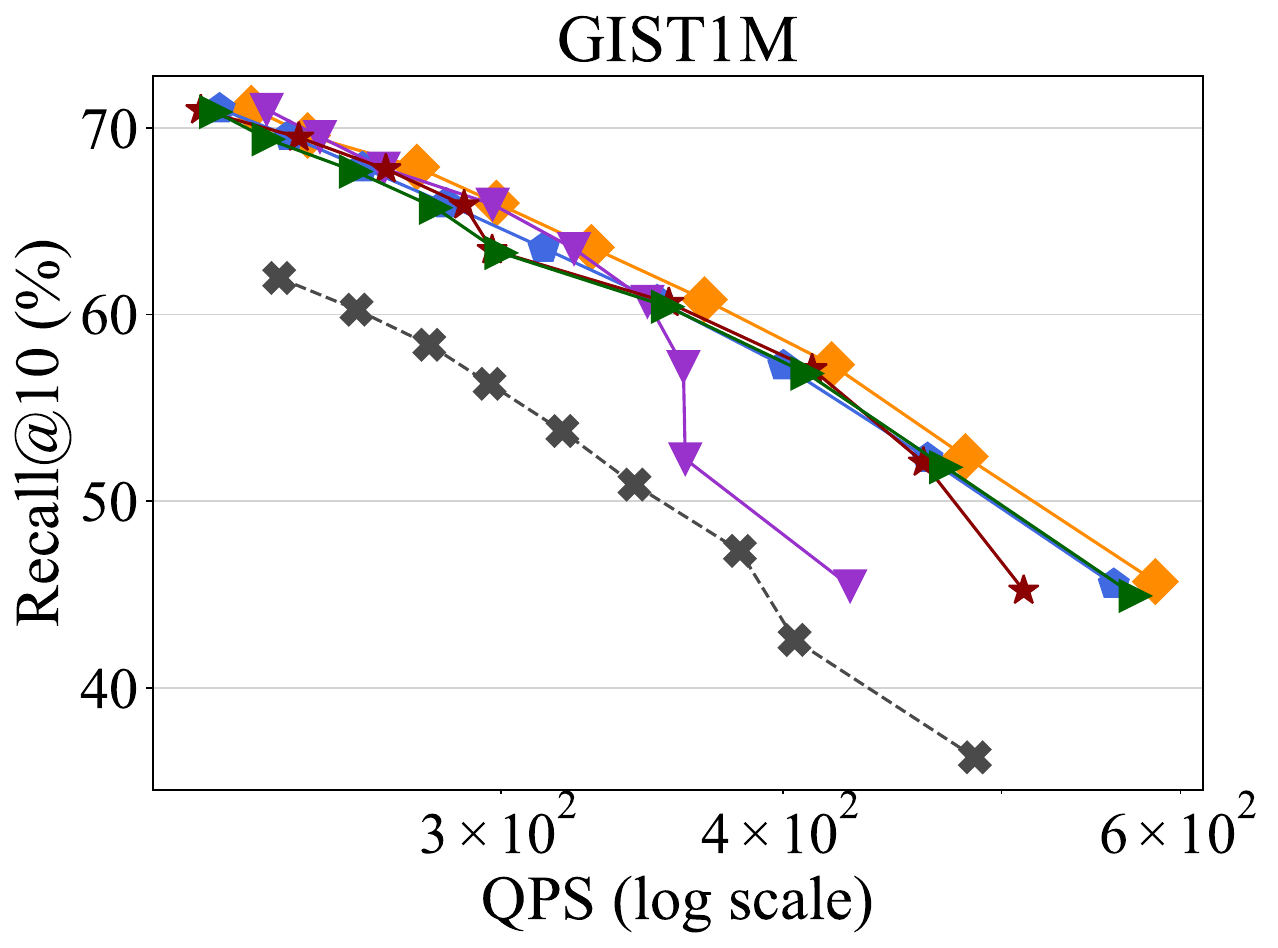}}
		\label{subfig:gist-960-euclidean_K10_var4}}
	\subfloat[][{\scriptsize GloVe-200 Recall@10}]{
		\scalebox{0.18}[0.18]{\includegraphics{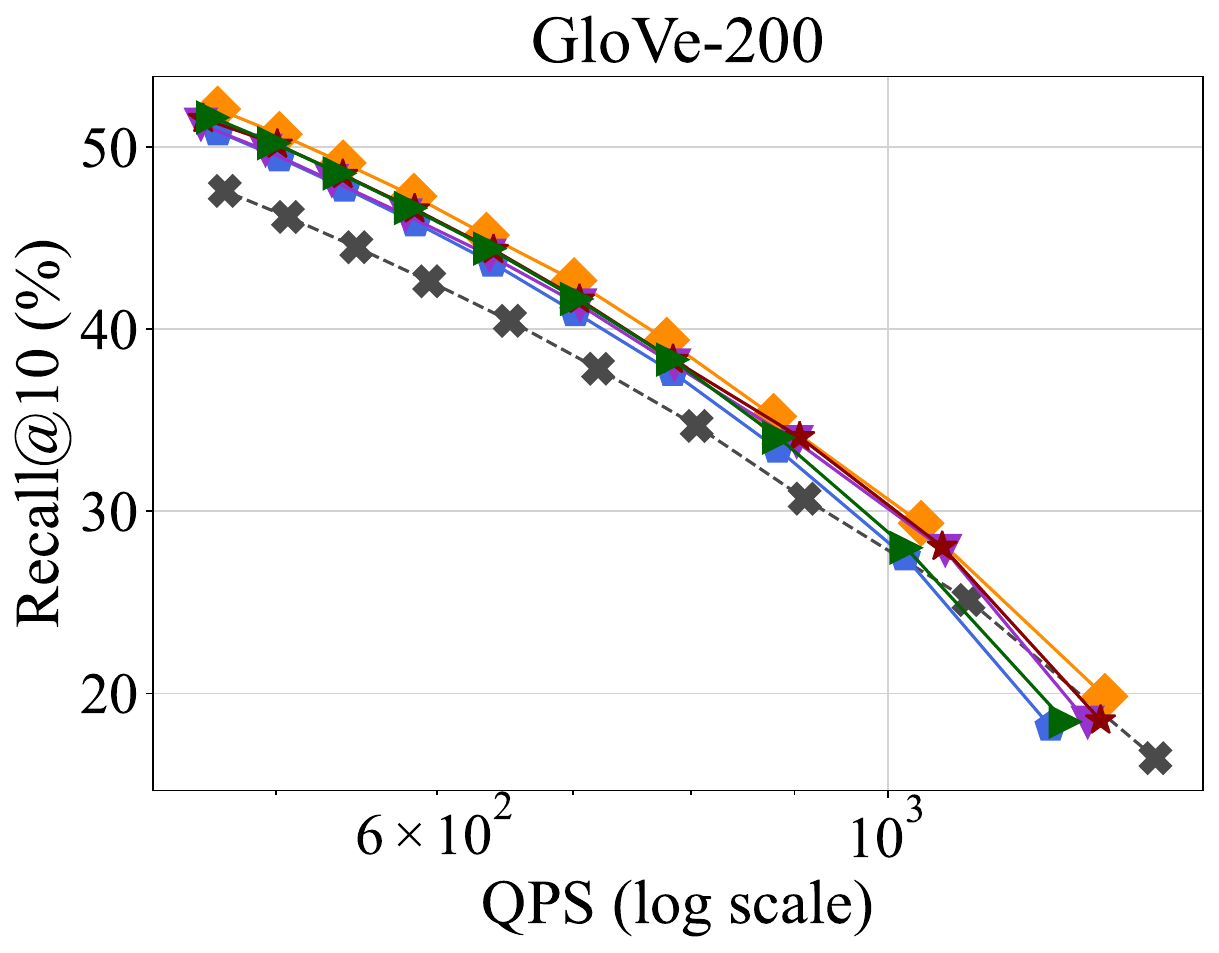}}
		\label{subfig:glove-200-angular_K10_var4}}
	\caption{\small Time-Accuracy Trade-off (Varying EnhanceGraph-specific generation parameter $\omega$).}
	\label{fig:var4}
	\vspace{-2ex}
\end{figure}

\noindent \underline{\textit{Results of Varying Generate Parameter $\omega$.}}
As shown in Figure \ref{fig:var4}, we demonstrate the effect of updating the conjugate graph using generated search logs collected with different parameters $\omega$. The VAMANA curve is a fixed reference because $\omega$ only affects the generated search logs used to update $G'$, while the VAMANA-SC curves vary $\omega$. For the simple dataset, GloVe-200, the $\omega$ setting has less impact on the final performance. However, for the more challenging dataset GIST1M, a smaller $\omega$ setting allows the conjugate graph to produce larger recall gains. This trend is consistent with the observation that smaller $\omega$ values place the constructed cases closer to Voronoi-cell boundaries, where local-optimum cases are more likely to appear.

\noindent \underline{\textit{Results of Varying Construction Parameter $L_1$.}}
We vary the VAMANA construction parameter $L_1$ to test whether EnhanceGraph still helps when the base graph is built with different construction budgets. The comparison focuses on how the baseline VAMANA curves and the corresponding VAMANA-SC curves change with $L_1$. Larger $L_1$ improves the base graph and the construction-log candidates, while smaller $L_1$ leaves more local-optimum failures for search-log repair. The complete curves are shown in
	Appendix B.

\noindent \underline{\textit{Log-Source Ablation.}}
Figure~\ref{fig:r2_log_source_ablation_main} separates $S_H$ (historical search logs), $S_G$ (generated search logs), $C$, their pairwise combinations, and $SC=S_H+S_G+C$ while keeping the same VAMANA base graph. $S_H$ and $S_G$ use historical search logs and generated search logs, respectively; $C$ uses construction-log candidates that were pruned from the base graph; and $SC$ combines all three sources. The search-log variants mainly improve $\Delta$Recall@1 by repairing local-optimum failures, whereas $C$ contributes more to $\Delta$Recall@10 by adding pruned near-neighbor candidates. $SC$ combines these effects and gives the largest observed $\Delta$Recall values across the evaluated search budgets.

\noindent \underline{\textit{Degree Sensitivity.}}
We vary the maximum graph degree $R$ to check whether the improvement depends on the single degree budget used in the main comparison. The experiment compares Baseline, S, C, and SC over $R\in\{8,12,16,24,32,48\}$ under search budgets $L=100$ and $L=200$. SC gives the largest gains on sparse and moderate-degree graphs, and the gains decrease as the base graph becomes stronger. EnhanceGraph can supplement the graph after the construction stage while keeping the original graph usable, whereas rebuilding a denser graph requires additional memory resources and construction time. The complete curves are shown in Appendix A.

\noindent \underline{\textit{Results of Varying Update Parameter $L_2$.}}
As shown in Figure \ref{fig:var3}, we examine the impact of different update parameters $L_2$ on the conjugate graph. The VAMANA curve is fixed because $L_2$ is only used to collect/update local-global repair pairs in $G'$, while the VAMANA-SC curves vary $L_2$. We discover that when $L_2<100$, a significant turning point appears on the curve at a specific point, precisely when the search parameter $L = L_2$. This is because we have set the maximum search size to $L=100$. When $L > L_2$, the recall rate drops as QPS decreases, which corresponds to an increase in $L$. This happens because we only utilize the local optimum that is reached when searching for the list size $L = L_2$ to connect to the global optimum. For $L > L_2$, the search process may converge to a better local optimum, which might not have an edge connected to the global optimum. Thus, this affects the improvements brought about by the conjugate graph. However, this provides a new insight: our method can approach the optimization cap even with a small search parameter $L$.

\begin{figure}[t!]\centering
	\subfloat{
		\scalebox{0.2}[0.2]{\includegraphics{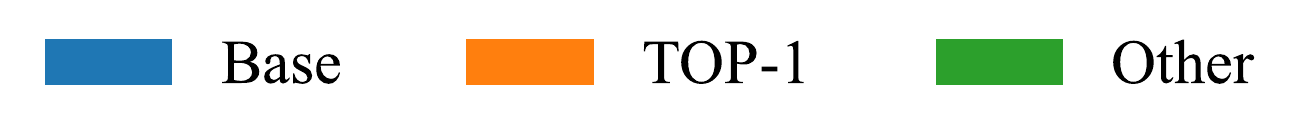}}}\hfill\\
	\vspace{-2ex}
	\addtocounter{subfigure}{-1}
	\figureBelowMargin
	
	\subfloat[][{\scriptsize VAMANA(GIST1M)}]{
		\scalebox{0.18}[0.18]{\includegraphics{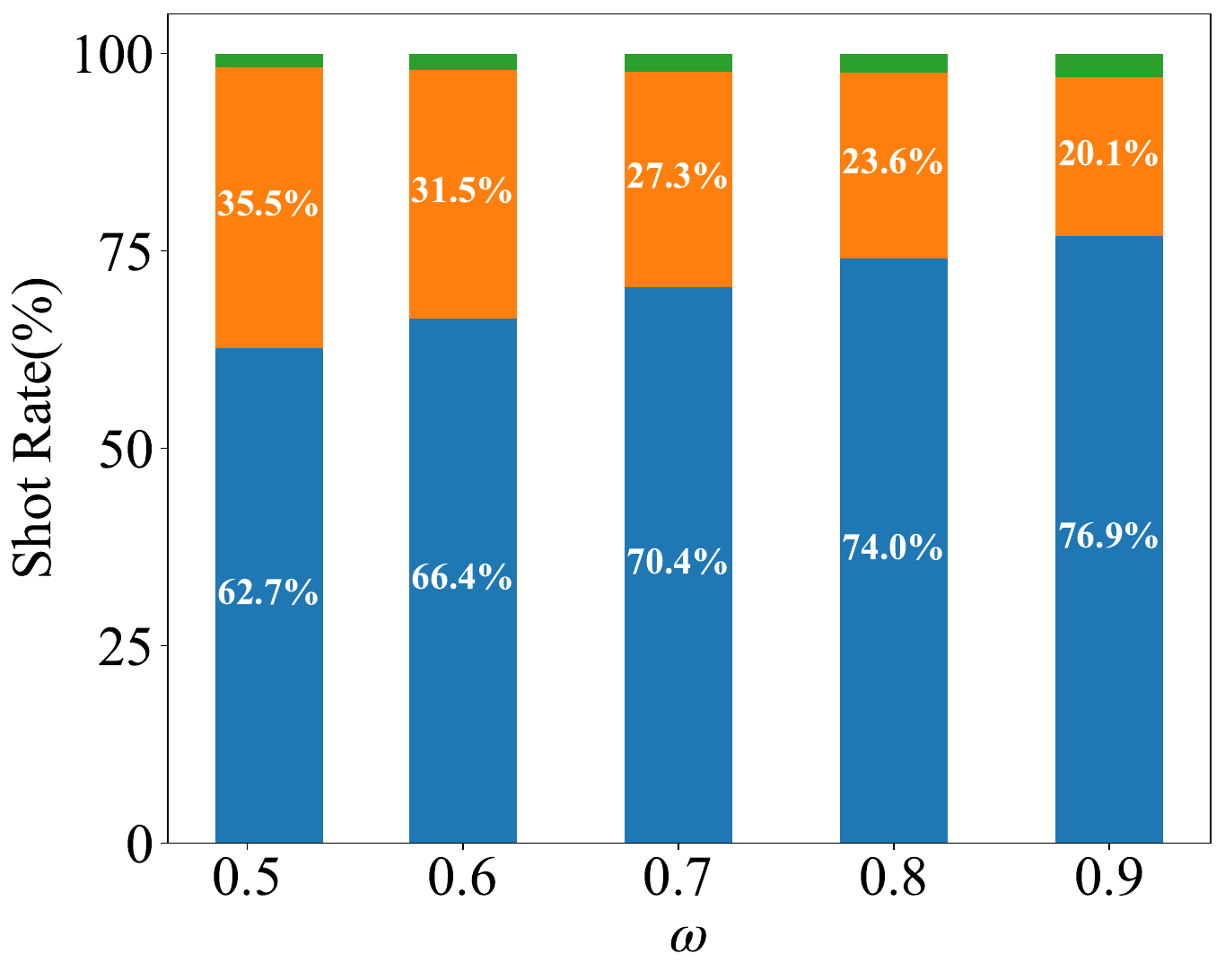}}
		\label{subfig:shot_rate_VAMANA_GIST1M}}
	\hspace{-1em}
	\subfloat[][{\scriptsize VAMANA(GLoVe-100)}]{
		\scalebox{0.18}[0.18]{\includegraphics{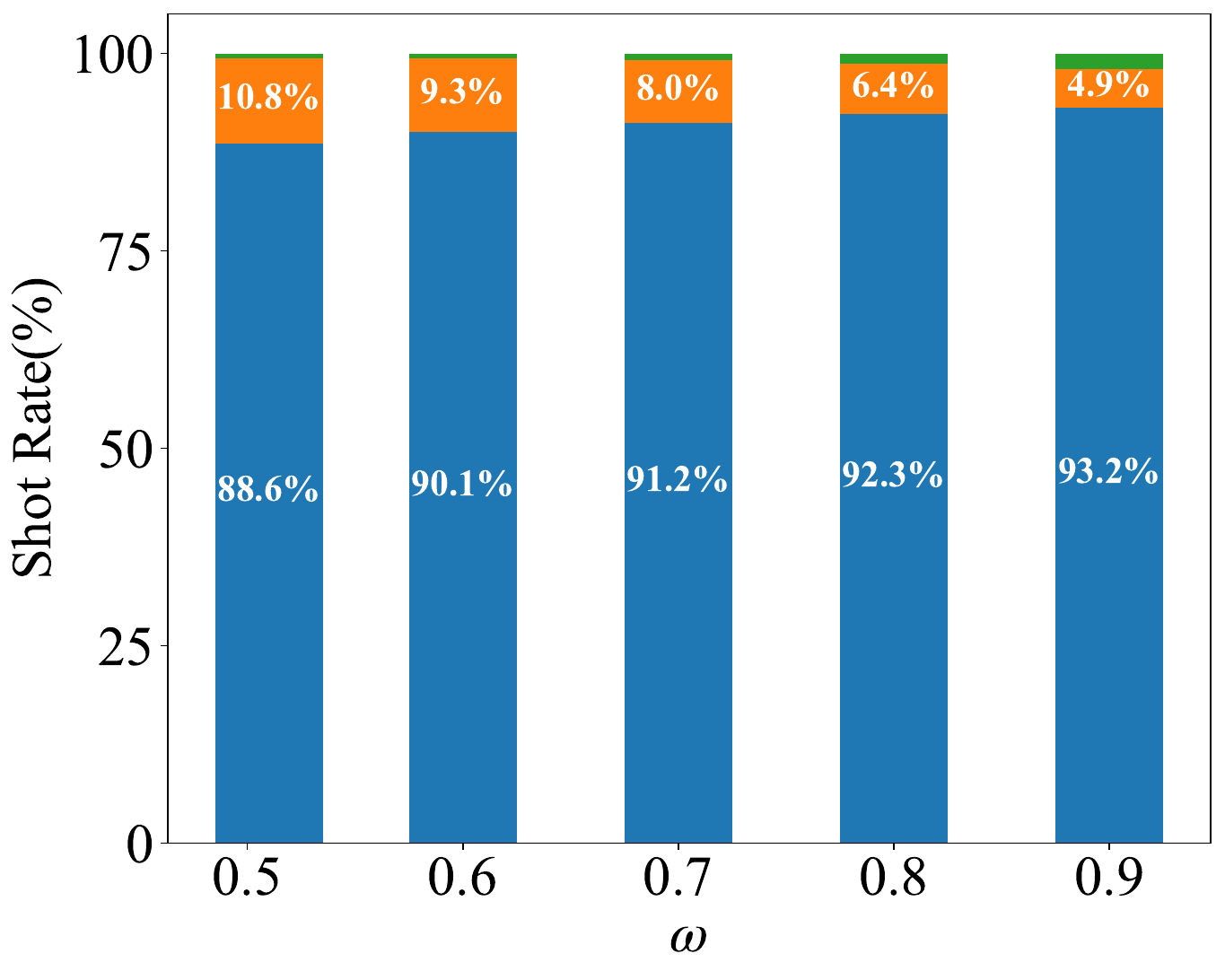}}
		\label{subfig:shot_rate_VAMANA_GLoVe-100}}
	\figureBelowMargin
	
	\subfloat[][{\scriptsize HNSW(GIST1M)}]{
		\scalebox{0.18}[0.18]{\includegraphics{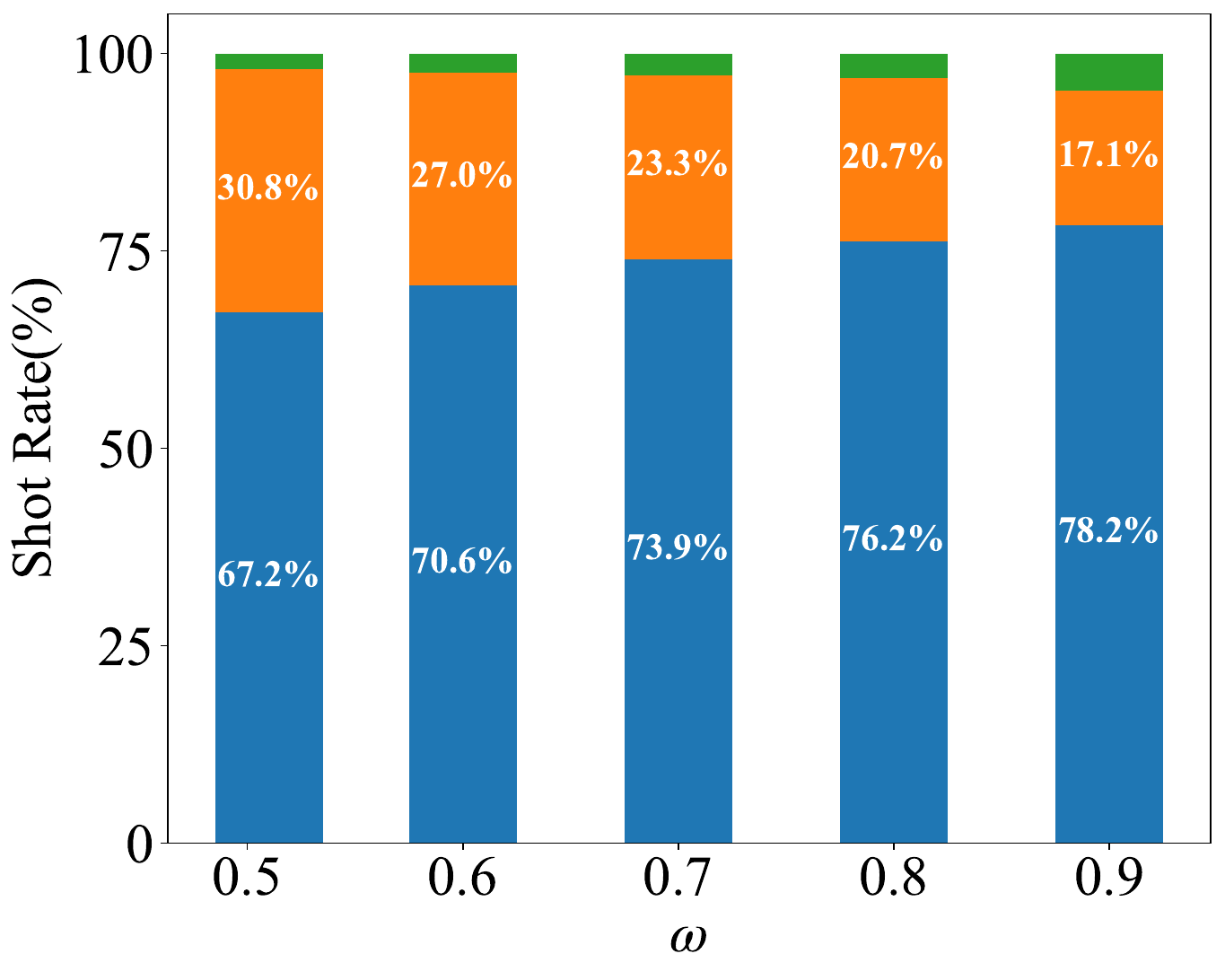}}
		\label{subfig:shot_rate_HNSW_GIST1M}}
	\hspace{-1em}
	\subfloat[][{\scriptsize HNSW(GLoVe-100)}]{
		\scalebox{0.18}[0.18]{\includegraphics{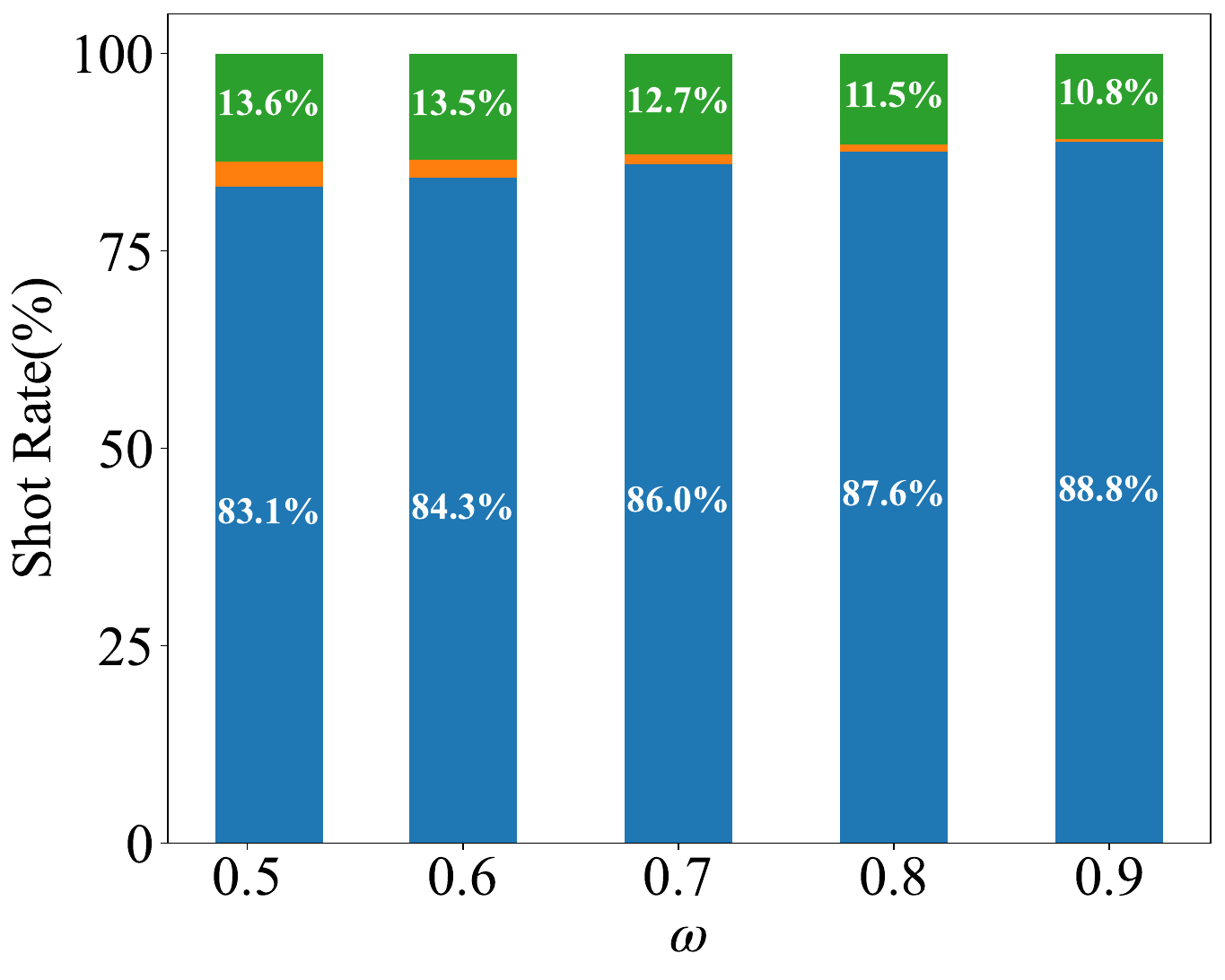}}
		\label{subfig:shot_rate_HNSW_GLoVe-100}}
	\caption{\small Shot Rate on Different Queries (Varying $\omega$)}
	\label{fig:shot_rate}
	\vspace{-2ex}
\end{figure}

\noindent \underline{\textit{Results of Shot Rate (Varying Generate Parameter $\omega$).}}
As shown in Figure \ref{fig:shot_rate}, we construct generated points by adjusting $\omega$ near the middle vertical line of each base point and its nearest neighbor, and then collect generated search logs by searching for their local optima with VAMANA and HNSW on GIST1M and GLoVe-100. The blue bars indicate the rate of reaching the global optimum, the orange bars represent the rate at which the local optimum is the base point's nearest neighbor, and the green bars show the rate of reaching other points. We find that the local optimum is either the base point or the base point's nearest neighbor in over 95\% of the points in VAMANA. Many points near the middle vertical line struggle to reach the global optimum (i.e., corresponding base point).

As $\omega$ increases, more points reach the global optimum in search process. Most points that do not reach the global optimum in VAMANA end up reaching its nearest neighbor, so many conjugate-graph edges are from the global optimum's nearest neighbor to the global optimum, implying that the number of local optimum is finite. In contrast, HNSW converges to other points more frequently because its hierarchy chooses different starting points in the bottom graph for each query, leading to varied convergence results.

\noindent \underline{\textit{Summary of the Experimental Results:}}
EnhanceGraph raises $Recall@1$ from $41.74\%$ to $93.42\%$ on GIST1M and exploits more than 90\% shared local optima in GLoVe-100. In the KNNG comparison, HNSW-SC reaches 93.24\% Recall@1 versus 49.41\% for KNNG on GIST1M, with lower build time (3420s vs. 16483s) and comparable space overhead (3817.11MB vs. 3802.11MB). Under high recall requirements, it reaches $99.9\%$ from $99.8\%$ with separately measured supplementary overhead.

\section{Related Work}
\label{sec:related}

\noindent\underline{\textit{Graph-based indexes.}}
The Relative Neighborhood Graph (RNG)~\cite{Godfried1980RNG} and Sparse Neighborhood Graph (SNG)~\cite{arya1993SNG} approximate the $k$-NNs graph and add shortcuts by using specific edge pruning strategy. The Diversified Proximity Graph (DPG)~\cite{Li2019DPG} is based on RNG and distributes edges over the angle. However, the edge pruning strategy of RNG is too tight to make it sufficient to be a MSNET~\cite{fu2019NSG}. The Monotonic Relative Neighborhood Graph (MRNG)~\cite{fu2019NSG} uses a looser strategy to reserve more navigation edges and ensure a Monotonic Decrease Path~\cite{fu2019NSG} from any source node to any target node.

Constructing theoretical structures such as RNG or MRNG incurs high cost, so engineering indexes approximate them to reduce construction cost.
The Navigating Spreading-out Graph (NSG)~\cite{fu2019NSG} uses MRNG-style edge pruning and limits the MSNET property by searching from a fixed node. Subramanya \textit{et al.} proposed Vamana~\cite{Jayaram2019diskann, Gollapudi2023filterDISKANN, jaiswal2022ooddiskann} as an approximation of NSG with a looser strategy. $\tau$-MG~\cite{peng2023taumg} is optimized based on MRNG and constrains both short and long edges. HNSW~\cite{Malkov2020hnsw} is a hierarchical graph index: the lowest level is a complete NSW, while higher levels quickly locate the highway near the query point. DEG~\cite{hezel2023deg} studies continuous base-graph refinement through edge replacement during construction.

\noindent\underline{\textit{Partition-based indexes.}}
Nowadays, various methods divide the data space into subspaces with specific properties, so vectors within a subspace can be searched efficiently. Tree-based methods like QD-tree~\cite{yang2020QDTree, Apple2023filter, xie2023sat} and KD-tree~\cite{kaiming2012kdtree, ram2019kdtree, Pinkham2020kdtree} create hierarchical structures whose leaves represent subspaces. Hash-based methods, like LSH~\cite{lv2007multi, andoni2015practical, zhao2023LSHAPG}, assign similar points to balanced buckets, and IVF~\cite{chen2010approximate,pag} uses clustering to partition buckets. However, boundary points may require searching many clusters simultaneously, and high dimensionality makes it challenging to utilize similarity information within delineated subspaces~\cite{koppen2000curse, verleysen2005curse}.

\noindent\underline{\textit{Compression-based indexes.}}
Computation for the distance of high dimensional data is a significant bottleneck for most existing methods. Recent studies compress the original data into a lower-dimensional space to reduce distance computation. Product Quantization (PQ)~\cite{jegou2010product, ge2013optimized} clusters vector segments and maps them to center IDs, allowing quick approximate distance computation using pre-computed center-to-center distances. ADsampling~\cite{gao2023adsample} combines stochastic mapping and early stop to provide correct distance-comparison results with high probability.

\section{Conclusion}
\enlargethispage{2\baselineskip}

\vspace{-1ex}
\label{sec:conclusion}
In this paper, we introduce EnhanceGraph to address the issue that construction and search logs have not been well utilized in existing graph-based indexes for ANNS. EnhanceGraph enhances deployed graph-based indexes with a log-driven auxiliary conjugate graph, which is consulted after the proximity-graph search to recover potential global optima and supplement missing $k$-NNs. This approach achieves the largest observed \textit{$Recall@1$} increase from 41.74\% to 93.42\% with limited supplementary-stage overhead.

\section{Acknowledgment}
This work is supported by New Generation Artificial Intelligence-National Science and Technology Major Project (No. 2025ZD0123000), Fundamental and Interdisciplinary Disciplines Breakthrough Plan of the Ministry of Education of china (No. JYB2025XDXM103), Shanghai Central and Local Science and Technology Development Fund Project (Grant No. YDZX20253100002004), Ant Group, Fundamental Research Funds for the Central Universities,  the New Cornerstone Science Foundation through the XPLORER PRIZE. 

\bibliographystyle{ieeetr}
\bibliography{add}
\balance

\newpage
\appendices
\section{Maximum-Degree Sensitivity}
\label{app:degree_sensitivity}

\begin{figure*}[th!]
	\centering
	\includegraphics[width=0.94\textwidth]{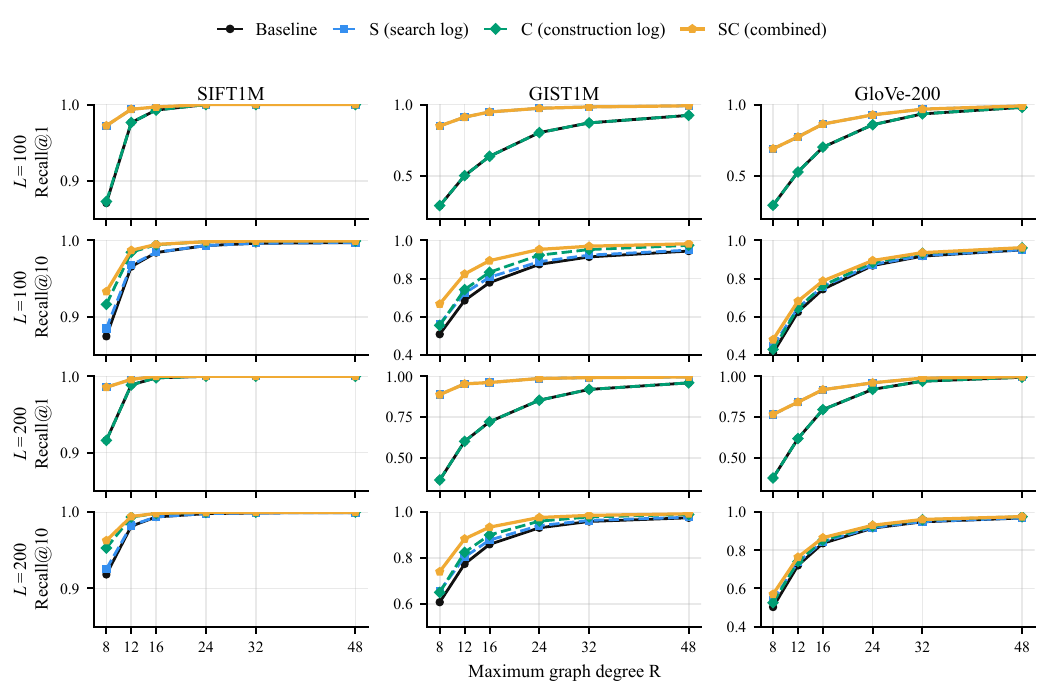}
	\caption{\small Recall sensitivity to the maximum graph degree $R$ at search budgets $L=100$ and $L=200$. Rows distinguish the two search budgets and the Recall@1/Recall@10 metrics; columns show the three datasets.}
	\label{fig:degree_sensitivity_l100}
	\label{fig:degree_sensitivity_l200}
	\vspace{-1.5ex}
\end{figure*}

We evaluate Baseline, S, C, and SC while varying the maximum graph degree over $R\in\{8,12,16,24,32,48\}$. Figure~\ref{fig:degree_sensitivity_l100} reports Recall@1 and Recall@10 at fixed search budgets $L=100$ and $L=200$ on SIFT1M, GIST1M, and GloVe-200. The gains are largest on sparse graphs and decrease as the base graph becomes stronger, while SC remains beneficial on the more challenging datasets.

At $L=100$, SC improves Recall@1/Recall@10 on GIST1M by $+0.170/+0.078$ at $R=24$ and by $+0.067/+0.038$ at $R=48$. On GloVe-200, the corresponding gains are $+0.069/+0.027$ at $R=24$ and $+0.010/+0.012$ at $R=48$. At $L=200$ and $R=24$, SC improves Recall@1/Recall@10 by $+0.133/+0.045$ on GIST1M and by $+0.040/+0.015$ on GloVe-200. Recall@1 on SIFT1M is already saturated at denser settings, leaving mainly Recall@10 gains.
\FloatBarrier

\FloatBarrier

\section{Additional Parameter Analyses}
\label{app:additional_parameters}

\begin{figure}[!h]
	\centering
	\subfloat{\makebox[\columnwidth][c]{\scalebox{0.31}[0.31]{\includegraphics{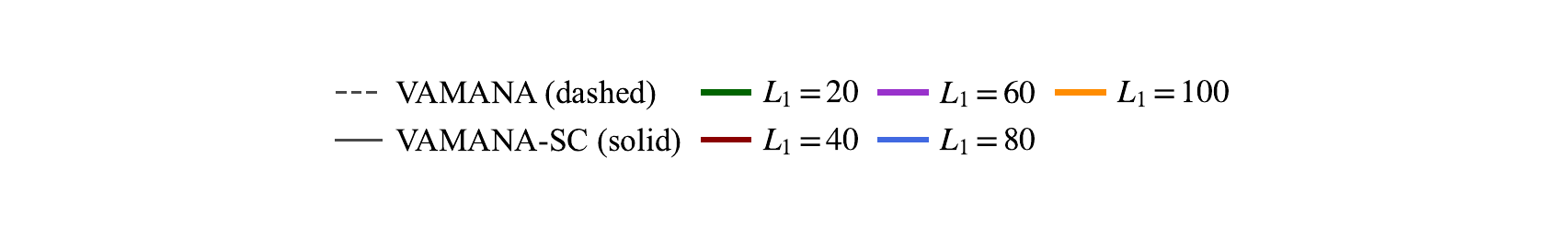}}}}\\[-1.5ex]
	\addtocounter{subfigure}{-1}
	\subfloat[][{\scriptsize GIST1M Recall@1}]{
		\scalebox{0.15}[0.15]{\includegraphics{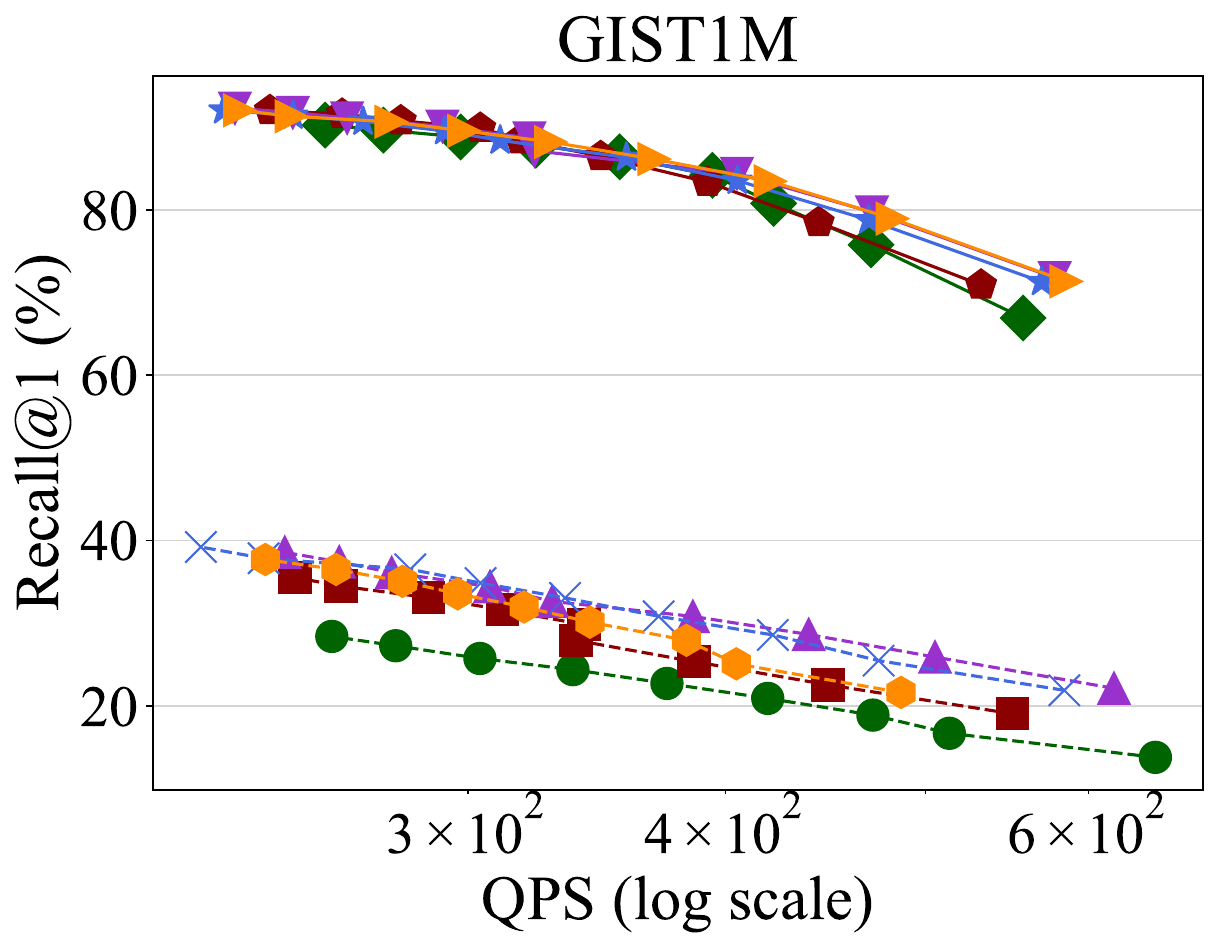}}
		\label{subfig:gist-960-euclidean_K1_var2}}
	\subfloat[][{\scriptsize GloVe-200 Recall@1}]{
		\scalebox{0.15}[0.15]{\includegraphics{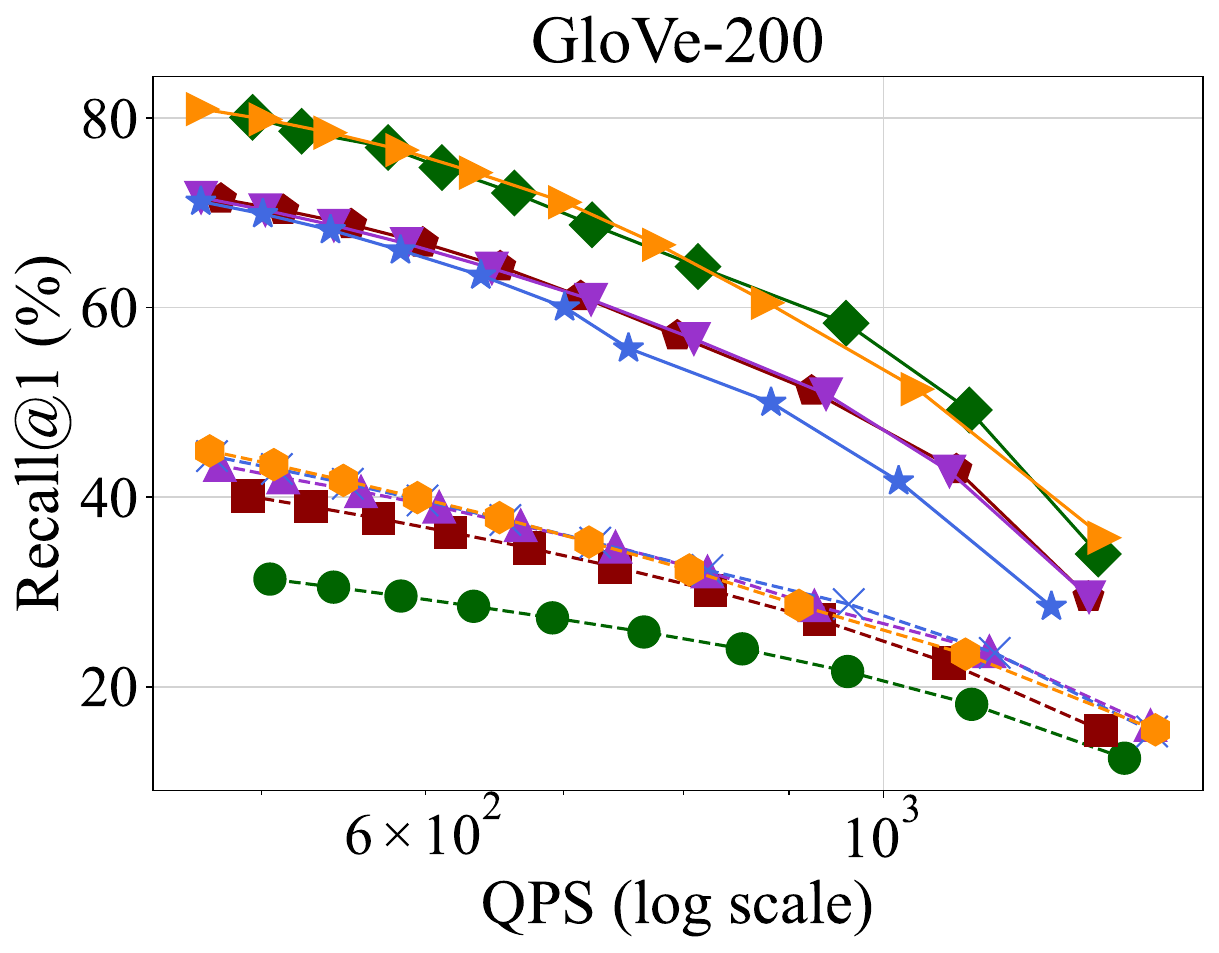}}
		\label{subfig:glove-200-angular_K1_var2}}\\[-0.5ex]
	\subfloat[][{\scriptsize GIST1M Recall@10}]{
		\scalebox{0.15}[0.15]{\includegraphics{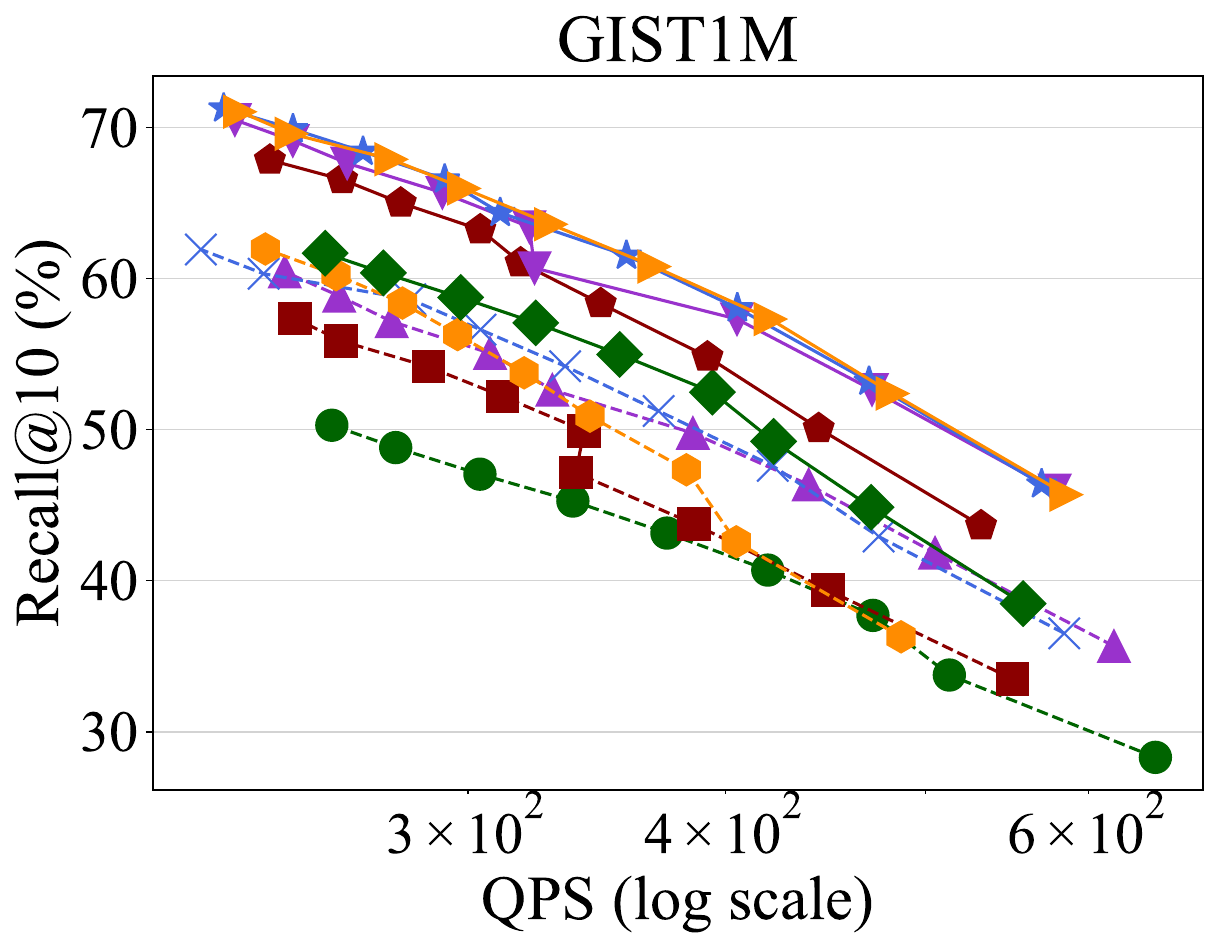}}
		\label{subfig:gist-960-euclidean_K10_var2}}
	\subfloat[][{\scriptsize GloVe-200 Recall@10}]{
		\scalebox{0.15}[0.15]{\includegraphics{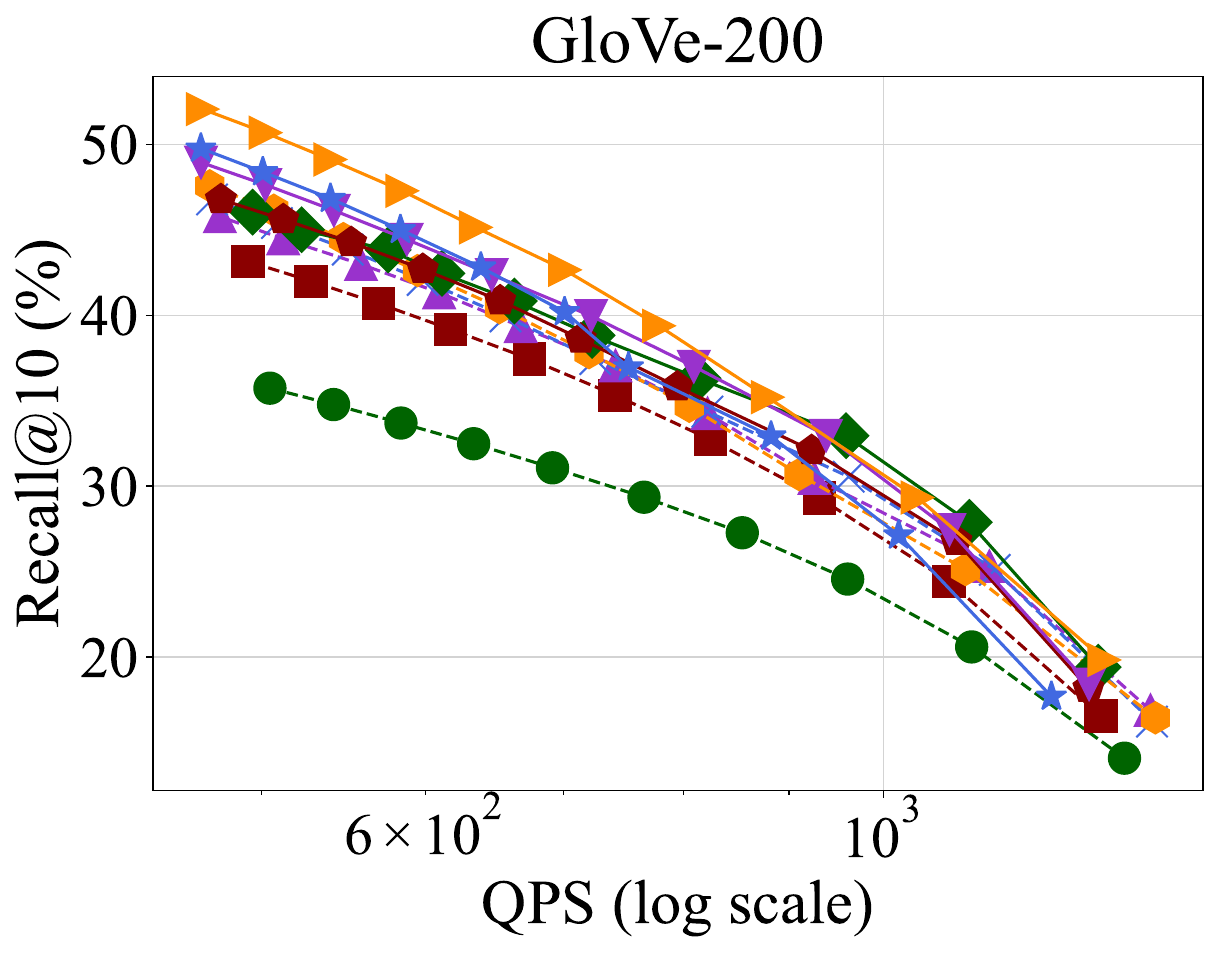}}
		\label{subfig:glove-200-angular_K10_var2}}
	\caption{\small Time-accuracy trade-off when varying the VAMANA construction parameter $L_1$.}
	\label{fig:var2}
	\vspace{-1.5ex}
\end{figure}

\noindent \underline{\textit{Construction Parameter $L_1$.}}
As shown in Figure~\ref{fig:var2}, a larger $L_1$ improves the quality of both the base VAMANA graph and the construction-log candidates retained in $G'$. The resulting union is closer to an accurate $k$-NN graph and therefore yields the clearest gain in Recall@10. At smaller $L_1$, the weaker base graph leaves more local-optimum failures, so the relative Recall@1 gain from the conjugate graph is larger.
\FloatBarrier

\section{\texorpdfstring{Supplementary-Stage Overhead}{Supplementary-Stage Overhead}}
\label{app:r2_supplementary_overhead}

\begin{figure}[!h]
	\centering
	\includegraphics[width=0.99\columnwidth]{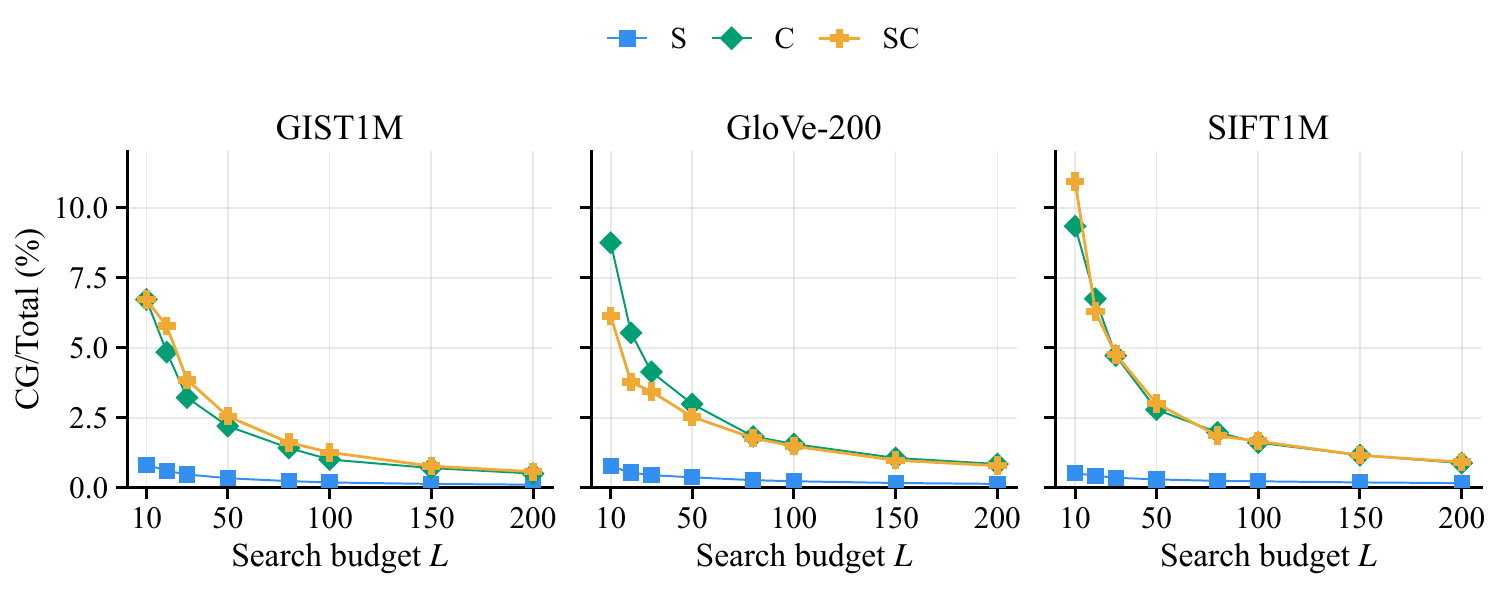}
	\caption{\small Supplementary-stage runtime as a percentage of total query time. S, C, and SC use search logs, construction logs, and both sources, respectively.}
	\label{fig:r2_supplementary_overhead}
	\vspace{-1.5ex}
\end{figure}

\begin{table}[!h]
	\begin{center}
		{
			\caption{\small SC overhead and benefited-query fractions at $L=100$.}\vspace{-2ex}
			\label{tab:r2_supplementary_overhead}
				\begin{tabular}{l|l|l|l|l}
					{\bf Dataset} & {\bf CG/Total} & {\bf Dist./Query} & {\bf Top1} & {\bf Top10} \\ \hline \hline
					GIST1M & 1.267\% & 3.787 & 16.39\% & 25.15\% \\
					GloVe-200 & 1.488\% & 3.208 & 7.00\% & 9.10\% \\
					SIFT1M & 1.664\% & 0.808 & 0.00\% & 1.33\% \\
					\hline
			\end{tabular}
		}
	\end{center}
\end{table}

We directly measure the supplementary conjugate-graph stage rather than inferring its cost from end-to-end QPS. Figure~\ref{fig:r2_supplementary_overhead} reports the stage-time fraction, CG/Total, across search budgets. Its relative cost decreases as the base-graph search budget $L$ increases. Search-log repair edges (S) are cheaper to check than the larger construction-log candidate sets used by C and SC.

At $L=100$, SC accounts for 1.267\%--1.664\% of total query time and introduces 0.808--3.787 additional distance computations per query. The benefited-query fractions are empirical observations on the historical/generated anchor-noise workload, not distribution-free guarantees. Across all evaluated budgets, one SIFT1M query (0.10\%) showed a lower Top10 score in several settings; no Top1 harmed query was observed.
\FloatBarrier

\end{document}